\documentclass[preprint,12pt]{elsarticle}

\usepackage{graphicx}
\usepackage{amssymb}
\usepackage{amsthm}

\journal{Journal of Computational Physics}

\begin{document}

\begin{frontmatter}

%\title{Wave propagation in 2D heterogeneous porous media with regular viscous efforts}

\title{Time domain numerical modeling of wave propagation in 2D heterogeneous porous media }

\author[M2P2]{Guillaume Chiavassa\corref{cor1}}
\ead{guillaume.chiavassa@centrale-marseille.fr}
\author[LMA]{Bruno Lombard}
\ead{lombard@lma.cnrs-mrs.fr}
\cortext[cor1]{Corresponding author. Tel.: +33 491 05 47 85.}
\address[M2P2]{M2P2, UMR 6181 - CNRS - Ecole Centrale Marseille,
Technop\^ole de Chateau-Gombert, 38 rue Fr\'ed\'eric Joliot-Curie, 13451 Marseille, France}
\address[LMA]{Laboratoire de M\'{e}canique et d'Acoustique, UPR 7051 CNRS, 31 chemin Joseph Aiguier, 13402 Marseille, France}

\begin{abstract}
This paper deals with the numerical modeling of wave propagation in porous media described by Biot's theory. The viscous efforts between the fluid and the elastic skeleton are assumed to be a linear function of the relative velocity, which is valid in the low-frequency range. The coexistence of propagating fast compressional wave and shear wave, and of a diffusive slow compressional wave, makes numerical modeling tricky. To avoid restrictions on the time step, the Biot's system is  splitted into two parts: the propagative part is discretized by a fourth-order ADER scheme, while the diffusive part is solved analytically. Near the material interfaces, a space-time mesh refinement is implemented to capture the small spatial scales related to the slow compressional wave. The jump conditions along the  interfaces are discretized by an immersed interface method. Numerical experiments and comparisons with exact solutions confirm  the accuracy of the numerical modeling. The efficiency of the approach is illustrated by simulations of multiple scattering.
\end{abstract}

\begin{keyword}
porous media \sep elastic waves \sep Biot's model \sep time splitting \sep finite difference methods \sep Cartesian grid \sep immersed interface method 
\MSC 35L50       % BVP hyperbolic
\sep 65M06       % finite-difference methods
\PACS 43.20.-Gp  % reflection, refraction, diffraction, scattering of elastic and poroelastic waves
\sep 46.40.-f    % vibration and mechanical waves

\end{keyword}

\end{frontmatter}

%% \linenumbers

%------------------------------------------------------------------------------------------

\section{Introduction}\label{SecIntro}

The propagation of waves in porous media has crucial implications in many areas, such as the characterization of industrial foams, spongious bones and petroleum rocks. The most widely used model describing the propagation of mechanical waves in a saturated porous medium was proposed by Biot in 1956. A major achievement in Biot's theory was the prediction of a second (slow) compressional wave, besides the (fast) compressional wave and the shear wave classically propagated in elastic media. 

Two regimes are distinguished, depending on the frequency of the waves. At frequencies smaller than a critical frequency $f_c$, the fluid flow inside the pores is of Poiseuille type, and the viscous efforts between the fluid and the solid depend linearly on the relative velocity. In this case, the slow compressional wave is almost static and highly attenuated \cite{BIOT56-A}. An adequate modeling of this diffusive mode remains a major challenge in real applications. At frequencies greater than $f_c$, inertial effects begin to dominate the shear forces, resulting in an ideal flow profile except in the viscous boundary layer, and the slow wave propagates \cite{BIOT56-B,MASSON06}. Experimental observations of the slow wave in the low-frequency range \cite{PLONA80} and in the high-frequency range \cite{CHANDLER81} have confirmed  Biot's theory. In the current study, we focus on the low-frequency range.

Until the 1990's, Biot's equations were mainly studied in the harmonic regime. Various time-domain methods have been proposed since, based on finite differences \cite{DAI95,ZENG01,WENZLAU09}, finite elements \cite{ZHAO05}, discontinuous Galerkin methods \cite{PUENTE08}, boundary elements \cite{SCHANZ01}, pseudospectral methods \cite{CARCIONE07} and spectral element methods \cite{MORENCY08}. A recent review of computational poroelasticity can be found in \cite{CARCIONE10}. Nevertheless, none of the methods proposed in the low-frequency range give a complete answer to the following difficulties:
\begin{itemize}
\item the viscous effects greatly influence  numerical stability, imposing a restrictive time step. In some physically relevant cases, computations cannot be carried out in a reasonable time;
\item the wavelength of the slow compressional wave is much smaller than that of the other waves. Consequently, one faces the following alternative: either a coarse grid well-suited to the fast wave is chosen, and the slow wave is badly discretized; either a fine mesh is used, and the computational cost increases dramatically;
\item maximum computational efficiency is obtained on a Cartesian grid; in counterpart, the interfaces are coarsely discretized, which yields spurious solutions. Alternatively, unstructured meshes adapted to the interfaces provide accurate description of geometries and jump conditions; however, the computational effort greatly increases, due to the cost of the mesh generation and to the CFL condition of stability.
\end{itemize}

The aim of the present study is to develop an efficient numerical strategy to remove these drawbacks. A time-splitting is used along with a fourth-order ADER scheme \cite{SCHWARTZKOPFF04} to integrate  Biot's equations. A flux-conserving space-time mesh refinement \cite{BERGER98} is implemented around the interfaces to capture the slow compressional wave. Lastly, an immersed interface method \cite{LOMBARD04,LOMBARD06} is developed to provide a subcell resolution of the interfaces and to accurately enforce the jump conditions between the different porous media. As illustrated by the simulations, the combination of these numerical methods highlights the importance of an accurate modeling of the slow wave.
 
This article, which generalizes a previous one-dimensional work \cite{CHIAVASSA10}, is organized as follows. Biot's model is briefly recalled in section 2. The numerical methods are described in section 3. Section 4 presents numerical experiments and comparisons with exact solutions. In section 5, conclusions are drawn and future perspectives are suggested.

%------------------------------------------------------------------------------------------
%------------------------------------------------------------------------------------------

\section{Physical modeling}\label{SecPhysMod}

\subsection{Biot's model}\label{SecPhysBiot}

Biot's model describes the propagation of mechanical waves in a porous medium consisting of a solid matrix saturated with fluid circulating freely through the pores \cite{BIOT56-A,BOURBIE87,CARCIONE07, CARCIONE10}. It is assumed that
\begin{itemize}
\item the wavelengths are large compared with the diameter of the pores;
\item the amplitudes of perturbations are small;
\item the elastic and isotropic matrix is fully saturated by a single fluid phase.
\end{itemize}
This model relies on 10 physical parameters: the density $\rho_f$ and the dynamic viscosity $\eta$ of the fluid; the density $\rho_s$ and the shear modulus $\mu$ of the elastic skeleton; the porosity $0<\phi<1$, the tortuosity $a\geq 1$, the absolute permeability $\kappa$, the Lam\'e coefficient $\lambda_f$ and the two Biot's coefficients $\beta$ and $m$ of the saturated matrix. The unknowns are the elastic and acoustic displacements ${\bf u}_s$ and ${\bf u}_f$, the  elastic stress tensor ${\bf \sigma}$, and the acoustic pressure $p$. In one hand, the constitutive laws  are:
\begin{equation}
\left\{
\begin{array}{l}
\displaystyle
{\bf \sigma}=\left(\lambda_f\,\mbox{ tr }{\bf \varepsilon}-\beta\,m\,\xi\right)\,{\bf I}+2\,\mu\, {\bf \varepsilon},\\
[8pt]
\displaystyle
p=m\,\left(-\beta\,\mbox{ tr }{\bf \varepsilon}+\xi\right),
\end{array}
\right.
\label{BiotComport}
\end{equation}
where ${\bf I}$ is the identity, $\xi$ is the rate of fluid change, and $\varepsilon$ is the strain tensor
\begin{equation}
\begin{array}{l}
\displaystyle
\xi=- {\bf \nabla}.\,\left(\phi\,({\bf u}_f-{\bf u}_s)\right),
\qquad
\displaystyle
{\bf \varepsilon}=\frac{\textstyle 1}{\textstyle 2}\,\left({\bf \nabla}\,{\bf u}_s+{\bf \nabla}\,{\bf u}_s^T\right).
\end{array}
\label{HPP}
\end{equation}
The symmetry of ${\bf \sigma}$ in (\ref{BiotComport}) implies compatibility conditions between spatial derivatives of ${\bf \varepsilon}$, leading to the Beltrami-Michell equation \cite{RICE76,COUSSY95}
\begin{equation}
\begin{array}{l}
\displaystyle
\frac{\textstyle \partial^2 \,\sigma_{12}}{\textstyle \partial \,x\,\partial\,y}
=\theta_0\,\frac{\textstyle \partial^2 \,\sigma_{11}}{\textstyle \partial \,x^2}
+\theta_1\,\frac{\textstyle \partial^2 \,\sigma_{22}}{\textstyle \partial \,x^2}
+\theta_2\,\frac{\textstyle \partial^2 \,p}{\textstyle \partial \,x^2}
+\theta_1\,\frac{\textstyle \partial^2 \,\sigma_{11}}{\textstyle \partial \,y^2}
+\theta_0\,\frac{\textstyle \partial^2 \,\sigma_{22}}{\textstyle \partial \,y^2}
+\theta_2\,\frac{\textstyle \partial^2 \,p}{\textstyle \partial \,y^2},\\
[10pt]
\displaystyle
\theta_0=-\frac{\textstyle \lambda_0}{\textstyle 4\,(\lambda_0+\mu)},\quad \theta_1=\frac{\textstyle \lambda_0+2\,\mu}{\textstyle 4\,(\lambda_0+\mu)},\quad \theta_2=\frac{\textstyle \mu \,\beta}{\textstyle 2\,(\lambda_0+\mu)},
\end{array}
\label{Barre}
\end{equation}
where 
$\lambda_0=\lambda_f-\beta^2\,m$ is the Lam\'e coefficient of the dry matrix. 

On the other hand, the conservation of momentum yields
\begin{equation}
\left\{
\begin{array}{l}
\displaystyle
\rho\,\frac{\textstyle \partial\,{\bf v}_s}{\textstyle \partial\,t}+\rho_f\,\frac{\textstyle \partial\,{\bf w}}{\textstyle \partial\,t}=\nabla {\bf \sigma},\\
[8pt]
\displaystyle
\rho_f\,\frac{\textstyle \partial\,{\bf v}_s}{\textstyle \partial\,t}+\rho_w\,\frac{\textstyle \partial\,{\bf w}}{\textstyle \partial\,t}+\frac{\textstyle \eta}{\textstyle \kappa}\,{\bf w}=-\nabla p,
\end{array}
\right.
\label{BiotMomentum}
\end{equation}
where ${\bf v_s}=\frac{\partial {\bf u}_s}{\partial t}=(v_{s1},\,v_{s2})^T$ is the elastic velocity, and ${\bf w}=\phi\,\frac{\partial }{\partial \,t}\,({\bf u}_f-{\bf u}_s)=(w_1,\,w_2)^T$ is the filtration velocity. To be valid, the second equation of (\ref{BiotMomentum}) requires that the spectrum of the waves lies mainly in the low-frequency range, involving frequencies lower than 
\begin{equation}
f_c=\frac{\textstyle \eta\,\phi}{\textstyle 2\,\pi\,a\,\kappa\,\rho_f}.
\label{Fc}
\end{equation}
If $f\geq f_c$, more sophisticated models are required \cite{BIOT56-B,LU05}. In practice, the viscosity of the fluid is always non-zero; nevertheless, considering $\eta=0$ can be relevant for two reasons:
\begin{itemize}
\item if $f\gg f_c$, the viscous forces are smaller than the inertial forces \cite{FENG83a,MORENCY08} and can be neglected to a first approximation;
\item the exact solutions of poro-elastodynamic equations are computed more accurately if the saturating fluid is inviscid, which is attractive to validate the numerical methods.
\end{itemize}

%------------------------------------------------------------------------------------------

\subsection{Evolution equations}\label{SecPhysEvol}

A velocity-stress formulation is followed: from (\ref{BiotComport}) and (\ref{BiotMomentum}), we obtain the system of PDEs 
\begin{equation}
\left\{
\begin{array}{l}
\displaystyle 
\frac{\textstyle \partial \,v_{s1}}{\textstyle \partial \,t}-\frac{\textstyle \rho_w}{\textstyle \chi}\left(\frac{\textstyle \partial\,\sigma_{11}}{\textstyle \,\partial\, x}+\frac{\textstyle \partial\,\sigma_{12}}{\textstyle \,\partial\, y}\right)-\frac{\textstyle \rho_f}{\textstyle \chi}\,\frac{\textstyle \partial\,p}{\textstyle \,\partial\, x}=\frac{\textstyle \rho_f}{\textstyle \chi}\,\frac{\textstyle \eta}{\textstyle \kappa}\,w_1,\\
[12pt]
\displaystyle 
\frac{\textstyle \partial \,v_{s2}}{\textstyle \partial \,t}-\frac{\textstyle \rho_w}{\textstyle \chi}\left(\frac{\textstyle \partial\,\sigma_{12}}{\textstyle \,\partial\, x}+\frac{\textstyle \partial\,\sigma_{22}}{\textstyle \,\partial\, y}\right)-\frac{\textstyle \rho_f}{\textstyle \chi}\,\frac{\textstyle \partial\,p}{\textstyle \,\partial\, y}=\frac{\textstyle \rho_f}{\textstyle \chi}\,\frac{\textstyle \eta}{\textstyle \kappa}\,w_2,\\
[12pt]
\displaystyle 
\frac{\textstyle \partial \,w_{1}}{\textstyle \partial \,t}+\frac{\textstyle \rho_f}{\textstyle \chi}\left(\frac{\textstyle \partial\,\sigma_{11}}{\textstyle \,\partial\, x}+\frac{\textstyle \partial\,\sigma_{12}}{\textstyle \,\partial\, y}\right)+\frac{\textstyle \rho}{\textstyle \chi}\,\frac{\textstyle \partial\,p}{\textstyle \,\partial\, x}=-\frac{\textstyle \rho}{\textstyle \chi}\,\frac{\textstyle \eta}{\textstyle \kappa}\,w_1,\\
[12pt]
\displaystyle 
\frac{\textstyle \partial \,w_{2}}{\textstyle \partial \,t}+\frac{\textstyle \rho_f}{\textstyle \chi}\left(\frac{\textstyle \partial\,\sigma_{12}}{\textstyle \,\partial\, x}+\frac{\textstyle \partial\,\sigma_{22}}{\textstyle \,\partial\, y}\right)+\frac{\textstyle \rho}{\textstyle \chi}\,\frac{\textstyle \partial\,p}{\textstyle \,\partial\, y}=-\frac{\textstyle \rho}{\textstyle \chi}\,\frac{\textstyle \eta}{\textstyle \kappa}\,w_2,\\
[12pt]
\displaystyle
\frac{\textstyle \partial \,\sigma_{11}}{\textstyle \partial \,t}-(\lambda_f+2\,\mu)\,\frac{\textstyle \partial \,v_{s1}}{\textstyle \partial \,x}-\beta\,m\,\frac{\textstyle \partial \,w_{1}}{\textstyle \partial \,x}-\lambda_f\,\frac{\textstyle \partial \,v_{s2}}{\textstyle \partial \,y}-\beta\,m\,\frac{\textstyle \partial \,w_{2}}{\textstyle \partial \,y}=f_{\sigma_{11}},\\
[12pt]
\displaystyle
\frac{\textstyle \partial \,\sigma_{12}}{\textstyle \partial \,t}-\mu\,\left(\frac{\textstyle \partial \,v_{s2}}{\textstyle \partial \,x}+\frac{\textstyle \partial \,v_{s1}}{\textstyle \partial \,y}\right)=f_{\sigma_{12}},\\
[12pt]
\displaystyle
\frac{\textstyle \partial \,\sigma_{22}}{\textstyle \partial \,t}-\lambda_f\,\frac{\textstyle \partial \,v_{s1}}{\textstyle \partial \,x}-\beta\,m\,\frac{\textstyle \partial \,w_{1}}{\textstyle \partial \,x}-(\lambda_f+2\,\mu)\,\frac{\textstyle \partial \,v_{s2}}{\textstyle \partial \,y}-\beta\,m\,\frac{\textstyle \partial \,w_{2}}{\textstyle \partial \,y}=f_{\sigma_{22}},\\
[12pt]
\displaystyle
\frac{\textstyle \partial \,p}{\textstyle \partial \,t}+m\,\left(\beta\,\frac{\textstyle \partial \,v_{s1}}{\textstyle \partial \,x}+\frac{\textstyle \partial \,w_1}{\textstyle \partial \,x} +\beta\,\frac{\textstyle \partial \,v_{s2}}{\textstyle \partial \,y}+\frac{\textstyle \partial \,w_2}{\textstyle \partial \,y}\right)=f_p,
\end{array}
\right.
\label{LC}
\end{equation}
where $f_{\sigma_{11}}$, $f_{\sigma_{12}}$, $f_{\sigma_{22}}$ and $f_p$ are force densities,  $\rho_w=\frac{a}{\phi}\,\rho_f$, $\rho=\phi\,\rho_f+(1-\phi)\,\rho_s$, and $\chi =\rho\,\rho_w-\rho_f^2>0$. Setting 
\begin{equation}
\begin{array}{l}
{\bf U}=(v_{s1},\,v_{s2},\,w_1,\,w_2,\,\sigma_{11},\,\sigma_{12},\,\sigma_{22},\,p)^T,\, \\
[6pt]
{\bf F}=(0,\,0,\,0,\,0,\,f_{\sigma_{11}},\,f_{\sigma_{12}},\,f_{\sigma_{22}},\,f_p)^T,
\end{array}
\label{DefU}
\end{equation}
equations (\ref{LC}) are written as a first-order non-homogeneous linear system 
\begin{equation}
\frac{\textstyle \partial}{\textstyle \partial\,t}\,{\bf U}+{\bf A}\,\frac{\textstyle \partial}{\textstyle \partial\,x}\,{\bf U}+{\bf B}\,\frac{\textstyle \partial}{\textstyle \partial\,y}\,{\bf U}=-{\bf S}\,{\bf U}+{\bf F},
\label{SystHyp}
\end{equation}
where ${\bf A}$, ${\bf B}$ and ${\bf S}$ are $8\times8$ real matrices detailled in  \ref{AnnexeSystem}.  

The energy of poroelastic waves can be deduced from (\ref{LC}). Without any source terms (${\bf F =0}$) and setting ${\bf C}\,\varepsilon=\sigma +\beta \,p\,{\bf I}$, it is proven in \cite{THESE_EZZIANI} that 
\begin{equation}
E(t)=\frac{1}{2} \int_{\mathbb{R}^2}(\rho {\bf v_s}^2 + {\bf C}\,\varepsilon : \varepsilon) \ dS + \frac{1}{2} \int_{\mathbb{R}^2}( \rho_{w}  {\bf w}^2 + \frac{1}{m} p^2) \ dS + \int_{\mathbb{R}^2} \rho_{f} {\bf v_s . w} \ dS
\label{DefEnergie}
\end{equation}
is an energy that satisfies
\begin{equation}
\frac{\textstyle d\,E}{\textstyle d\,t}=-\int_{\mathbb{R}^2}\frac{\textstyle \eta}{\textstyle \kappa}\,{\bf w}^2\,dS.
\label{VariationNRJ}
\end{equation}
Consequently, $E$ is conserved when the viscous effects are neglected ($\eta=0$) and is a decreasing function otherwise. 

%------------------------------------------------------------------------------------------

\subsection{Heterogeneous media}\label{SecPhysInterface}

\begin{figure}[h!]
\centering
\includegraphics[scale=0.75]{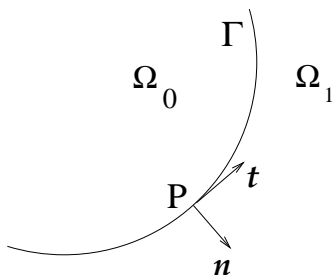}
\caption{Interface $\Gamma$ between two poroelastic media $\Omega_0$ and $\Omega_1$}. 
\label{FigInterface}
\end{figure}

The physical parameters defined in section \ref{SecPhysBiot} are piecewise constant and can be discontinuous across interfaces. In what follows, we will focus on two domains $\Omega_0$ and $\Omega_1$, which are separated by a stationary interface $\Gamma$ described by a parametric equation $(x(\tau),\,y(\tau))$ (figure \ref{FigInterface}). At any point $P$ on $\Gamma$, the unit tangential vector ${\bf t}$ and the unit normal vector ${\bf n}$ are 
\begin{equation}
{\bf t}=
\frac{\textstyle 1}{\textstyle \sqrt{x^{'2}+y^{'2}}}\,
\left(
x^{'},\,\,y^{'}
\right)^T,\qquad
{\bf n}=
\frac{\textstyle 1}{\textstyle \sqrt{x^{'2}+y^{'2}}}\,
\left(
y^{'},\,-x^{'}
\right)^T.
\label{NT}
\end{equation}
The derivatives $x^{'}=\frac{d\,x}{d\,\tau}$ and $y^{'}=\frac{d\,y}{d\,\tau}$ are assumed to be continuous everywhere along $\Gamma$, and to be differentiable as many times as required. 

The evolution equations (\ref{LC}) must be completed by a set of jump conditions. The simple case of perfect bonding and perfect hydraulic contact along $\Gamma$ is considered here, modeled by the jump conditions \cite{GUREVICH99}:
\begin{equation}
[{\bf v_s}]= {\bf 0},\quad
[{\bf w}.{\bf n}]=0, \quad
[{\bf \sigma}.{\bf n} ]= {\bf 0},\quad
[p]=0.
\label{JC}
\end{equation}
Enforcing these conditions is one of the main objective of the immersed interface method presented in section \ref{SecEsim}.

%------------------------------------------------------------------------------------------

\subsection{Dispersion analysis}\label{SecPhysDispersion}

\begin{figure}[htbp]
\begin{center}
\begin{tabular}{cc}
(a) & (b)\\
\includegraphics[scale=0.33]{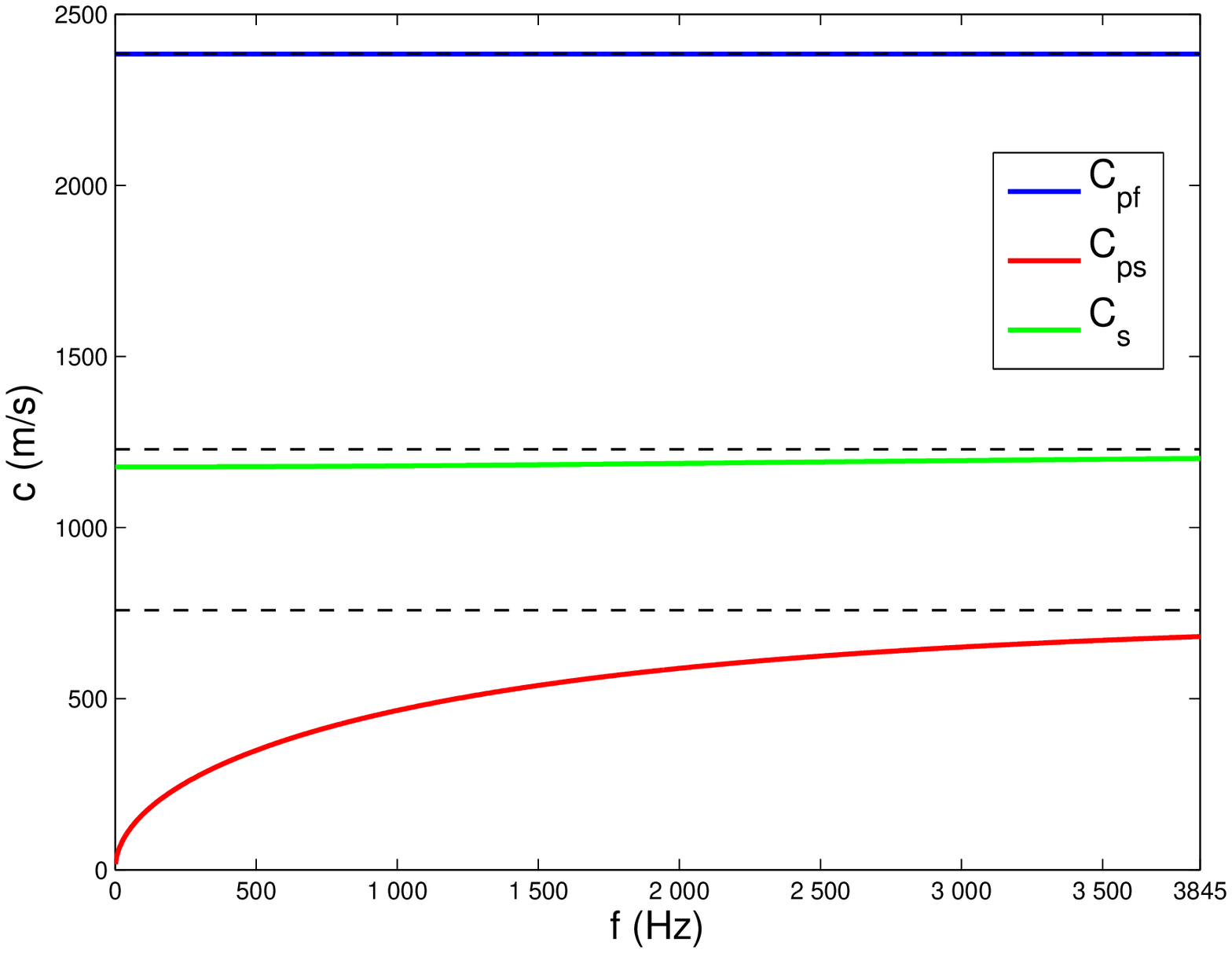} &
\includegraphics[scale=0.33]{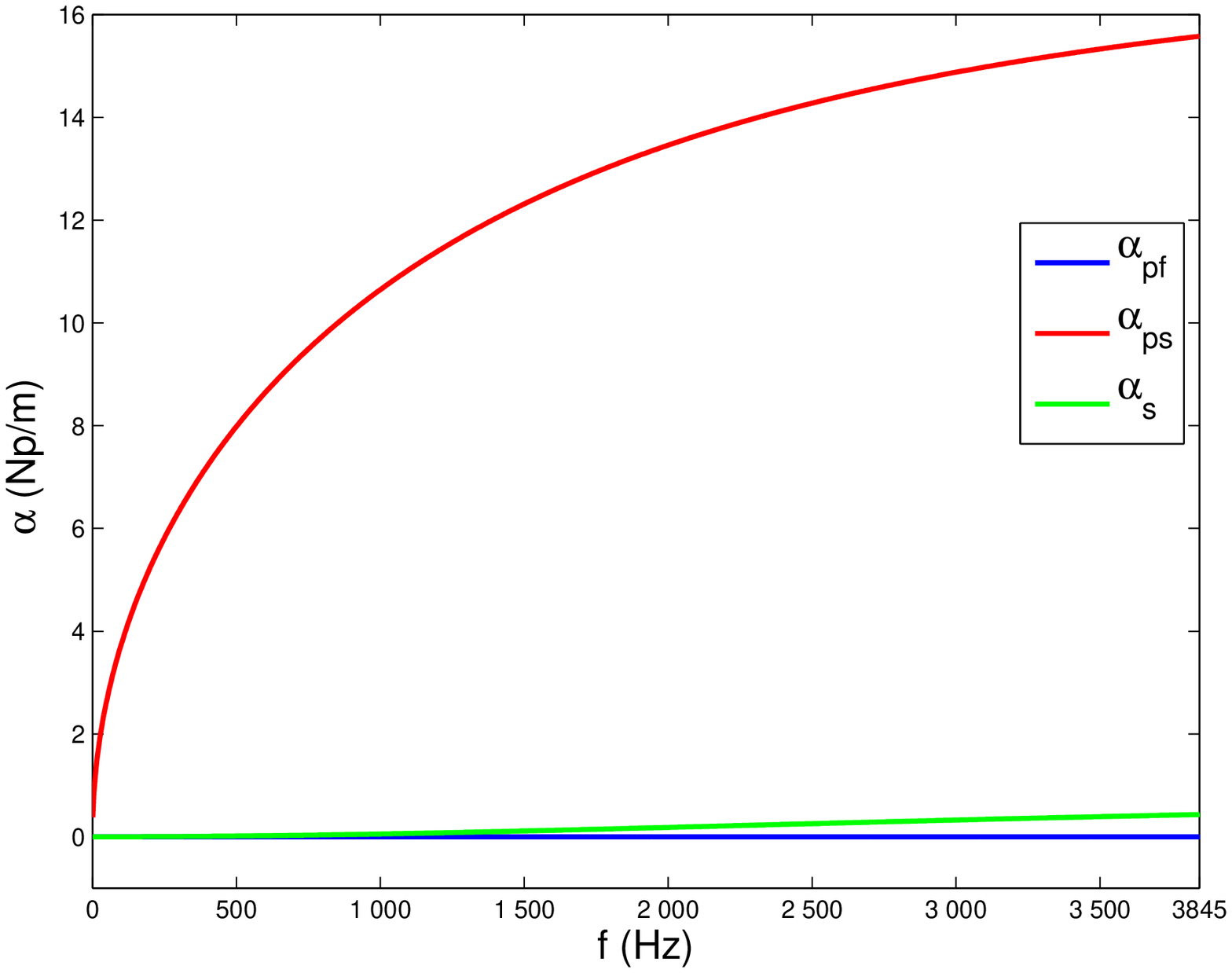} 
\end{tabular}
\end{center}
\caption{Phase velocities (a) and attenuations (b) of the solutions to Biot's model corresponding to the porous medium $\Omega_0$ in table \ref{TabParametres}. $pf$: fast compressional wave; $ps$: slow compressional wave; $s$: shear wave. In (a), the horizontal dotted lines refer to the eigenvalues $\overline{c}_{pf}$, $\overline{c}_{ps}$ and $\overline{c}_{s}$ of ${\bf A}$ and ${\bf B}$.}
\label{FigDispersion}
\end{figure}

The eigenvalues of ${\bf A}$ and ${\bf B}$ in (\ref{SystHyp}) are real: $\pm \overline{c}_{pf}$, $\pm \overline{c}_{ps}$, $\pm \overline{c}_s$, and 0 (multiplicity 2), where $\overline{c}_{pf}>\max(\overline{c}_s,\,\overline{c}_{ps})>0$. If $\eta\neq 0$, the spectral radius $R({\bf S})=\frac{\eta}{\kappa}\,\frac{\rho}{\chi}$ can be very large and then the system (\ref{SystHyp}) is stiff. 

A plane wave ${\bf d}\,e^{i(\omega\,t-{\bf k}.\,{\bf r})}$ is injected in (\ref{LC}), where ${\bf k}=k\,{\bf e}$ and ${\bf d}$ are the wavevector and the polarization, respectively; ${\bf r}$ is the position, $\omega=2\,\pi\,f$ is the angular frequency and $f$ is the frequency. If ${\bf d}$ is collinear with ${\bf k}$, the dispersion relation of compressional waves is obtained:
\begin{equation}
\begin{array}{l}
\displaystyle
A\,k^4+B(\omega)\,k^2+C(\omega)=0,\\
[8pt]
\displaystyle
A=\kappa\,m\left(\lambda_f+2\,\mu-\beta^2\,m\right),\quad C(\omega)=\chi\,\kappa\,\omega^4-i\,\eta\,\rho\,\omega^3,\\
[8pt]
\displaystyle
B(\omega)=-\kappa\left(\left(\lambda_f+2\,\mu\right)\,\rho_w+m\left(\rho-2\,\beta\,\rho_f\right)\right)\omega^2+i\,\eta\left(\lambda_f+2\,\mu\right)\omega,%\\
%[8pt]
%\displaystyle
%C(\omega)=\chi\,\kappa\,\omega^4-i\,\eta\,\rho\,\omega^3,
\end{array}
\label{DispersionP}
\end{equation}
where the roots $\pm k_{pf}$ and $\pm k_{ps}$ satisfy $0<\Re\mbox{e}\left\{k_{pf}\right\}<\Re\mbox{e}\left\{k_{ps}\right\}$. If ${\bf d}$ is orthogonal with ${\bf k}$, the dispersion relation of the shear wave is obtained:
\begin{equation}
\begin{array}{l}
\displaystyle
k=\frac{\textstyle 1}{\textstyle \sqrt{\mu}}\,\left(\frac{\textstyle A\,C-B^2}{\textstyle C}\right)^{1/2},\\
[8pt]
\displaystyle
A=\omega^2\,\left(\rho+\phi\,\rho_f(a-2)\right)-i\,\omega\,\phi^2\,\frac{\textstyle \eta}{\textstyle \kappa},\\
[8pt]
\displaystyle
B=-\omega^2\,\phi\,\rho_f\,(a-1)+i\,\omega\,\phi^2\,\frac{\textstyle \eta}{\textstyle \kappa},\\
[8pt]
\displaystyle
C=\omega^2\,\phi\,\rho_f\,a-i\,\omega\,\phi^2\,\frac{\textstyle \eta}{\textstyle \kappa},
\end{array}
\label{DispersionS}
\end{equation}
where the roots are $\pm k_{s}$, $\Re\mbox{e}\left\{k_{s}\right\}>0$. Based on (\ref{DispersionP}) and (\ref{DispersionS}), the phase velocities $c=\omega/\Re\mbox{e}\left\{k\right\}$ and the attenuations $\alpha=\Im\mbox{m}\left\{k\right\}$ of each wave are defined. In the remainder of this article,  
the subscripts $pf$, $ps$ and $s$ denote fast compressional, slow compressional and shear waves, respectively. 

The phase velocities $c_{pf}(f)$, $c_{ps}(f)$ and $c_s(f)$ are monotonically increasing functions, tending asymptotically towards the eigenvalues $\overline{c}_{pf}$, $\overline{c}_{ps}$ and $\overline{c}_{s}$. If $\eta=0$, the three waves are non-dispersive and non-dissipative, and the  energy of poroelastic waves (\ref{DefEnergie}) is conserved. If $\eta\neq0$, the fast compressional wave and the shear wave are weakly dispersive and dissipative. The slow compressional wave, however, is highly modified by the viscosity of the saturating fluid. If $f \ll f_c$, then $c_{ps}(f) \ll\overline{c}_{ps}$, and the slow compressional wave tends towards a static diffusive mode \cite{CHANDLER81}. At greater frequencies, $c_{ps}$ is larger but the attenuation increases. These properties are summarized in figure \ref{FigDispersion}.

In the low-frequency range, the direct contribution of the slow compressional wave to the overall wave propagation processes is therefore negligible when considering an homogeneous medium. However, the influence of the slow wave becomes crucial in heterogeneous media \cite{BOURBIE87}. The slow compressional wave, generated during the interaction between the propagative waves and the scatterers, remains localized around the interfaces. Consequently, this slow wave has a major influence on the balance equations at the interfaces, modifying crucially the behavior of fast compressional and shear diffracted waves. An accurate computation of the slow wave is therefore necessary, as shown in the numerical tests of  section \ref{SecNumExp}.

%------------------------------------------------------------------------------------------
%------------------------------------------------------------------------------------------

\section{Numerical modeling}\label{SecNumMod}
 
\subsection{Numerical scheme}\label{SecSplitting}

To integrate the system (\ref{SystHyp}), a uniform grid is introduced; the spatial mesh sizes are $\Delta\,x, \Delta\,y$, and the time step is $\Delta\,t$. A straightforward discretization of (\ref{SystHyp}) by an explicit time scheme leads to the following stability condition: 
\begin{equation}
\Delta t\leq \min \left(\frac{\textstyle \Theta\,\Delta x}{\textstyle \overline{c}_{pf}},\,\frac{\textstyle 2}{\textstyle R({\bf S})}\right),
\label{CFLdirect}
\end{equation}
where $\Theta$ is obtained by a Von-Neumann analysis of stability when ${\bf S}={\bf 0}$. In (\ref{CFLdirect}), the bound induced by the spectral radius of ${\bf S}$ can be very restrictive. In sandstone saturated with bitumen, for example, the maximal CFL number is roughly $\overline{c}_{pf}\,\Delta t/\Delta x\approx 10^{-12}\ll \Theta$, which is intractable for  computations. 

We follow here a more efficient strategy based on second-order Strang's splitting \cite{LEVEQUE07}, solving alternatively the hyperbolic system
\begin{equation}
\frac{\textstyle \partial}{\textstyle \partial\,t}\,{\bf U}+{\bf A}\,\frac{\textstyle \partial}{\textstyle \partial\,x}\,{\bf U}+{\bf B}\,\frac{\textstyle \partial}{\textstyle \partial\,y}\,{\bf U}={\bf 0},
\label{SplittingA}
\end{equation}
and then the diffusive system with a source term
\begin{equation}
\frac{\textstyle \partial}{\textstyle \partial\,t}\,{\bf U}=-{\bf S}\,{\bf U}+{\bf F}.
\label{SplittingB}
\end{equation}
The linear system (\ref{SplittingA}) is solved by applying any scheme for hyperbolic systems, giving  ${\bf U}_{i,j}^{n+1/2}$. In the numerical experiments performed in section \ref{SecNumExp}, a fourth-order ADER scheme is used \cite{SCHWARTZKOPFF04}, which involves a centered stencil of 25 nodes. On Cartesian grids, this scheme amounts to a fourth-order Lax-Wendroff scheme \cite{LORCHER05}. It is dispersive of order 4 and dissipative of order 6, and its stability limit is $\Theta=1$ \cite{STRIKWERDA99,HDR-LOMBARD}. Other single-grid schemes can be used without any restrictions.

Since the physical parameters do not vary with time, the diffusive system (\ref{SplittingB}) is solved exactly. For simplicity, null force density is taken: ${\bf F}={\bf 0}$. In this case, $p$ and ${\bf \sigma}$ are unchanged, whereas the velocities become ($k=1,\,2$)
\begin{equation}
\begin{array}{l}
\displaystyle
v_k^{n+1}=v_k^{n+1/2}+\frac{\rho_f}{\rho}\left(1-e^{-\frac{\eta}{\kappa}\,\frac{\rho}{\chi}\,T}\right)w_k^{n+1/2},\\
[8pt]
\displaystyle
w^{n+1}_k=e^{-\frac{\eta}{\kappa}\,\frac{\rho}{\chi}\,T}\,w_k^{n+1/2},
\end{array} 
\label{ODEexact}
\end{equation} 
where $T$ depends on the time step (see section \ref{SecAlgo}). The splitting (\ref{SplittingA})-(\ref{SplittingB}) along with exact integration (\ref{ODEexact}) recovers the optimal condition of stability: $\overline{c}_{pf}\,\Delta\,t/\Delta\,x\leq \Theta$. 

Since the matrices ${\bf A}$ and ${\bf B}$ do not commute with ${\bf S}$, the theoretical order of convergence falls from 4 to 2 when the viscosity is non-negligible. Using a fourth-order accurate scheme such as ADER 4 is nevertheless advantageous, compared with a second-order scheme such as Lax-Wendroff: the stability limit is improved, and numerical artifacts (dispersion, attenuation, anisotropy) are greatly reduced.

In \cite{PUENTE08}, the authors notice that the first-order splitting does not lead to a correct representation of the slow mode at low frequencies. Nevertheless, the  numerous one-dimensional examples provided in  \cite{CHIAVASSA10} demonstrate that the second-order splitting accurately represents the static mode when a sufficient number of discretization points per wavelength is used.  
 This can be obtained by using a local space-time refinement presented in the following section.

%------------------------------------------------------------------------------------------

\subsection{Mesh refinement}\label{SecAMR}

The slow wave has much smaller spatial scales of evolution than the wavelength of the other waves. A very fine grid is therefore required to account for its evolution. Since the use of a fine uniform grid on the whole computational domain is out of reach, grid refinement provides a good alternative. In addition, the slow wave remains localized near the interfaces (section \ref{SecPhysDispersion}), and hence grid refinement is necessary only around these places. Lastly, even if the slow wave propagates ($\eta=0$) the property $\overline{c}_{pf}\gg \overline{c}_{ps}$ is usually satisfied: consequently, a fine mesh near the interface is still useful to perform accurate extrapolations, as required by the immersed interface method (section \ref{SecEsim}).  
 
We adopt here a space-time mesh refinement approach based on flux conservation \cite{BERGER84,BERGER98}, which is more naturally coupled to the flux-conserving scheme developed to solve (\ref{SplittingA}). The refined zones are rectangular Cartesian patches with mesh sizes $\Delta\,x\,/\,q, \Delta\,y\,/\,q$, where the integer $q$ is the refinement factor. To reduce the cpu time and to limit the numerical dispersion on the coarse grid, a local time step $\Delta\,t\,/\,q$ is used \cite{RODRIGUEZ05,THESE_RODRIGUEZ}. When one time step is done on the coarse grid, $q$ time substeps are done on the refined zone. The extrapolated values required to couple coarse and fine grids are obtained by linear interpolation in space and time on the numerical values at the surrounding nodes \cite{HDR-LOMBARD}. In the case of the Lax-Wendroff scheme applied to the scalar advection equation, the stability of the coupling is proven in \cite{BERGER85} whatever $q$.

The additional cost induced by mesh refinement can become prohibitive, both concerning the memory requirements and the computational time, because of the $q$ substeps inside one time step. The value of $q$ must be estimated carefully in terms of the physical parameters. For this purpose, the wavelengths $\lambda_{pf}(f_0)=c_{pf}(f_0)/f_0$ and $\lambda_{ps}(f_0)=c_{ps}(f_0)/f_0$ are deduced from the dispersion analysis, where $f_0$ is the central frequency of the source. The number of fine grid nodes per wavelength of the slow compressional wave and the number of coarse grid nodes per wavelength of the fast compressional wave must then be roughly equal:
\begin{equation}
\frac{\lambda_{ps}(f_0)}{\Delta x\,/\,q} \approx \frac{\lambda_{pf}(f_0)}{\Delta x}\, \quad \Rightarrow\, \quad
q\approx \frac{c_{pf}(f_0)}{c_{ps}(f_0)}.
\label{factq}
\end{equation} 

%------------------------------------------------------------------------------------------

\subsection{Immersed interface method}\label{SecEsim}

The discretization of the interfaces requires special care. A straightforward stair-step discretization of the interfaces introduces a first-order geometrical error and yields spurious numerical diffractions. In addition, the jump conditions (\ref{JC}) are not enforced numerically if no special treatment is applied. Lastly, the smoothness requirements to solve (\ref{SplittingA}) are not satisfied, decreasing the convergence rate of the ADER scheme.

To remove these drawbacks while maintaining the efficiency of Cartesian grid methods, we adapt an immersed interface method previously developed in acoustics and elastodynamics \cite{PIRAUX01,LOMBARD04,LOMBARD06,HDR-LOMBARD}. At the {\it irregular points} where the ADER's stencil crosses the interface $\Gamma$, the scheme will use {\it modified values} of the solution, instead of the usual numerical values. The modified values are extrapolations, based on the local geometry of $\Gamma$ and on $r$ successive derivatives of the jump conditions (\ref{JC}). The
parameter $r$ is discussed at the end of this section. 

\begin{figure}[htbp]
\begin{center}
\includegraphics[scale=0.75]{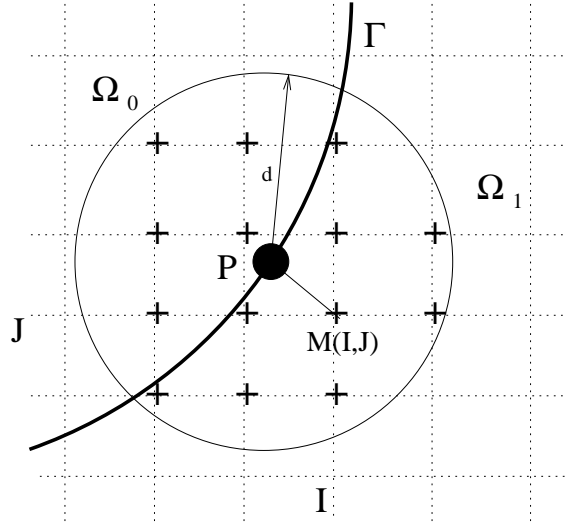} 
\caption{Irregular point $M(x_I,\,y_J)\in \Omega_1$ and its orthogonal projection $P$ onto $\Gamma$. The grid nodes used to compute ${\bf U}_{I,J}^*$ are inside the circle with radius $d$ and centered on $P$; they are denoted by ${\bf +}$.}
\label{Patate}
\end{center}
\end{figure}

Let us consider a point $M(x_I,\,y_J)\in \Omega_1$ and its orthogonal projection $P$ onto $\Gamma$ (figure \ref{Patate}). The algorithm to build the modified value at $M$ is divided into four steps.\\

\noindent
{\bf Step 1: high-order interface conditions}.\\
On the side $\Omega_k$ ($k=0,\,1$), the boundary values of the spatial derivatives of ${\bf U}$ up to the $r$-th order are put in a vector ${\bf U}^r_k$ with $n_v=4\,(r+1)\,(r+2)$ components:
\begin{equation}
\begin{array}{l}
\displaystyle
{\bf U}^r_k=\lim_{M\rightarrow P,\,M\in\Omega_k}
\left(
{\bf U}^T,
...,\,
\frac{\textstyle \partial^l}{\textstyle \partial\, x^{l-m}\,\partial\,y^m}\,{\bf U}^T,
...,\,
\frac{\textstyle \partial^r}{\textstyle \partial\,y^r}\,{\bf U}^T
\right)^T,
\label{Ur}
\end{array}
\end{equation}
where $l=0,\,...,\,r$ and $m=0,\,...,\,l$. Following this formalism, the zero-th order jump conditions (\ref{JC}) are written 
\begin{equation}
{\bf C}_1^0\,{\bf U}_1^0={\bf C}_0^0\,{\bf U}_0^0,
\label{JC0}
\end{equation}
where the matrices of the jump conditions ${\bf C}_k^0$ depend on the local geometry of $\Gamma$:
\begin{equation}
{\bf C}_k^0(\tau)=
\left(
\begin{array}{cccccccc}
1 & 0 & 0 & 0 & 0 & 0 & 0 & 0 \\
0 & 1 & 0       & 0 & 0 & 0 & 0 & 0 \\
0 & 0 & y^{'} & -x^{'} & 0 & 0 & 0 & 0 \\
0 & 0 & 0 & 0 & y^{'} & -x^{'} & 0 & 0 \\
0 & 0 & 0 & 0 & 0 & y^{'} & -x^{'} & 0 \\
0 & 0 & 0 & 0 & 0 & 0 & 0 & 1 \\
\end{array}
\right).
\end{equation}
The jump condition (\ref{JC0}) is differentiated with respect to time $t$, and then the time derivatives are replaced by spatial derivatives thanks to the conservation law (\ref{SplittingA}). For example, we obtain
\begin{equation}
\begin{array}{lll}
\displaystyle
\frac{\textstyle \partial}{\textstyle \partial\,t}\,({\bf C}_0^0\,{\bf U}_0^0)
=
-{\bf C}_0^0\,{\bf A}_0\,\frac{\textstyle \partial}{\textstyle \partial\,x}\,{\bf U}_0^0-{\bf C}_0^0\,{\bf B}_0\,\frac{\textstyle \partial}{\textstyle \partial\,y}\,{\bf U}_0^0,
\end{array}
\label{JC1a}
\end{equation}
where ${\bf A}_0$ and ${\bf B}_0$ are the matrices in $\Omega_0$. The jump condition (\ref{JC0}) is also differentiated in terms of $\tau$. Taking advantage of the chain-rule, we obtain e.g.
\begin{equation}
\begin{array}{lll}
\displaystyle
\frac{\textstyle d}{\textstyle d\,\tau}\,({\bf C}_0^0\,{\bf U}_0^0)
=
\displaystyle\left(\frac{\textstyle d}{\textstyle d\,\tau}\,{\bf C}_0^0\right)\,{\bf U}_0^0+{\bf C}_0^0\left(x^{'}\frac{\textstyle \partial}{\textstyle \partial\,x}\,{\bf U}_0^0+y^{'}\frac{\textstyle \partial}{\textstyle \partial\,y}\,{\bf U}_0^0\right).
\end{array}
\label{JC1b}
\end{equation}
From (\ref{JC0}), (\ref{JC1a}) and (\ref{JC1b}), we build matrices ${\bf C}_k^1$ such that ${\bf C}_1^1\,{\bf U}_1^1={\bf C}_0^1\,{\bf U}_0^1$, which provides first-order jump conditions. By iterating this process $r$ times, $r$-th order interface conditions are obtained
\begin{equation}
{\bf C}_1^r\,{\bf U}_1^r={\bf C}_0^r\,{\bf U}_0^r,
\label{JCr}
\end{equation}
where ${\bf C}_k^r$ are $n_c\times n_v$ matrices ($k=0,\,1$), and $n_c=3\,(r+1)\,(r+2)$. The computation of matrices ${\bf C}_k^r$ is a tedious task when $r\geq 2$, that can be greatly simplified using computer algebra tools. \\

\noindent
{\bf Step 2: high-order Beltrami-Michell equations}.\\
The equation (\ref{Barre}) is satisfied anywhere in a poroelastic medium. Under sufficient smoothness requirements, it can be differentiated with respect to $x$ and $y$, as many times as required:
\begin{equation}
\begin{array}{l}
\displaystyle
\frac{\textstyle \partial^j \,\sigma_{12}}{\textstyle \partial \,x^{j-i-1}\,\partial\,y^{i+1}}
=\displaystyle
\theta_0\,\frac{\textstyle \partial^j \,\sigma_{11}}{\textstyle \partial \,x^{j-i}\,\partial\,y^i}
+\theta_1\,\frac{\textstyle \partial^j \,\sigma_{22}}{\textstyle \partial \,x^{j-i}\,\partial\,y^i}
+\theta_2\,\frac{\textstyle \partial^j \,p}{\textstyle \partial \,x^{j-i}\,\partial\,y^i}\\
[12pt]
\displaystyle
\hspace{2.8cm}
+\theta_1\,\frac{\textstyle \partial^j \,\sigma_{11}}{\textstyle \partial \,x^{j-i-2}\,\partial\,y^{i+2}}
+\theta_0\,\frac{\textstyle \partial^j \,\sigma_{22}}{\textstyle \partial \,x^{j-i-2}\,\partial\,y^{i+2}}
+\theta_2\,\frac{\textstyle \partial^j \,p}{\textstyle \partial \,x^{j-i-2}\,\partial\,y^{i+2}},
\end{array}
\label{BarreK}
\end{equation}
where $j\geq 2$ and $i=0,\cdots,\,j-2$. The equations (\ref{BarreK}) are also satisfied along $\Gamma$. They can be used therefore to reduce the number of independent components in ${\bf U}_k^r$. For this purpose, we define the vectors ${\bf V}_k^r$ such that 
\begin{equation}
{\bf U}_k^r = {\bf G}_k^r\,{\bf V}_k^r, 
\label{UGV}
\end{equation}
where ${\bf G}_k^r$ are $n_v \times (n_v-n_b)$ matrices, and $n_b=r\,(r-1)/2$ if $r\geq 2$, $n_b=0$ otherwise. Based on (\ref{Ur}) and (\ref{BarreK}), an algorithm to compute the non-zero components of ${\bf G}_k^r$ is proposed in  \ref{AnnexeBeltrami}.\\

\noindent
{\bf Step 3: high-order boundary values}.\\
Based on (\ref{JCr}) and (\ref{UGV}), the vectors of independent boundary values satisfy
\begin{equation}
{\bf S}_1^r\,{\bf V}_1^r={\bf S}_0^r\,{\bf V}_0^r,
\label{CGV}
\end{equation}
where ${\bf S}_k^r={\bf C}_k^r\,{\bf G}_k^r$ are $ n_c\times (n_v-n_b)$ matrices. Since the system (\ref{CGV}) is underdetermined, the solution is not unique, and hence it can be written 
\begin{equation}
{\bf V}_1^r=\left(\left({\bf S}_1^r\right)^{-1}\,{\bf S}_0^r\,|\,{\bf K}_{{\bf S}_1^r}\right)
\left(
\begin{array}{c}
\displaystyle
{\bf V}_0^r\\
[8pt]
\displaystyle
{\bf \Lambda}^r
\end{array}
\right),
\label{SVD}
\end{equation}
where $({\bf S}_1^r)^{-1}$ is the least-squares pseudo-inverse of ${\bf S}_1^r$, ${\bf K}_{{\bf S}_1^r}$ is the matrix filled with the kernel of ${\bf S}_1^r$, and ${\bf \Lambda}^r$ is a set of $n_v-n_c-n_b$ Lagrange multipliers that represent the coordinates of ${\bf V}_1^r$ onto the kernel. A singular value decomposition of ${\bf S}_1^r$ is used to build $({\bf S}_1^r)^{-1}$ and the kernel ${\bf K}_{{\bf S}_1^r}$ \cite{NRC}.\\

\noindent
{\bf Step 4: construction of modified values}.\\
 Let ${\bf \Pi}_{ij}^r$ be the matrix of $r$-th order 2D Taylor expansions  
\begin{equation}
{\bf \Pi}_{i,j}^r=\left(1,...,
\frac{\textstyle 1}{\textstyle l \,!\,(l-m)\,!}\,(x_i-x_P)^{l-m}(y_j-y_P)^m,...,
\frac{\textstyle (y_j-y_P)^r}{\textstyle r\,!}\right)\,{\bf I}_8,
\label{Taylor}
\end{equation}
where ${\bf I}_8$ is the $8 \times 8$ identity matrix, $l=0,...,\,r$ and $m=0,...,\,l$. The modified value at $(x_I,\,y_J)$ is a smooth extension of the solution on the other side of $\Gamma$ (figure \ref{Patate}), and it writes:
\begin{equation}
{\bf U}_{I,J}^*={\bf \Pi}_{I,J}^r\,{\bf U}_0^r={\bf \Pi}_{I,J}^r\, {\bf G}_0^r \, {\bf V}_0^r.
\label{SolMod}
\end{equation}
The vector ${\bf V}_0^r$ in (\ref{SolMod}) remains to be estimated in terms of the boundary conditions and of the numerical values at surrounding grid points. For this purpose, we consider the disc ${\cal D}$ centered at $P$ with a radius $d$, that contains $N_d$ grid points. At the grid points of ${\cal D}\cap\Omega_0$, $r$-th order Taylor expansion of the solution at $P$ gives
\begin{equation}
\begin{array}{lll}
{\bf U}(x_i,y_j,t_n) &=& {\bf \Pi}_{i,j}^r\,{\bf U}_0^r+{\cal O}(\Delta\,x^{r+1}),\\
[10pt]
&=& {\bf \Pi}_{i,j}^r \,{\bf G}_0^r \, \left({\bf 1}\,|\,{\bf 0}\right)
\left(
\begin{array}{c}
{\bf V}_0^r\\
[8pt]
{\bf \Lambda}^r
\end{array}
\right)
+{\cal O}(\Delta\,x^{r+1}).
\end{array}
\label{Taylor1}
\end{equation}
At the grid points of ${\cal D}\cap\Omega_1$, $r$-th order Taylor expansion of the solution at $P$ and the boundary conditions (\ref{SVD}) give 
\begin{equation}
\begin{array}{lll}
{\bf U}(x_i,y_j,t_n) &=& {\bf \Pi}_{i,j}^r\,{\bf U}_1^r+{\cal O}(\Delta\,x^{r+1}),\\
[10pt]
&=& {\bf \Pi}_{i,j}^r\,{\bf G}_1^r \, \left(\left({\bf S}_1^r\right)^{-1}\,{\bf S}_0^r\,|\,{\bf K}_{{\bf S}_1^r}\right)
\left(
\begin{array}{c}
{\bf V}_0^r\\
[8pt]
{\bf \Lambda}^r
\end{array}
\right)
+{\cal O}(\Delta\,x^{r+1}).
\end{array}
\label{Taylor2}
\end{equation}
Equations (\ref{Taylor1}) and (\ref{Taylor2}) are written in the matrix form 
\begin{equation}
\left(
{\bf U}(.,\,t_n)
\right)_{\mathcal D}
={\bf M}
\left(
\begin{array}{c}
{\bf V}_0^r\\
[8pt]
{\bf \Lambda}^r
\end{array}
\right)
+
\left(
\begin{array}{c}
{\cal O}(\Delta\,x^{r+1})\\
\vdots\\
{\cal O}(\Delta\,x^{r+1})
\end{array}
\right),
\label{Taylor3}
\end{equation}
where ${\bf M}$ is a convenient $8\,N_d \times (2\,n_v-2\,n_b-n_c)$ matrix. To ensure that the system (\ref{Taylor3}) is overdetermined, the radius $d$ of the disc is chosen to satisfy  
\begin{equation}
\varepsilon(d,\,r)=\frac{\textstyle 8\,N_d}{\textstyle 2\,n_v-2\,n_b-n_c}\geq 1.
\label{RayonPatate}
\end{equation}
Exact values in (\ref{Taylor3}) are replaced by numerical ones, and the Taylor rests are removed. The least-squares inverse of ${\bf M}$ is denoted by ${\bf M}^{-1}$. The  Lagrange multipliers ${\bf \Lambda}^k$ are accounted in the construction of ${\bf M}$, but are not involved in the definition
of the modified value (\ref{SolMod}). As a consequence, they can be removed and the $(n_v-n_b)\times 8\,N_d$ restriction $\overline{{\bf M}^{-1}}$ of ${\bf M}^{-1}$ is defined by
\begin{equation}
{\bf V}_0^r=\overline{{\bf M}^{-1}}
\,\left(
{\bf U}^n
\right)_{\mathcal D}.
\label{Taylor4}
\end{equation}
Lastly, the modified value follows from (\ref{SolMod}) and (\ref{Taylor4}):
\begin{equation}
{\bf U}_{I,J}^*={\bf \Pi}_{I,J}^r\,{\bf G}_0^r \, \overline{{\bf M}^{-1}}\,\left(
{\bf U}^n
\right)_{\mathcal D}.
\label{UIJ*}
\end{equation}
\hspace{0.5cm}

\begin{table}[htbp]
\begin{center}
\begin{tabular}{|l|l|}  
\hline
Quantity & Size\\
\hline
\hline
$n_v$ & $4\,(r+1)\,(r+2)$ \\
$n_c$ & $3\,(r+1)\,(r+2)$ \\
$n_b$ & $r\,(r-1)\,/\,2$ if $r\geq 2$, 0 else\\
\hline
${\bf C}_k^r$ & $n_c \times n_v$ \\
${\bf G}_k^r$ & $n_v \times (n_v-n_b)$ \\
${\bf S}_k^r$ & $n_c \times (n_v-n_b)$ \\
${\bf \Pi}_k^r$ & $8 \times n_v$\\
${\bf 1}$     & $(n_v-n_b) \times (n_v-n_b)$\\
${\bf 0}$     & $(n_v-n_b) \times (n_v-n_c-n_b)$\\
${\bf M}$     & $8\,N_d \times (2\,n_v-2\,n_b-n_c)$\\
$\overline{{\bf M}^{-1}}$ & $(n_v-n_b) \times 8\,N_d$\\
\hline
\end{tabular}
\caption{Quantities involved in the computation of the modified values (section \ref{SecEsim}).}
\label{TabMatrices}
\end{center}
\end{table}

\noindent
{\bf Comments and practical details}. \\
\begin{enumerate}
\item A similar algorithm is applied at each irregular point along $\Gamma$. The sizes of the matrices involved are summarized in table \ref{TabMatrices}. Since the jump conditions do not vary with time, the evaluation of the matrices in (\ref{UIJ*}) is done during a preprocessing step. Only small matrix-vector products are therefore required at each time step. After optimization of the computer codes, this additional cost is made negligible, lower than 1\% of the time-marching.

\item The matrix ${\bf M}$ in (\ref{Taylor3}) depends on the subcell position of $P$ inside the mesh and on the jump conditions at $P$, involving the local geometry and the curvature of $\Gamma$ at $P$. Consequently, all these insights are incorporated in the modified value (\ref{UIJ*}), and hence in the scheme.

\item \label{EstimeVarEps} The simulations indicate that overestimation of $\varepsilon$ in (\ref{RayonPatate}) has a crucial influence on the stability of the immersed interface method. Various strategies can be used to ensure (\ref{RayonPatate}), for instance an adaptive choice of $d$ depending on the local geometry of $\Gamma$ at $P$. We adopt here a simpler strategy, based on a constant radius $d$. Taking $r=2$, numerical experiments have shown that $d=3.2\,\Delta\,x$ is a good candidate, while $d=4.5\,\Delta\,x$ is used when $r=3$. In this case, we obtain typically $N_d \approx 20$ and $\varepsilon \approx 4$. 

\item The order $r$ plays an important role on the accuracy of the coupling between the immersed interface method and a $s$-th order scheme. If $r\geq s$, then a $s$-th order local truncation error is obtained at the irregular points. This condition can be slightly relaxed: $r=s-1$ still ensures a $s$-th order overall accuracy \cite{GUSTAFSSON75}. As a consequence, a fourth-order ADER scheme ($s=4$) requires a third-order immersed interface method ($r=3$) to maintain fourth-order convergence.

\item A GKS analysis of stability has been performed in 1D in the case of an inviscid saturating fluid \cite{HDR-LOMBARD}. Extending this approach to 2D problems with viscous saturating fluids is out of reach. Various numerical experiments, however, indicate the stability of the method under the usual CFL condition (section \ref{SecSplitting}), if two requirements are satisfied: (i) the number of grid nodes used for extrapolations is sufficiently large, as stated in point \ref{EstimeVarEps}; (ii) the Beltrami-Michell equations (\ref{UGV}) are used.
\end{enumerate} 

%-------------------------------------------------------

\subsection{Summary of the algorithm}\label{SecAlgo}

The numerical strategy presented in this section couples three numerical methods: a finite difference numerical scheme with splitting (section \ref{SecSplitting}), a space-time mesh refinement (section \ref{SecAMR}), and an immersed interface method (section \ref{SecEsim}). To clarify the interactions between these methods, the global algorithm is summarized as follows:

{\ttfamily
\noindent
$\triangleright$ Preprocessing
\begin{itemize}
\item[-] Detection of irregular grid points
\item[-] Computation of extrapolation matrices in (\ref{UIJ*})
\item[-] Initialization of the solution at $t=0$
\item[-] Diffusive step (\ref{ODEexact}) where $T=\Delta \,t\,/\,2$ on the coarse grid
\item[-] Diffusive step (\ref{ODEexact}) where $T=\Delta \,t\,/\,(2\,q)$ on the refined grids
\end{itemize}
$\triangleright$ Time iterations
\begin{itemize}
\item Coarse grid:
\begin{itemize}
\item[-] Computation of modified values (\ref{UIJ*}) if present
\item[-] Solving the propagative step (\ref{SplittingA}) 
\item[-] Diffusive step (\ref{ODEexact}) where $T=\Delta \,t$
\end{itemize}
\item  On each refined grid, $q$ subtime iterations:
\begin{itemize}
\item[-] Space-time interpolations at the grid boundaries
\item[-] Computation of modified values (\ref{UIJ*}) 
\item[-] Solving the propagative step (\ref{SplittingA})
\item[-] Diffusive step (\ref{ODEexact}) where $T=\Delta \,t\,/\,q$ 
\end{itemize}
\end{itemize}
$\triangleright$ End of time iterations
\begin{itemize}
\item[-] Diffusive step (\ref{ODEexact}) where $T=\Delta \,t\,/\,2$ on the coarse grid
\item[-] Diffusive step (\ref{ODEexact}) where $T=\Delta \,t\,/\,(2\,q)$ on the refined grids
\end{itemize}
}

%------------------------------------------------------------------------------------------
%------------------------------------------------------------------------------------------

\section{Numerical experiments}\label{SecNumExp}

\subsection{Configurations}

Five tests are proposed along this section. In Test 1, the convergence order of the ADER scheme coupled with the immersed interface method is measured. Test 2 illustrates the different kind of waves in homogeneous media, and also the influence of the local space-time refinement. Test 3 investigates the numerical stability of the global algorithm. Diffraction of a plane wave by one (Test 4) and four (Test 5) cylindrical scatterers illustrates the accuracy and the physical relevance of the proposed numerical methods.

The physical parameters given in table \ref{TabParametres} correspond to Cold Lake sandstone and shale saturated with water \cite{DAI95}, respectively. In some experiments, an inviscid saturating fluid is artificially considered: $\eta=0$ Pa.s, the other parameters being unchanged. As recalled in section \ref{SecPhysBiot}, this limit-case has physical significance only in the high-frequency range. It is mainly addressed here for a numerical purpose.

\begin{table}[htbp]
\begin{center}
\begin{tabular}{|l|ll|}
\hline  
Parameters & $\Omega_0$ & $\Omega_1$                            \\
\hline
\hline
$\rho_s$ (kg/m$^3$)        & 2650            & 2211             \\
$\mu$ (Pa)                 & $2.926\,10^9$   & $3.539\,10^9$    \\
$\rho_f$ (kg/m$^3$)        & 1040            & 1040             \\ 
$\eta$ (Pa.s)              & $1.5\,10^{-3}$  & $10^{-3}$        \\
$\phi$                     & 0.335           & 0.05             \\
$a$                        & 2               & 2                \\
$\kappa$ (m$^2$)           & $10^{-11}$      & $5.\,10^{-12}$   \\
$\lambda_f$ (Pa)           & $6.1425\,10^9$  & $4.689\,10^9$    \\
$\beta$                    & $0.9558$        & $0.0527$         \\
$m$ (Pa)                   & $6.491\,10^9$   & $9.852\,10^9$    \\
\hline
$\overline{c}_{pf}$ (m/s)  & 2384.1          & 2350.4           \\
$\overline{c}_{ps}$ (m/s)  & 758.9           &  486.4           \\
$\overline{c}_s$ (m/s)     & 1229.0          & 1290.0           \\
$f_c$ (Hz)                 & 3844.9          & 765.1            \\
\hline
\end{tabular}
\caption{Physical parameters of the matrix ($\Omega_0$) and of the scatterer ($\Omega_1$), corresponding to sandstone and shale saturated with water, respectively.}
\label{TabParametres}
\end{center}
\end{table}

Once the spatial mesh sizes $\Delta\,x$ and $\Delta\,y$ are chosen on the coarse grid, the time step follows from the CFL number in $\Omega_0$: $\overline{c}_{pf0}\,\Delta\,t/\max{(\Delta\,x,\Delta\,y)}=0.95<1$. If $\eta\neq 0$, the maximum CFL number induced by (\ref{CFLdirect}) is equal to 0.5: consequently, the simulations done here with the splitting (\ref{SplittingA})-(\ref{SplittingB}) are twice faster than with unsplitted methods.

The grids are excited by two means: either a plane fast compressional wave, either a point source  that generates cylindrical waves. Details of the excitation method are given in  \ref{SecSources}. In the case of an incident plane wave, the exact expression given in  \ref{SecSources} is also enforced numerically on the edges of the computational domain. No special attention is paid to simulate outgoing waves, for instance with Perfectly-Matched Layers \cite{MARTIN08,ZENG01}. In all the presented tests, the size of the domain and the number of iterations are chosen to avoid the spurious reflections of diffracted waves with the outer frontiers. 

%------------------------------------------------------------------------------------------

\subsection{Test 1: convergence measurements}\label{SecNumTest1}

\begin{figure}[h]
\begin{center}
\begin{tabular}{cc}
(a) & (b)\\
\includegraphics[scale=0.33]{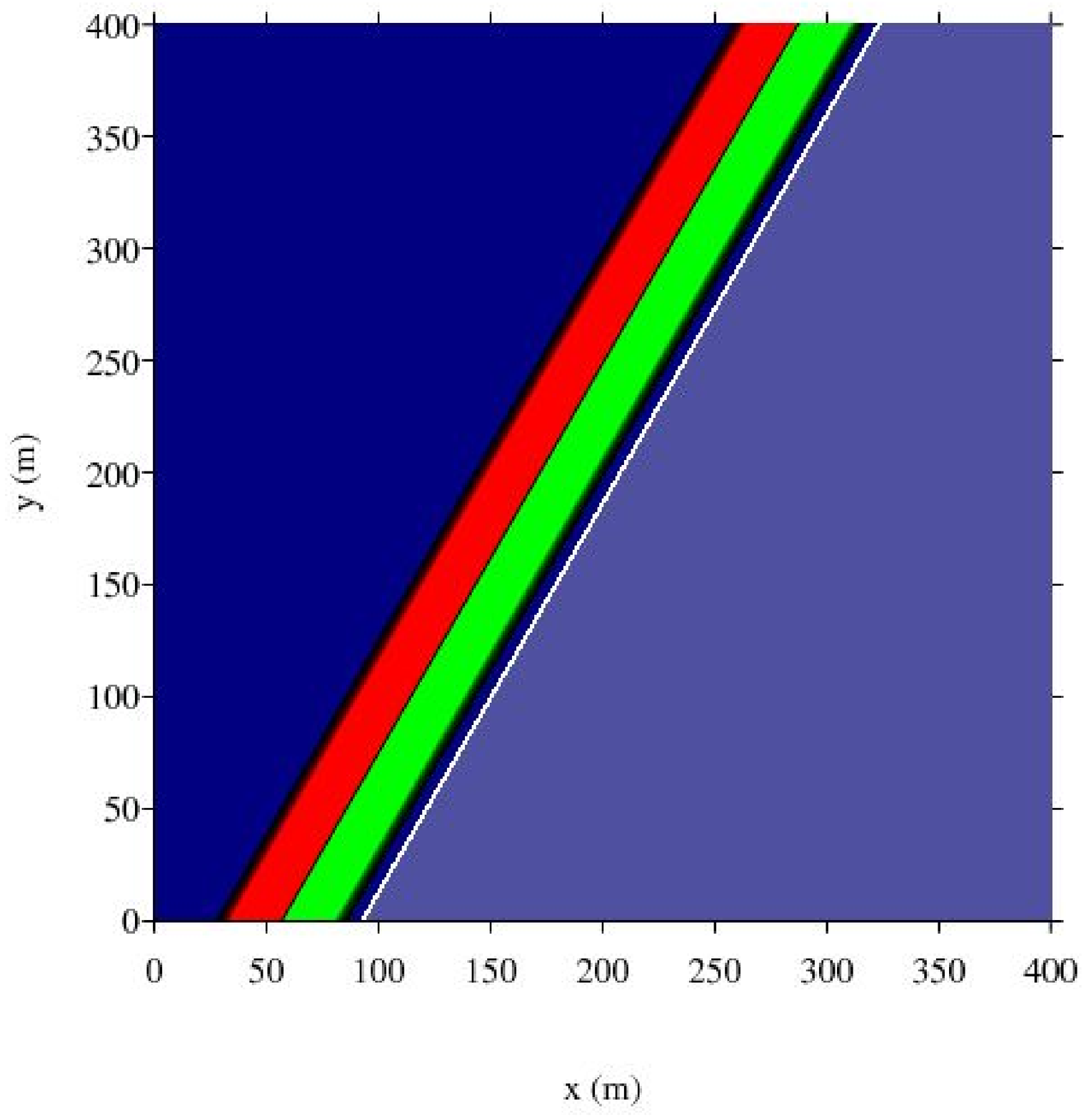}&
\includegraphics[scale=0.33]{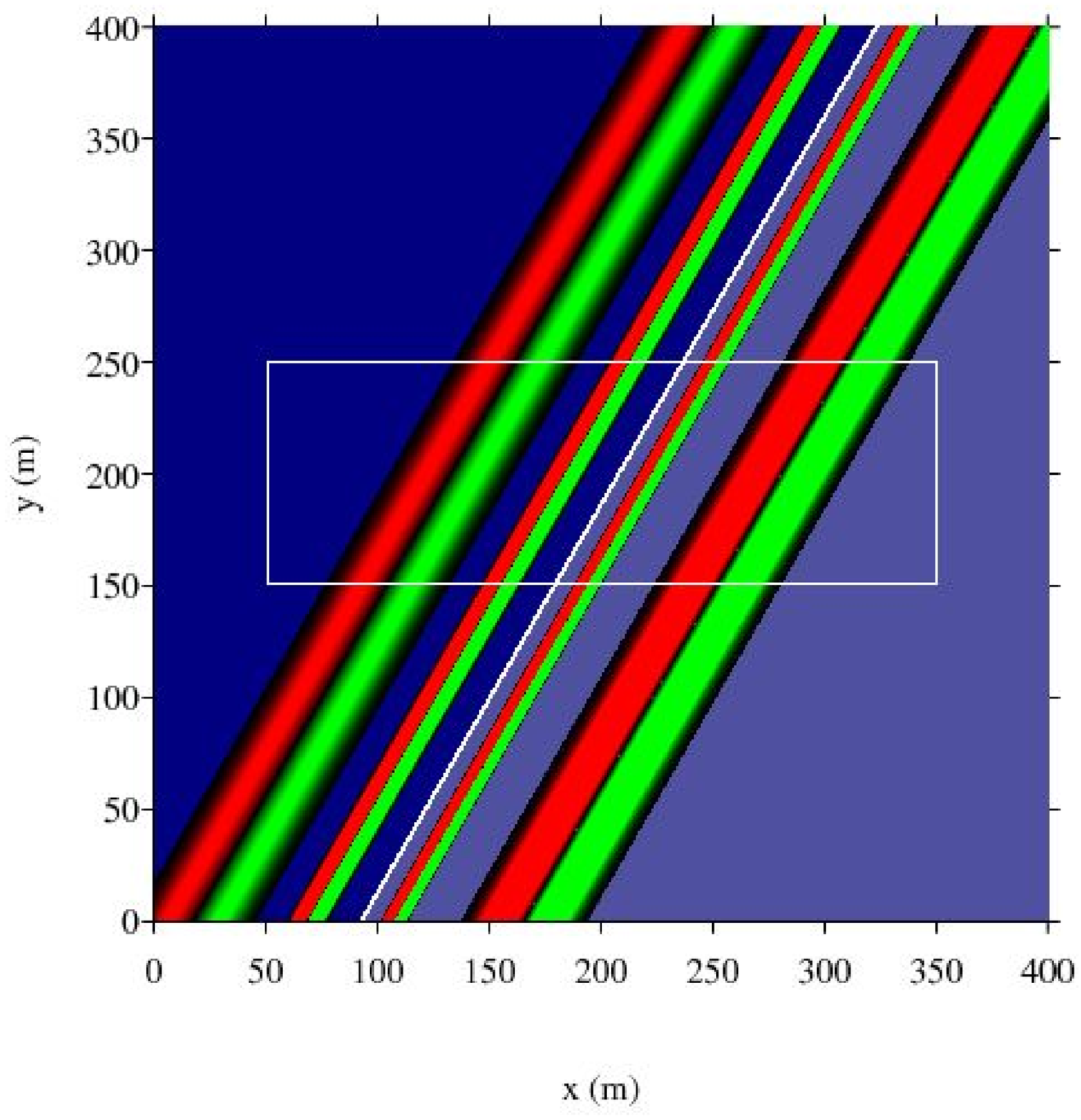}
\end{tabular}
\end{center}
\caption{Test 1. Snapshots of $p$ at the initial instant (a) and at the instant of measure (b). The white rectangle denotes the zone where convergence errors are measured.}
\label{FigTest1Cartes}
\end{figure}

In Test 1, we focus on the coupling between the ADER scheme (section \ref{SecSplitting}) and the immersed interface method (section \ref{SecEsim}). For this purpose, we consider a domain $[0,\,400]$ m$^2$ cut by a plane interface with slope 60 degrees. The saturating fluids are inviscid: exact solutions can be computed very accurately without Fourier synthesis, and splitting errors of the scheme are avoided. The source is plane (\ref{OndePlaneFourier}), with parameters: $\gamma=10^{-3}$, $f_0=40$ Hz, $t=3.3\,10^{-2}$ s, and $\theta=-30$ degrees (figure \ref{FigTest1Cartes}-a). Consequently, the incident wave propagates normally to the interface, leading to a 1-D configuration; from a numerical point of view, however, the problem is fully bidimensional. 

The computations are done on a uniform grid of $N\times N$ points, during $N/4$ time steps. Comparisons with the exact values of the pressure $p$ are done on the subdomain  $[50,\,350]\mbox{ m} \times[150,\,250]$ m, in order to avoid spurious effects induced by the edges of the computational domain (figure \ref{FigTest1Cartes}-b). The measures involve reflected and transmitted fast and slow compressional waves generated by the interface (no shear wave is generated in 1D); these waves are highly sensitive to the discretization of the jump conditions.

\begin{table}[h]
\begin{center}
\begin{small}
\begin{tabular}{l|ll|ll|ll|ll}
$N$   &  $r=0$         & order     & $r=1$          & order   & $r=2$          & order  & $r=3$          & order    \\
\hline
400    & $4.894\,10^{0}$     & {\bf -}   & $6.527\,10^{0}$    & {\bf -} & $6.107\,10^{0}$     & {\bf -} & $ 5.067\,10^{0}$    & {\bf -}     \\
800    & $1.247\,10^{0}$     & {\bf 1.973}   & $1.667\,10^{0}$    & {\bf 1.961} & $1.065\,10^{0}$  & {\bf 2.520} & $ 8.642\,10^{-1}$ & {\bf 2.552}     \\
1200   & $6.520\,10^{-1}$    & {\bf 1.599}   & $6.758\,10^{-1}$    & {\bf 2.242} & $2.273\,10^{-1}$  & {\bf 3.809} & $ 1.936\,10^{-1}$ & {\bf 3.690}     \\
1600   & $4.770\,10^{-1}$     & {\bf 1.086}   & $3.617\,10^{-1}$ & {\bf 2.173} & $7.995\,10^{-2}$  & {\bf 3.632} & $ 6.835\,10^{-2}$ & {\bf 3.619}     \\
2000   & $3.818\,10^{-1}$     & {\bf 0.998}   & $2.254\,10^{-1}$    & {\bf 2.119} & $3.344\,10^{-2}$  & {\bf 3.906} & $ 2.888\,10^{-2}$ & {\bf 3.861}     \\
2400   & $3.128\,10^{-1}$     & {\bf 1.093}   & $1.509\,10^{-1}$    & {\bf 2.201} & $1.564\,10^{-2}$  & {\bf 4.168} & $ 1.414\,10^{-2}$ & {\bf 3.917}     \\
2800   & $2.696\,10^{-1}$     & {\bf 0.964}   & $1.097\,10^{-1}$    & {\bf 2.069} & $9.157\,10^{-3}$  & {\bf 3.473} & $ 7.697\,10^{-3}$ & {\bf 3.945}     \\
3200   & $2.390\,10^{-1}$     & {\bf 0.902}   & $8.431\,10^{-2}$    & {\bf 1.971} & $5.598\,10^{-3}$  & {\bf 3.685} & $ 4.504\,10^{-3}$ & {\bf 4.013}     \\
3600   & $2.115\,10^{-1}$     & {\bf 1.038}   & $6.573\,10^{-2}$    & {\bf 2.114} & $3.613\,10^{-3}$  & {\bf 3.718} & $ 2.840\,10^{-3}$ & {\bf 3.915}     \\
4000   & $1.928\,10^{-1}$     & {\bf 0.879}   & $5.362\,10^{-2}$    & {\bf 1.933} & $2.485\,10^{-3}$  & {\bf 3.552} & $ 1.866\,10^{-3}$ & {\bf 3.986}     \\ 
\end{tabular}
\end{small} 
\end{center}
\caption{Test 1. Convergence rate in $l_2$ norm. No immersed interface method $(r=0)$, linear $(r=1)$, quadratic $(r=2)$ or cubic $(r=3)$ immersed interface method.}
\label{TabConvergence}
\end{table}

\begin{figure}[htbp]
\begin{center}
\includegraphics[scale=0.45]{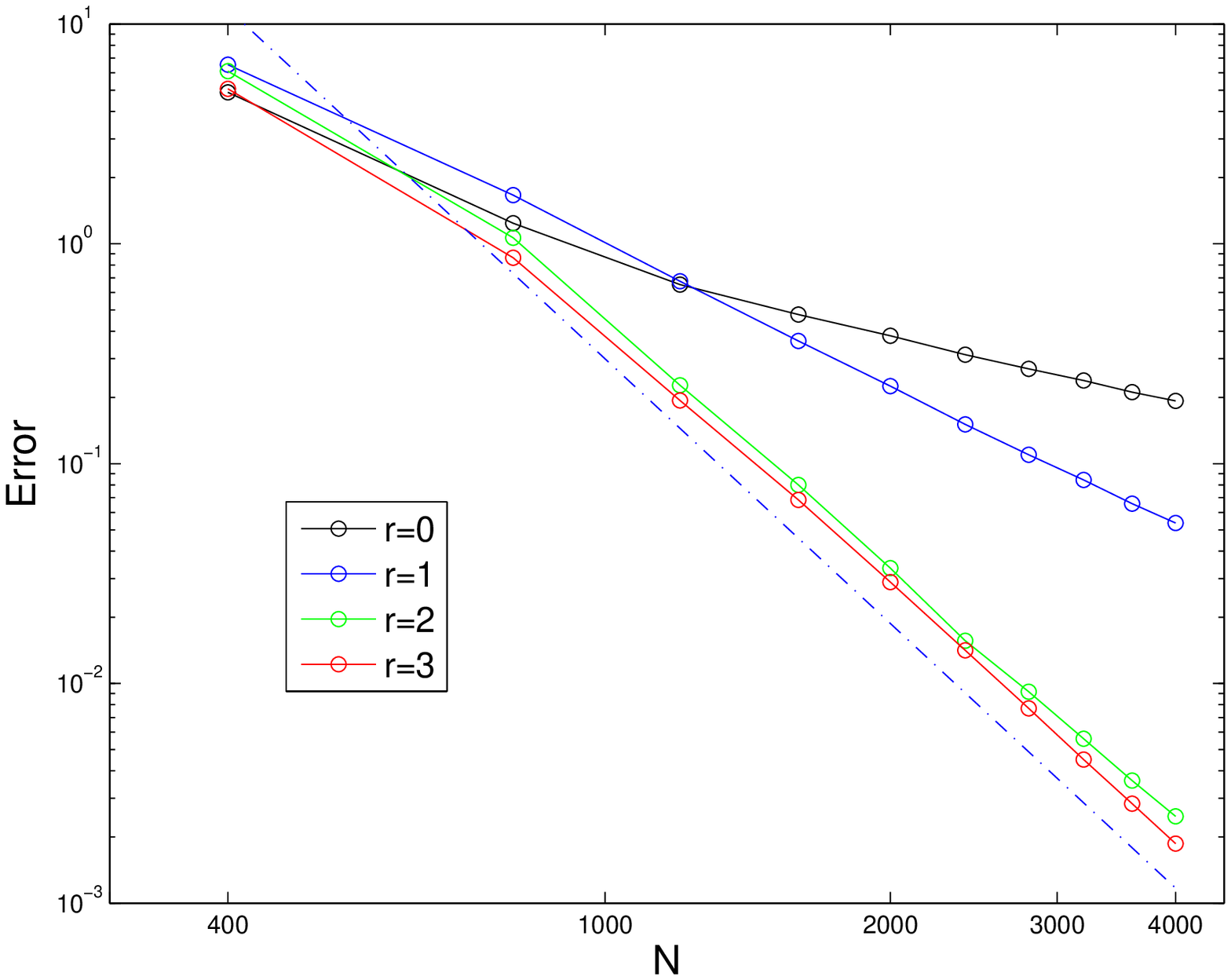}
\end{center}
\caption{Test 1. Error measured in $l_2$ norm versus the number of points $N_x$, with various order $r$ of the immersed interface method. Dotted line corresponds to 4-th order slope. }
\label{FigTest1Convergence}
\end{figure}

Errors in $l_2$ norm and convergence rates are reported in table \ref{TabConvergence} and drawn on figure \ref{FigTest1Convergence}. Various values of $r$ are investigated; $r=0$ means that no immersed interface method is applied; in this case, first-order accuracy is obtained. As stated in section \ref{SecEsim}, fourth-order accuracy is maintained if $r=4-1=3$, i.e if third-order extrapolations are used in the immersed interface method. In the present test case, $r=2$ is sufficient to obtain the same level of accuracy on a large range of grid size. Nevertheless, this could be untrue in other contexts, and hence we will always use $r=3$ in the following simulations. 

%------------------------------------------------------------------------------------------

\subsection{Test 2: mesh refinement}\label{SecNumTest2}

\begin{figure}[htbp]
\begin{center}
\begin{tabular}{cc}
$p$ & $p$\\
\includegraphics[scale=0.4]{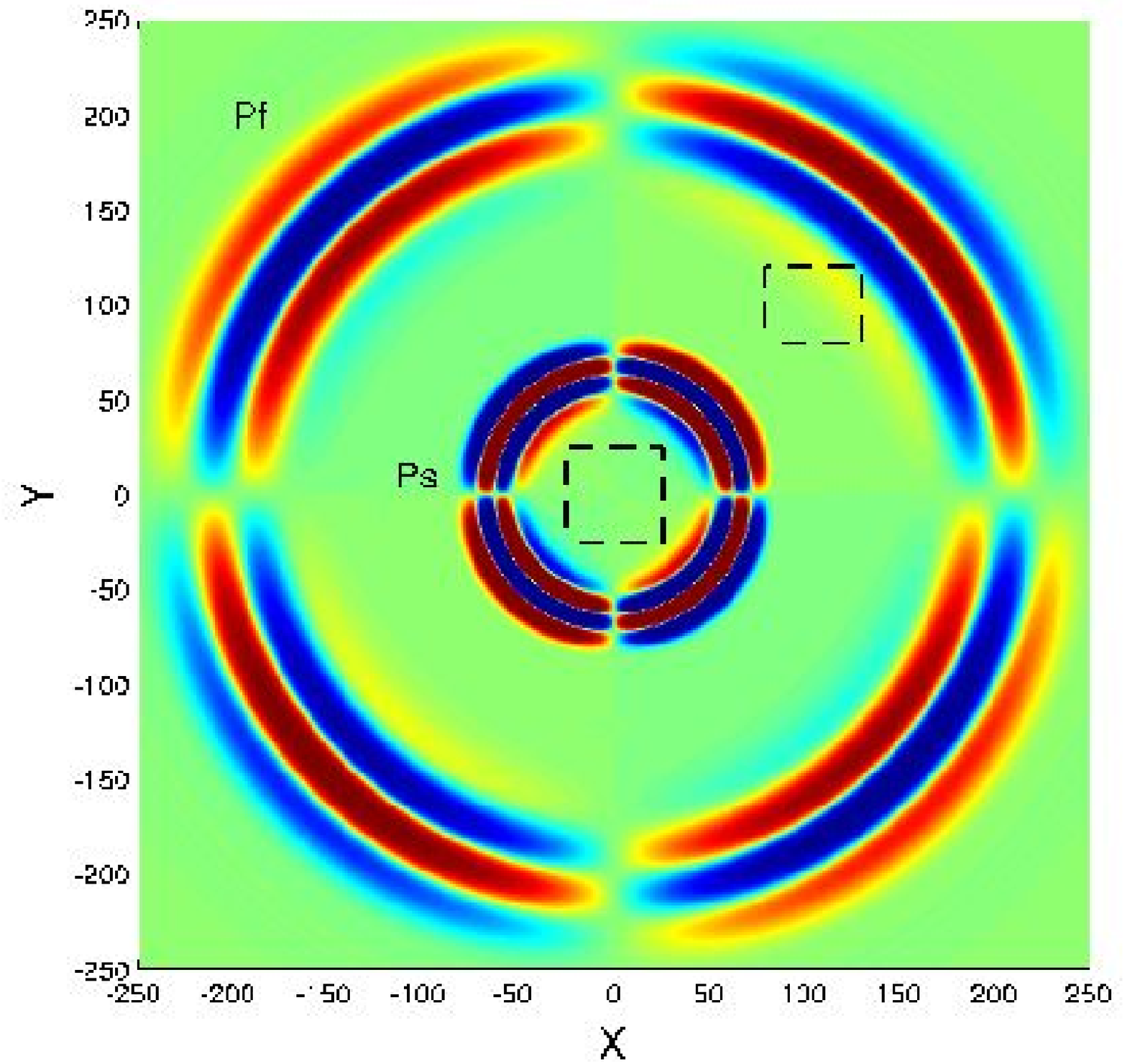}&
\includegraphics[scale=0.4]{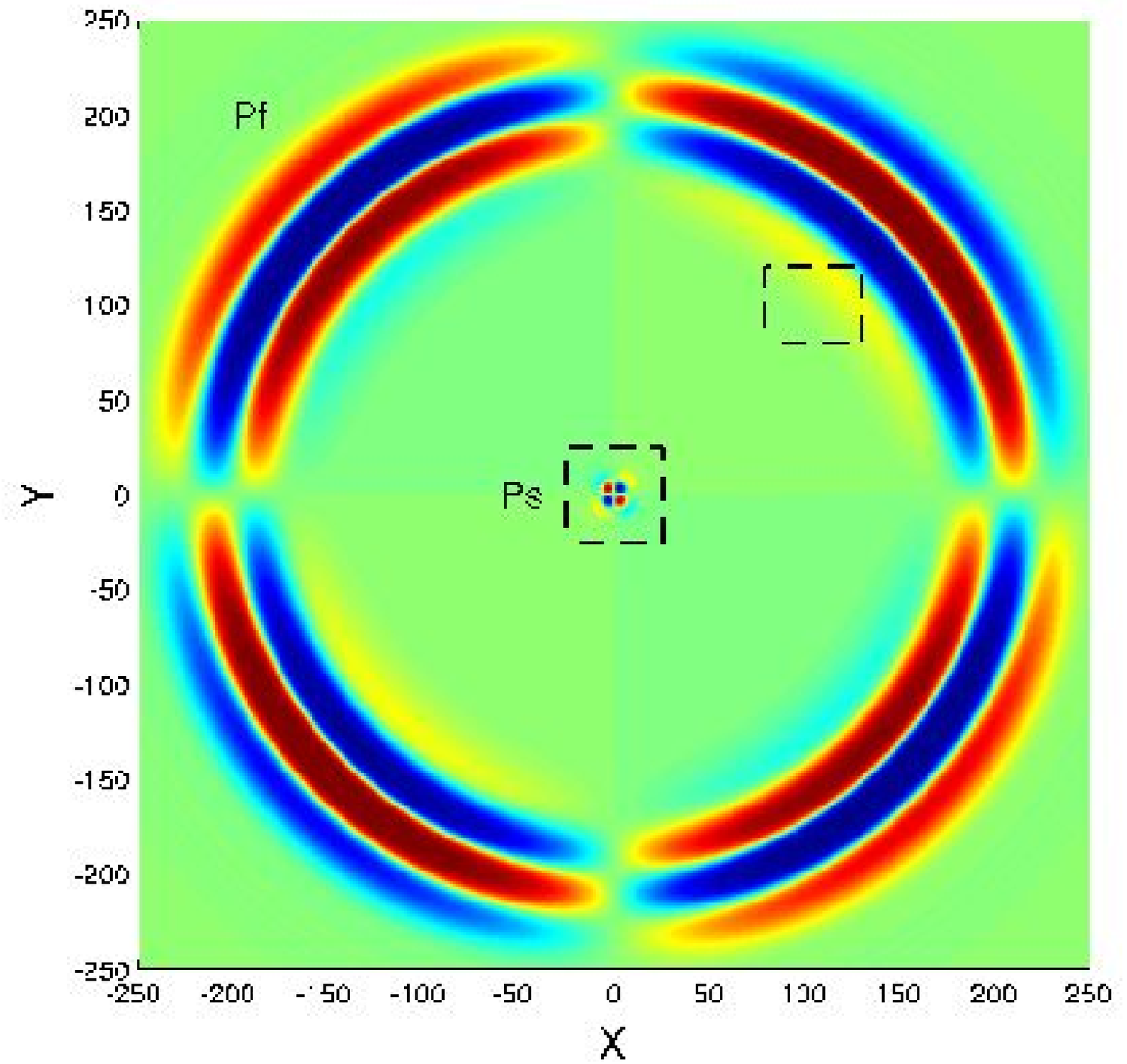}\\
$\sigma_{11}$& $\sigma_{11}$\\
\includegraphics[scale=0.4]{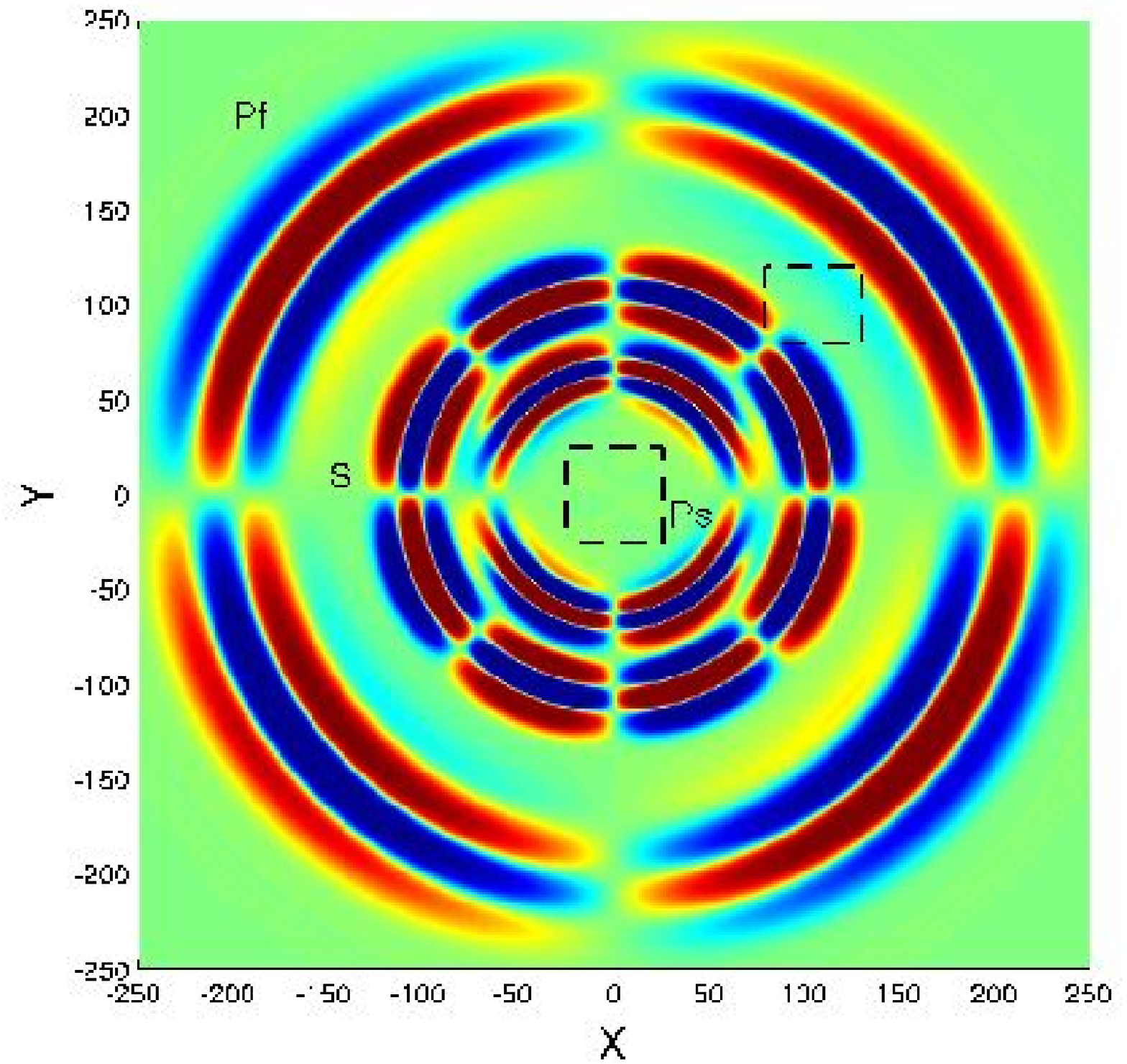}&
\includegraphics[scale=0.4]{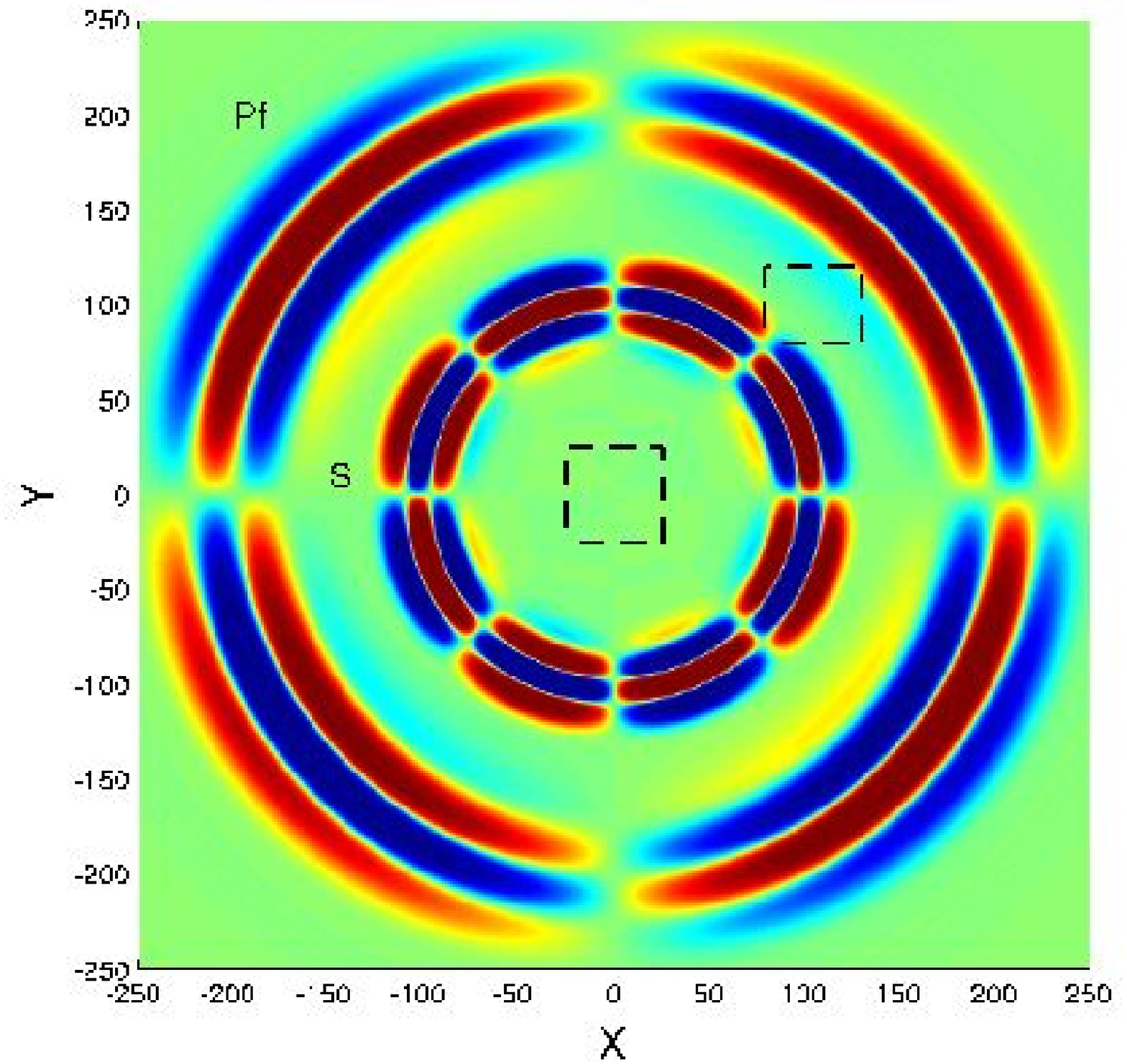}\\
$\sigma_{12}$& $\sigma_{12}$\\
\includegraphics[scale=0.4]{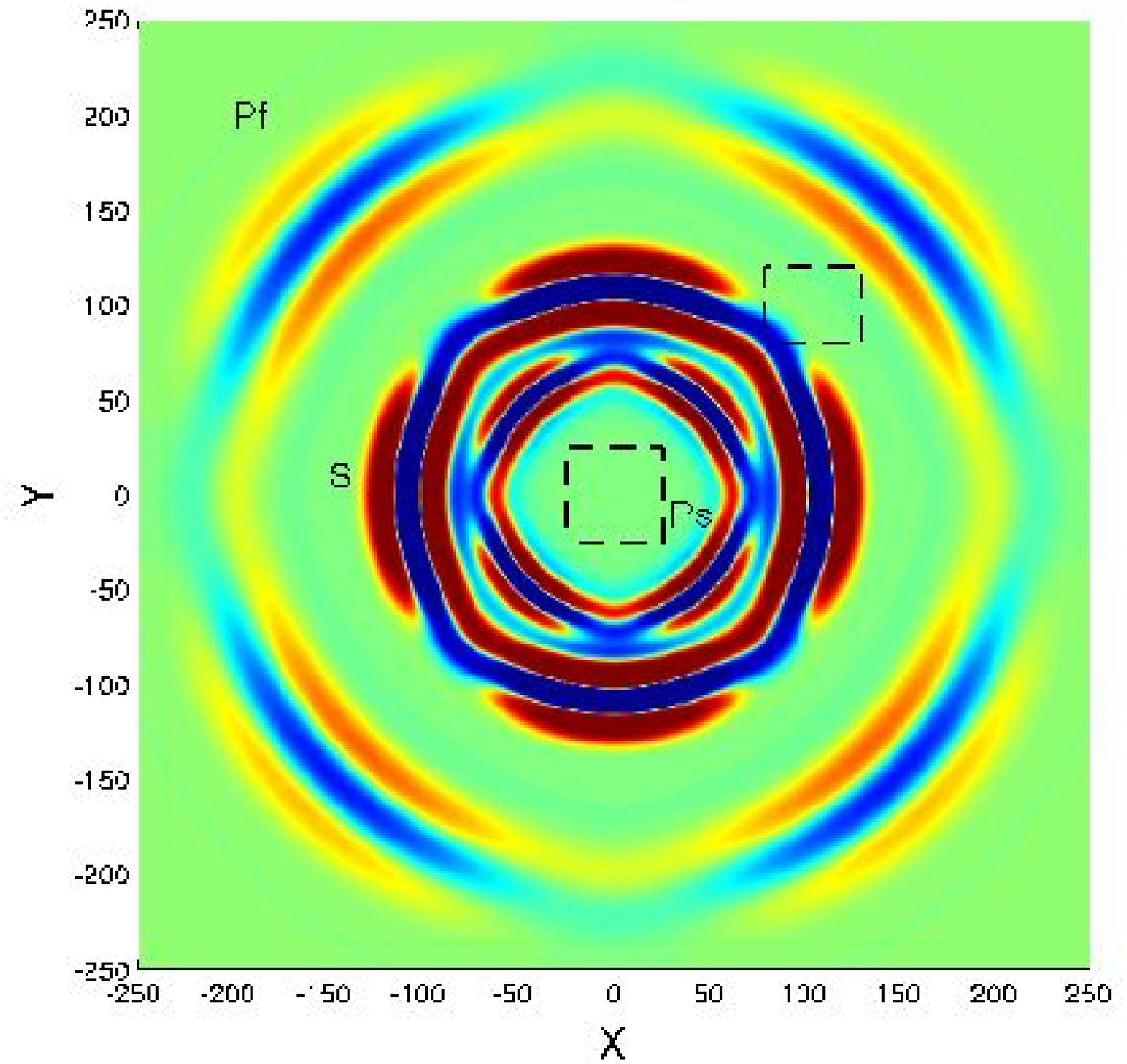}&
\includegraphics[scale=0.4]{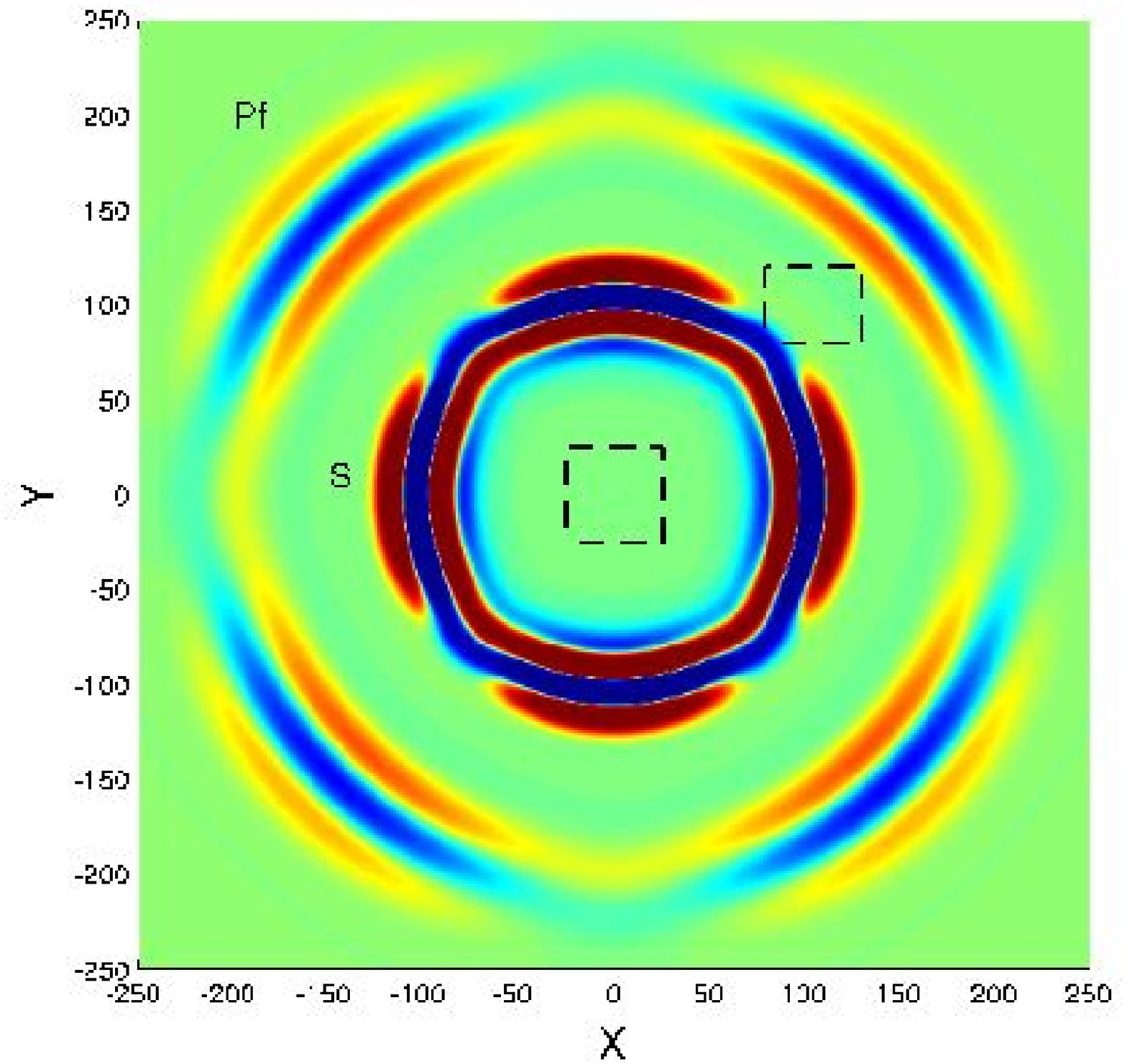}
\end{tabular}
\end{center}
\caption{Test 2. Snapshots of $p$, $\sigma_{11}$ and $\sigma_{12}$ of the fast ($p_f$), slow ($p_s$), and shear ($s$) waves emitted by a source point. Left column: viscosity $\eta=0$. Right column: viscous case. Dashed areas indicate the location of the refined grids ${\cal G}_1$ and ${\cal G}_2$.}
\label{carte_sourcepoint}
\end{figure}

In the second test, we focus on the coupling between the ADER scheme (section \ref{SecSplitting}) and the mesh refinement (section \ref{SecAMR}). For this purpose, a homogeneous medium $\Omega_0$ on a domain $[-250,\,250]$ m$^2$ is excited by the force density (\ref{SourcePonctuel}). The parameters of the source are: $x_s=y_s=0$, $f_0=40$ Hz, $\zeta=\frac{\overline{c}_{pf0}}{15\,f_0}$, $r_0=2\,\zeta$, and $\gamma=10^3$. The computational domain is discretized on a coarse mesh of $500^2$ points. Two locally refined areas are added: one around the source point ${\cal G}_1=[-25, 25]^2$, and one at ${\cal G}_2=[80, 130]\times [80, 120]$. Both grids are refined by a factor $q=5$, leading to $255^2$ points in ${\cal G}_1$ and $205\times 255$ points in ${\cal G}_2$.

\begin{figure}[htbp]
\begin{center}
\begin{tabular}{cc}
(a) & (b) \\
\includegraphics[scale=0.35]{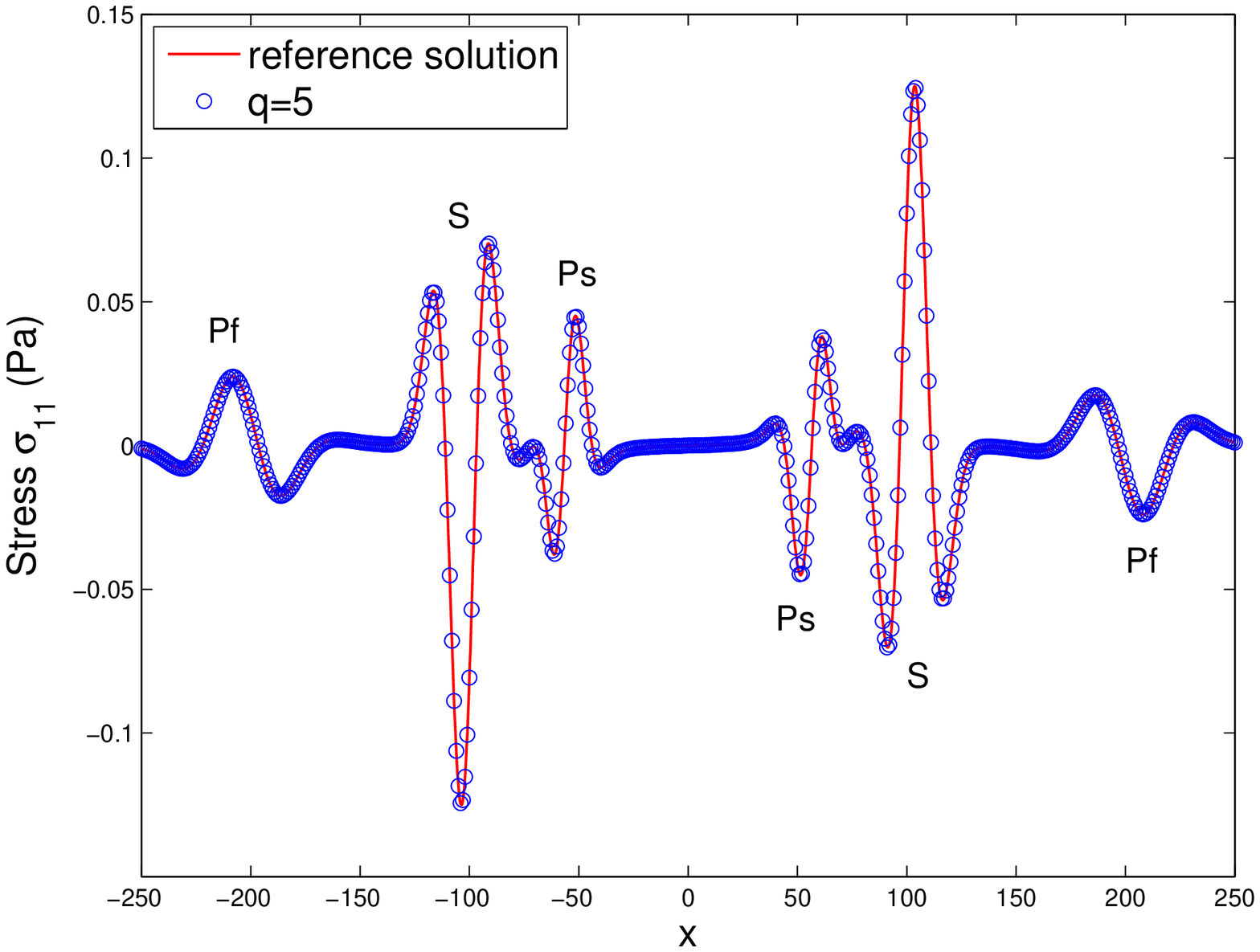}&
\includegraphics[scale=0.35]{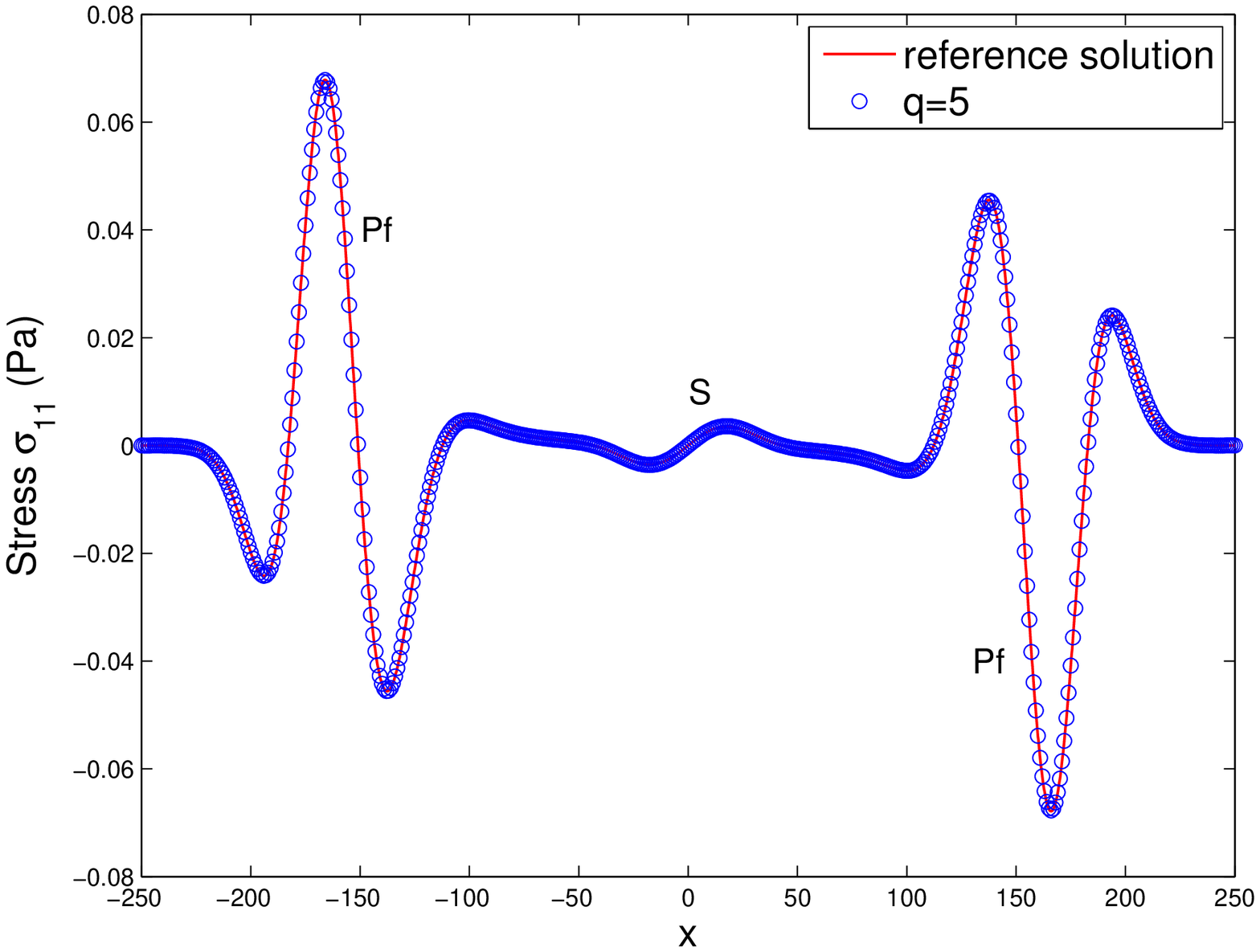}\\
(c) & (d) \\
\includegraphics[scale=0.35]{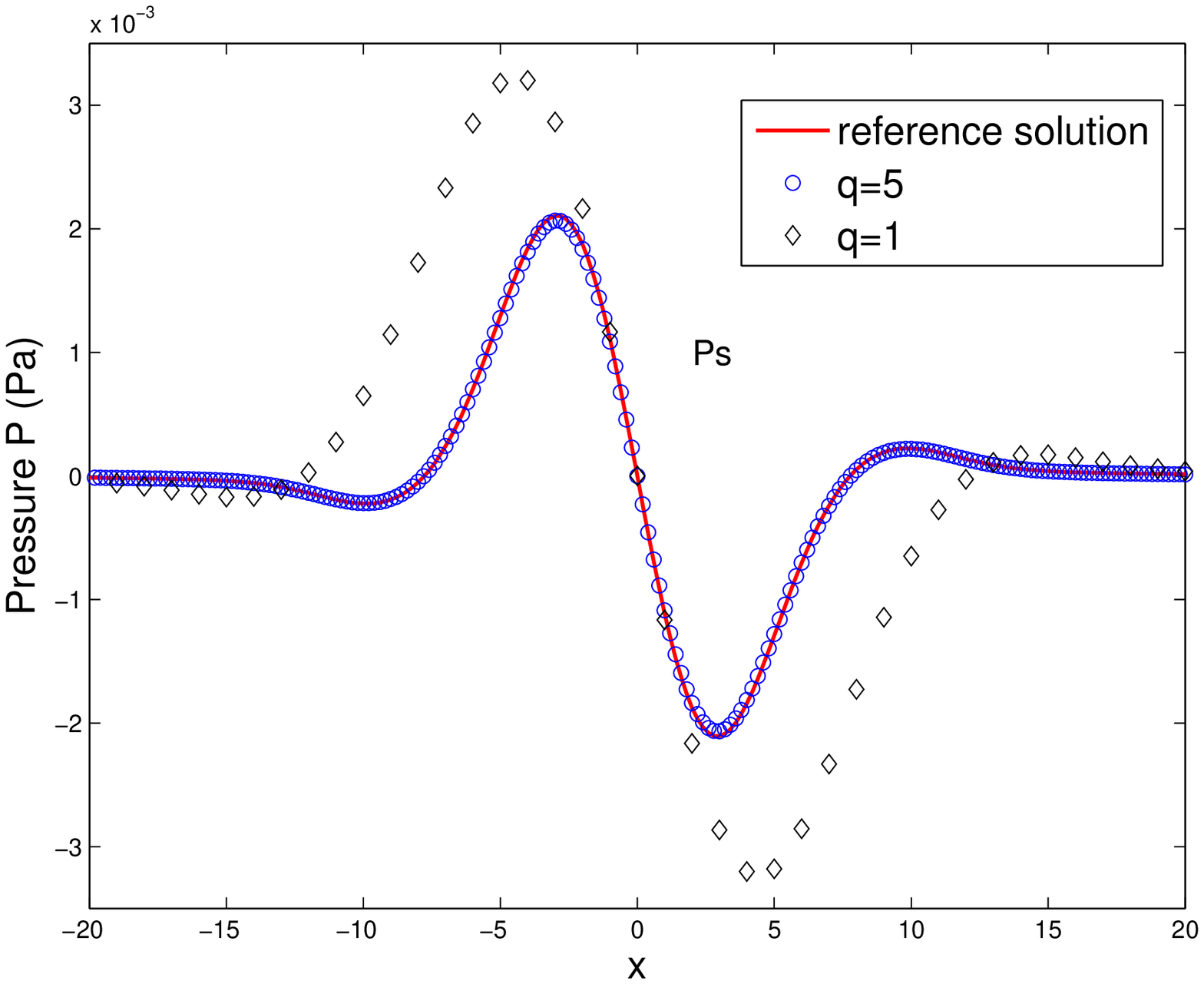}&
\includegraphics[scale=0.35]{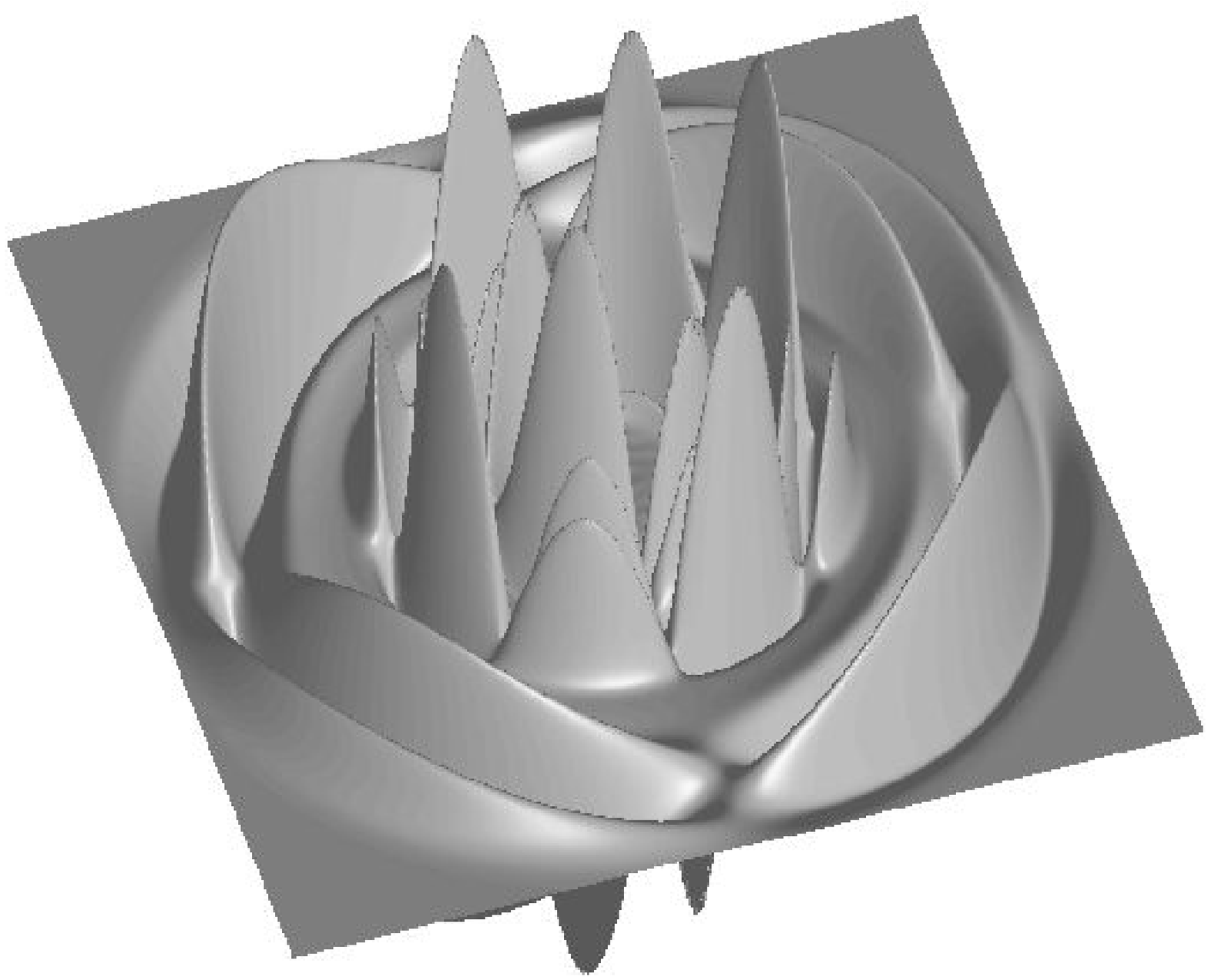}
\end{tabular}
\end{center}
\caption{Test 2 in the inviscid case, corresponding to the left column of figure \ref{carte_sourcepoint}. Stress $\sigma_{11}$ along lines $y=30$ m (a) and $y=130$ m (b). Comparison between the fine grid reference solution and the solution obtained with a coarse grid and two refined areas of factor $q=5$ (c). 3D representation of $\sigma_{11}$ (d).}
\label{coupe_sourcepoint}
\end{figure}

Figure \ref{carte_sourcepoint} shows snapshots of the fields after 280 iterations. At this time $t\simeq 4.4\,/\,f_0$, the fast compressional wave has also crossed  the grid ${\cal G}_2$. In the left column, one takes $\eta=0$, and consequently all the waves generated by the source propagate radially. In the right column, the viscosity is non zero, and the slow compressional wave remains localized around the source point. This wave is clearly observed on the pressure field, but is also present in the stress fields although not visible on the corresponding panels of figure \ref{carte_sourcepoint}.

In order to evaluate the influence of the refined areas, a reference solution is computed using a fine mesh $\Delta x=\Delta y=1/5$ m on the whole computational domain. As observed in figure \ref{coupe_sourcepoint}-(a),(b), the local space-time refinement applied in ${\cal G}_1$ and ${\cal G}_2$ does not lead to significant spurious reflections during the propagation of the waves. On the contrary and as expected, the refinement around the source point leads to a much better resolution of the slow wave in the viscous case: see figure \ref{coupe_sourcepoint}-(c).
 
In conclusion, mesh refinement coupled with the ADER scheme (with or without splitting)  accurately represents the behavior of the different poroelastic waves in an homogeneous medium, even though they present a spatially complex structure, as observed in figure \ref{coupe_sourcepoint}-(d). 

%------------------------------------------------------------------------------------------

\subsection{Test 3: stability of the complete algorithm}\label{SecNumTest3}

Some theoretical results based on GKS theory exist concerning the stability of immersed interface methods or local mesh refinement \cite{GUSTAFSSON75,BERGER85,TREFETHEN85,LIN91,BERGER98,HDR-LOMBARD}. However, these analyses have been done mainly in the case of one-dimensional model problems and basic numerical schemes. The present algorithm combines more sophisticated numerical methods in 2D, and hence GKS analysis is out of reach. The only reasonable way to confirm the stability of the full method is to perform numerical perturbation tests. 

For this purpose, we consider a heterogeneous porous media made up by a matrix $\Omega_0$ with a cylindrical scatterer $\Omega_1$ of radius $r=10$ m, centered in a domain $[-50, 50] \, \rm{m}^2$. In both $\Omega_0$ and $\Omega_1$, the viscosity of the saturating fluids are taken into account ($\eta \neq 0$). All the components of ${\bf U}$ (\ref{DefU}) are initialized randomly at each grid points, and the source ${\bf F}$ is set to zero. The computational domain $[-50, 50] \, \rm{m}^2$ is discretized by a coarse grid. The scatterer is included in a grid ${\cal G}_1$ of size $15\times 15 \ \rm{m^2}$ with a large refinement factor $q=9$. 

\begin{figure}[htbp]
\begin{center}
\includegraphics[scale=0.4]{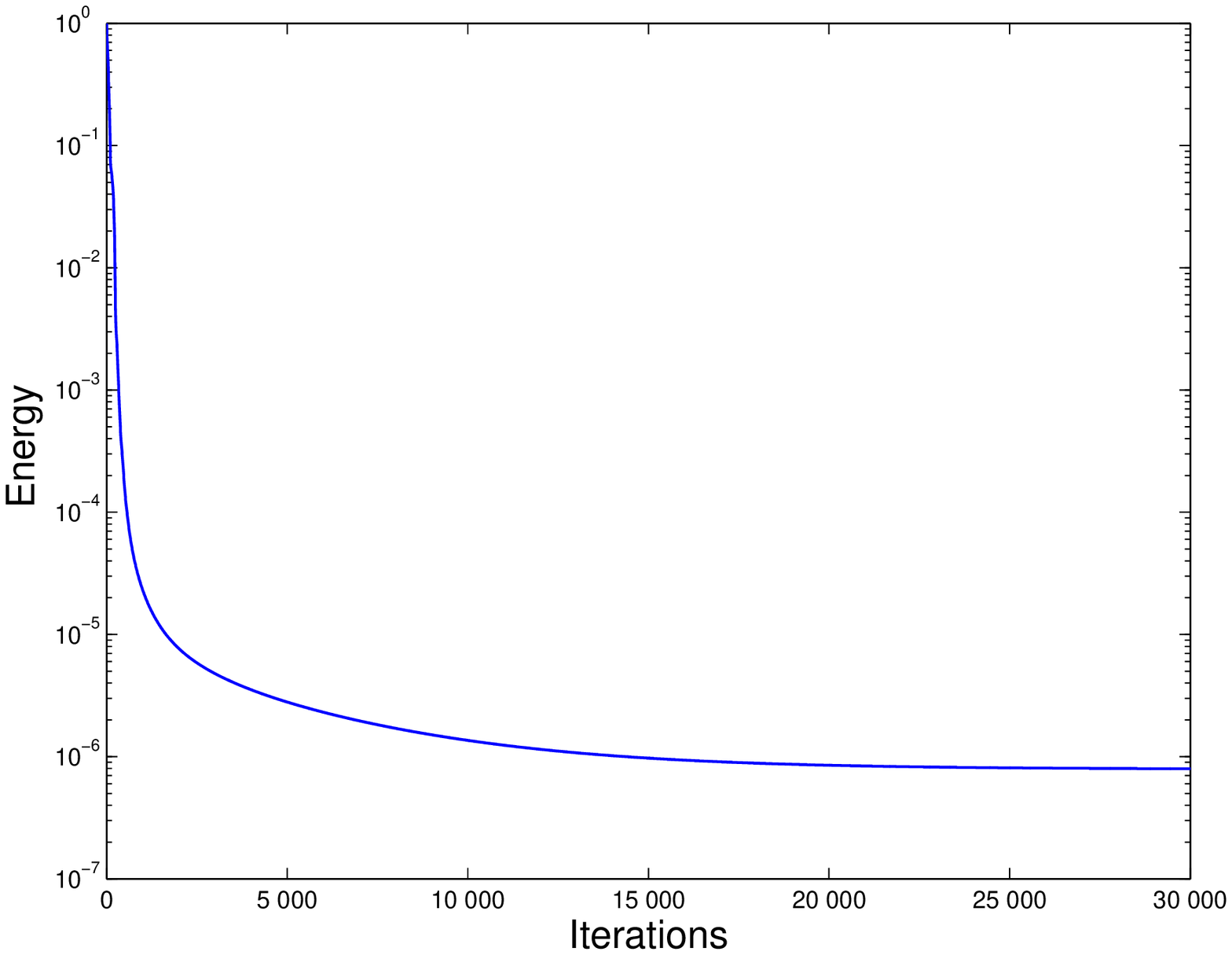}
\end{center}
\caption{Test 3. Time evolution of the computed total energy (\ref{DefEnergie}).}
\label{figStab}
\end{figure}

In figure \ref{figStab}, the energy of poroelastic waves (\ref{DefEnergie}) is displayed during $3\,10^4$ iterations. Theoretically, this energy should slowly decrease, depending on ${\bf w}$ in (\ref{VariationNRJ}). At the beginning of the simulations, a large decrease of $E$ is observed. It is logically induced by the random non-smooth initial field, which generates a large numerical dissipation. After roughly 1000 time steps, a smooth field is reached and  the mechanical energy slowly decreases. This confirms the stability of the full numerical method, involving the ADER scheme with splitting, mesh refinement, and  immersed interface method. 

%------------------------------------------------------------------------------------------

\subsection{Test 4: diffraction of a plane wave by a cylinder}\label{SecNumTest4}

\begin{figure}[htbp]
\begin{center}
\begin{tabular}{cc}
(a) & (b)\\
\includegraphics[scale=0.40]{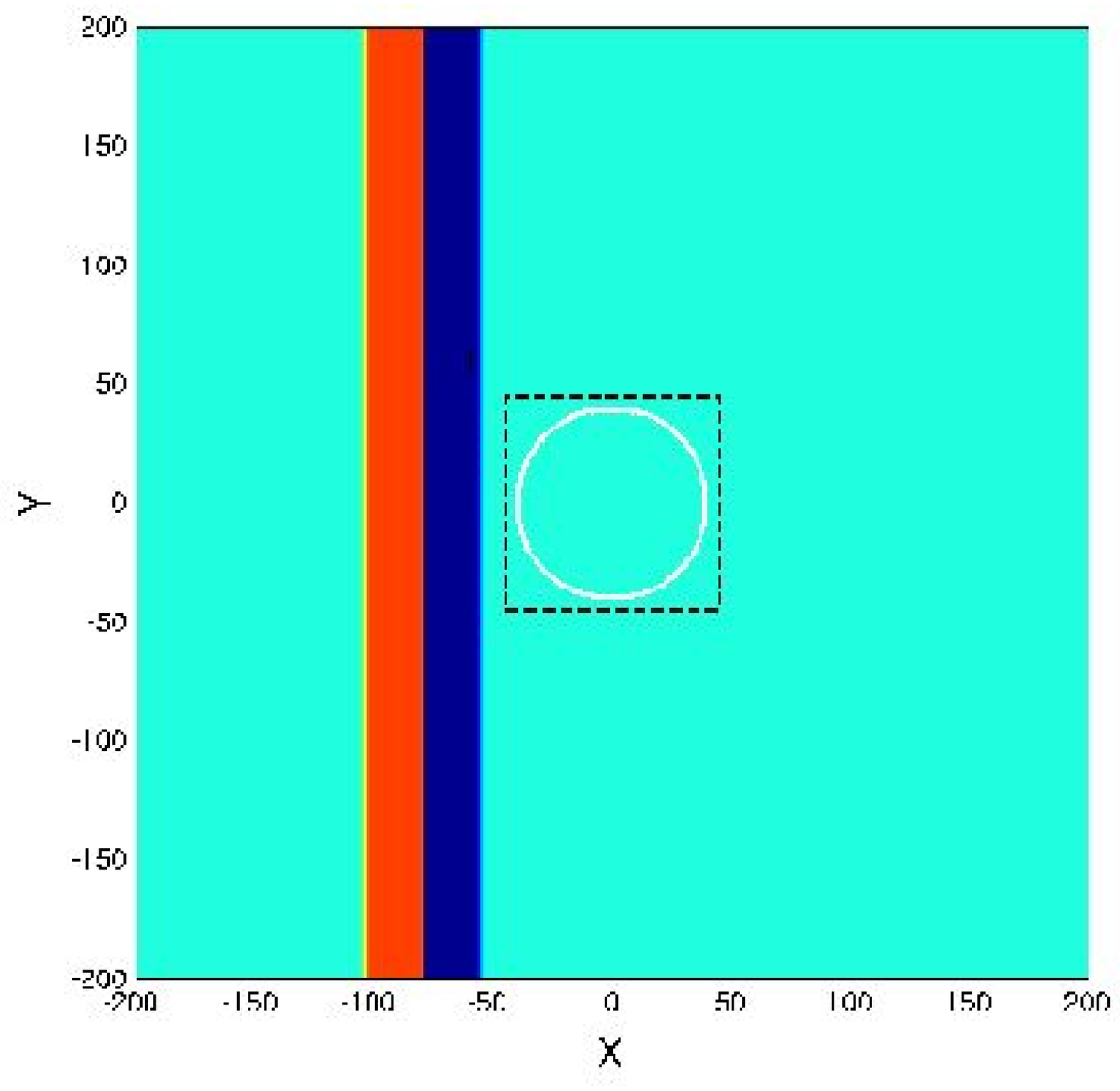}&
\includegraphics[height=5.7cm,width=6cm]{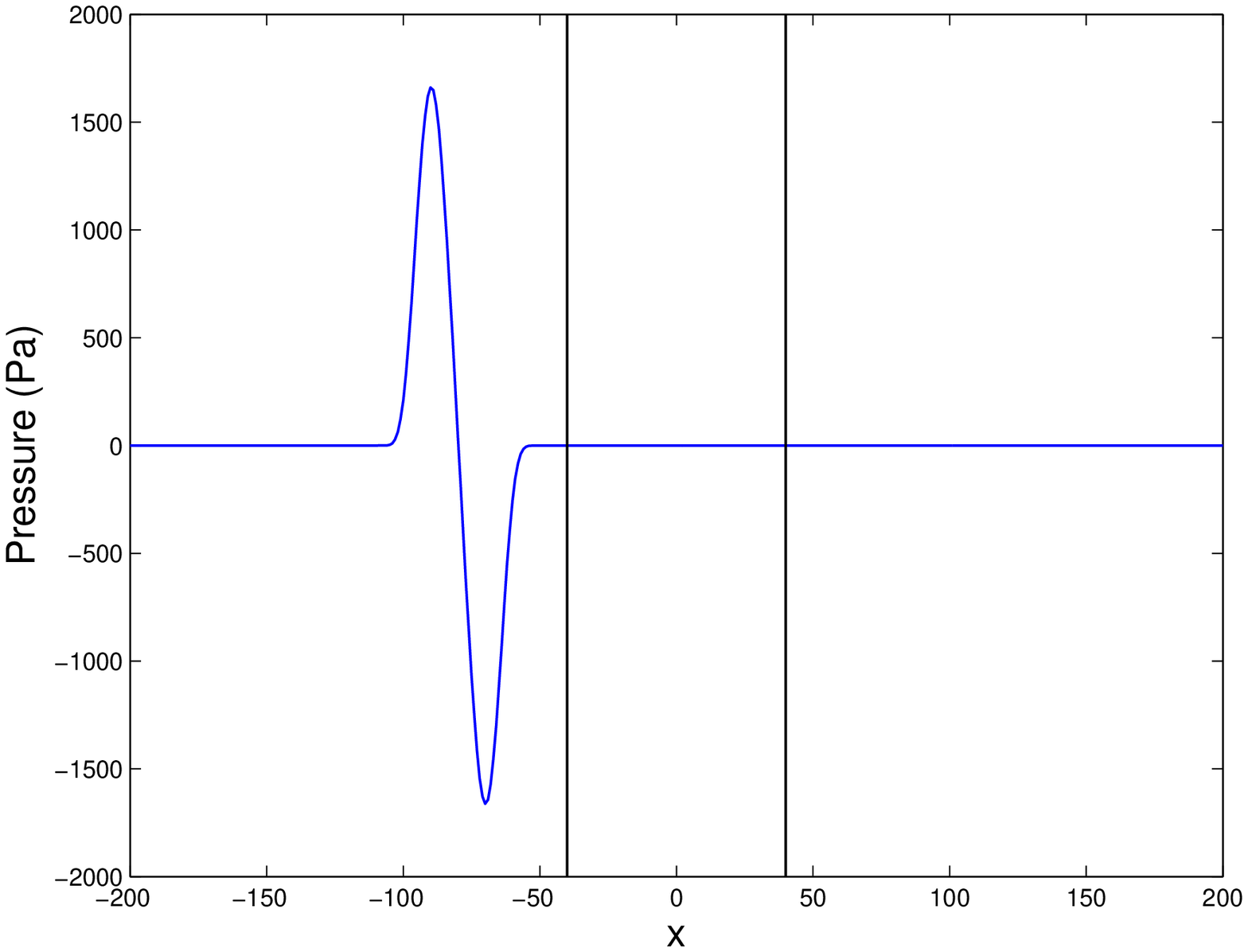}\\
(c) & (d)\\
\includegraphics[scale=0.40]{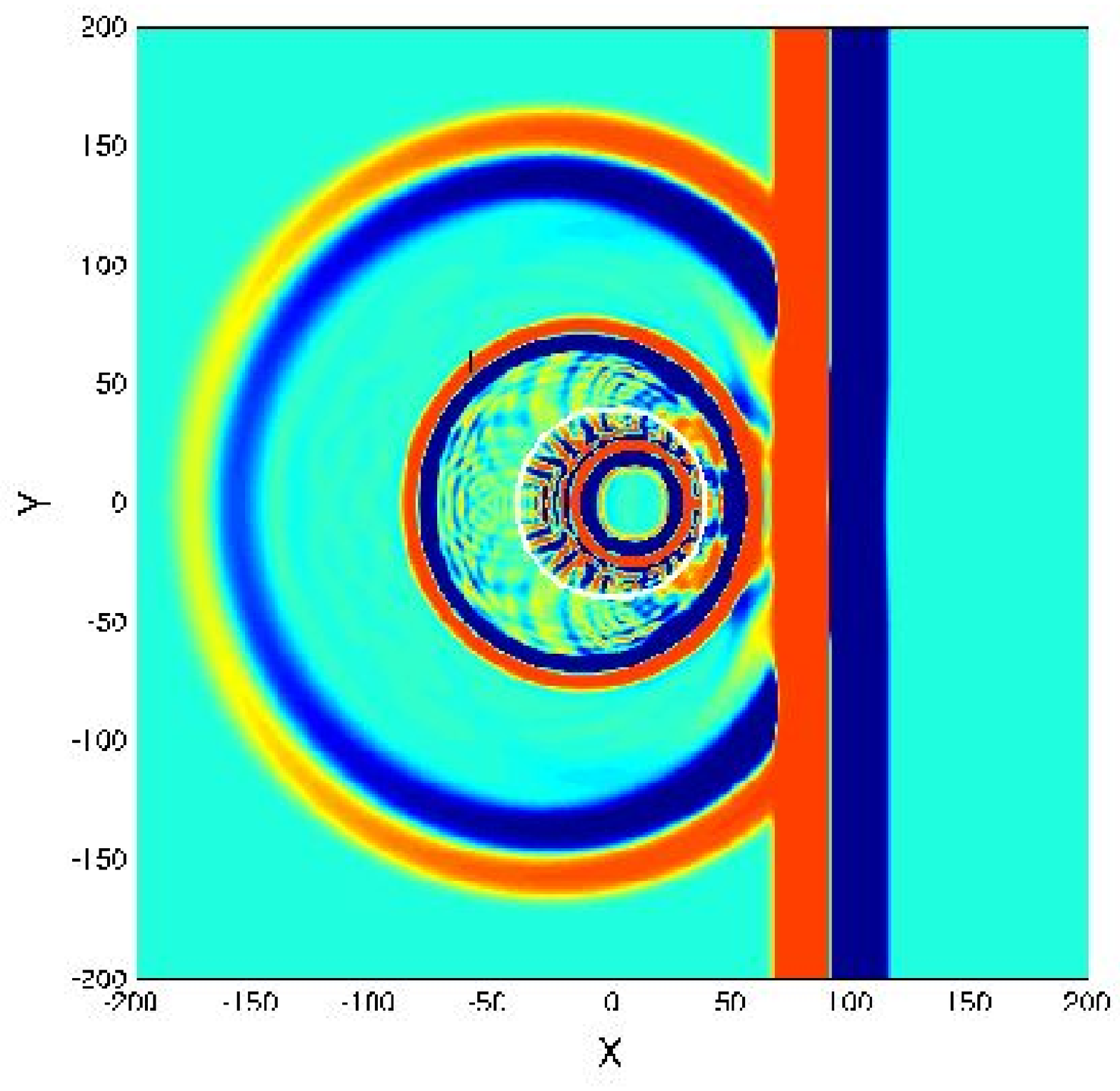}&
\includegraphics[scale=0.40]{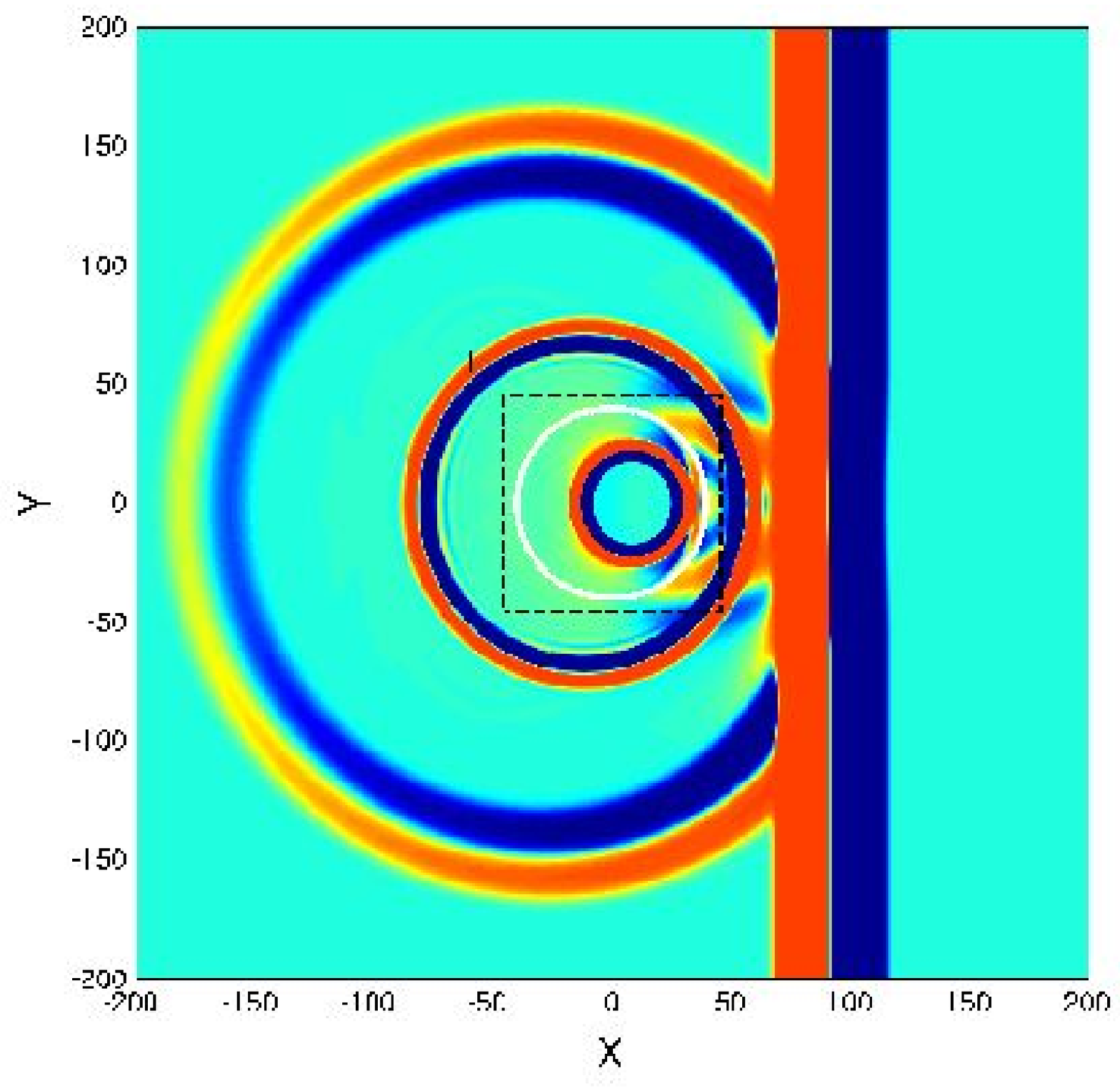}
\end{tabular}
\end{center}
\caption{Test 4, inviscid fluids. (a): $p$ at initial time. (b): $p$ along the line $y=0$; vertical lines denote the frontiers of the scatterer. (c-d): $p$ after 180 iterations. (c): no refinement nor interface method ($q=1$, $r=0$). (d): refinement factor $q=5$ and third order immersed interface method ($r=3$). Dotted square represents the frontiers of the refined domain. The cross denotes the location of the receiver. }
\label{FigCercleNonvisq}
\end{figure}

A cylindrical scatterer $\Omega_1$ of radius $40$ m is centered at point $(0,\,0)$ in the computational domain $[-200, 200] \, \rm{m}^2$. The source is a plane wave (\ref{OndePlaneFourier}) initially in medium $\Omega_0$, with parameters: $\gamma=1$, $f_0=40$ Hz, $t=-2.09\,10^{-2}$ s, and $\theta=0$ degree. The initial conditions are illustrated in figure \ref{FigCercleNonvisq} (a-b).

In figure \ref{FigCercleNonvisq}, the viscosity has been canceled in both media $\Omega_0$ and $\Omega_1$. In such a configuration, the analytical solution of (\ref{DefU}) can be computed using Fourier and Bessel expansions and is used to validate the simulations. The diffracted waves propagate with velocities $\overline{c}_{pf}, \overline{c}_{ps}$ and $\overline{c}_{s}$ given in table \ref{TabParametres}. To ensure the same number of points per wavelength for all the diffracted waves (see section \ref{SecAMR}), the computational domain is locally refined by inserting the cylinder in the grid ${\cal G}_1=[-45, 45]^2$ with a refinement factor $q=5$, deduced from relation (\ref{factq}) and table \ref{TabParametres}. The influence of this refinement combined with the immersed interface method is clearly visible in figure \ref{FigCercleNonvisq}.

\begin{figure}[htbp]
\begin{center}
\begin{tabular}{cc}
(a)&(b)\\
\includegraphics[scale=0.35]{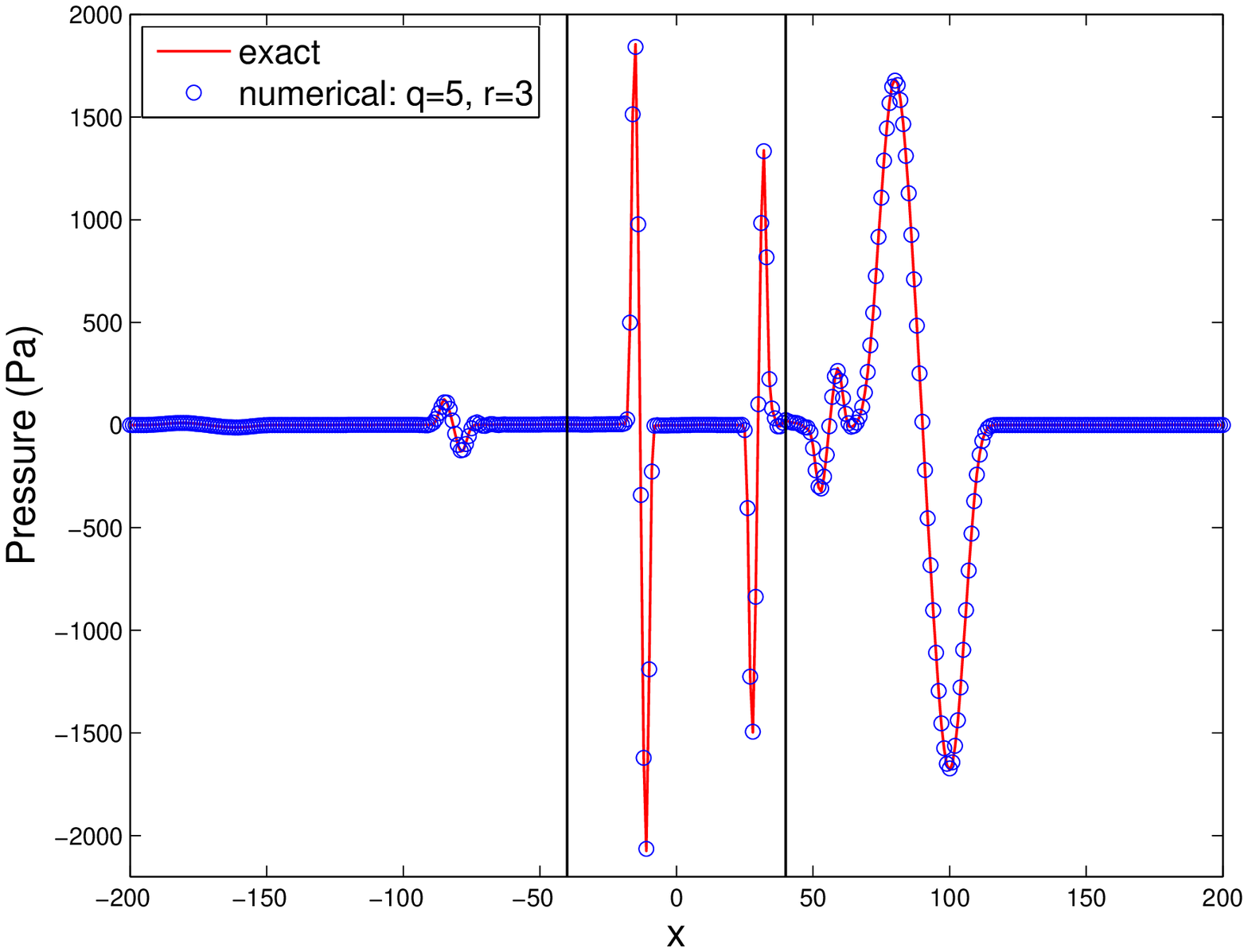}&
\includegraphics[scale=0.35]{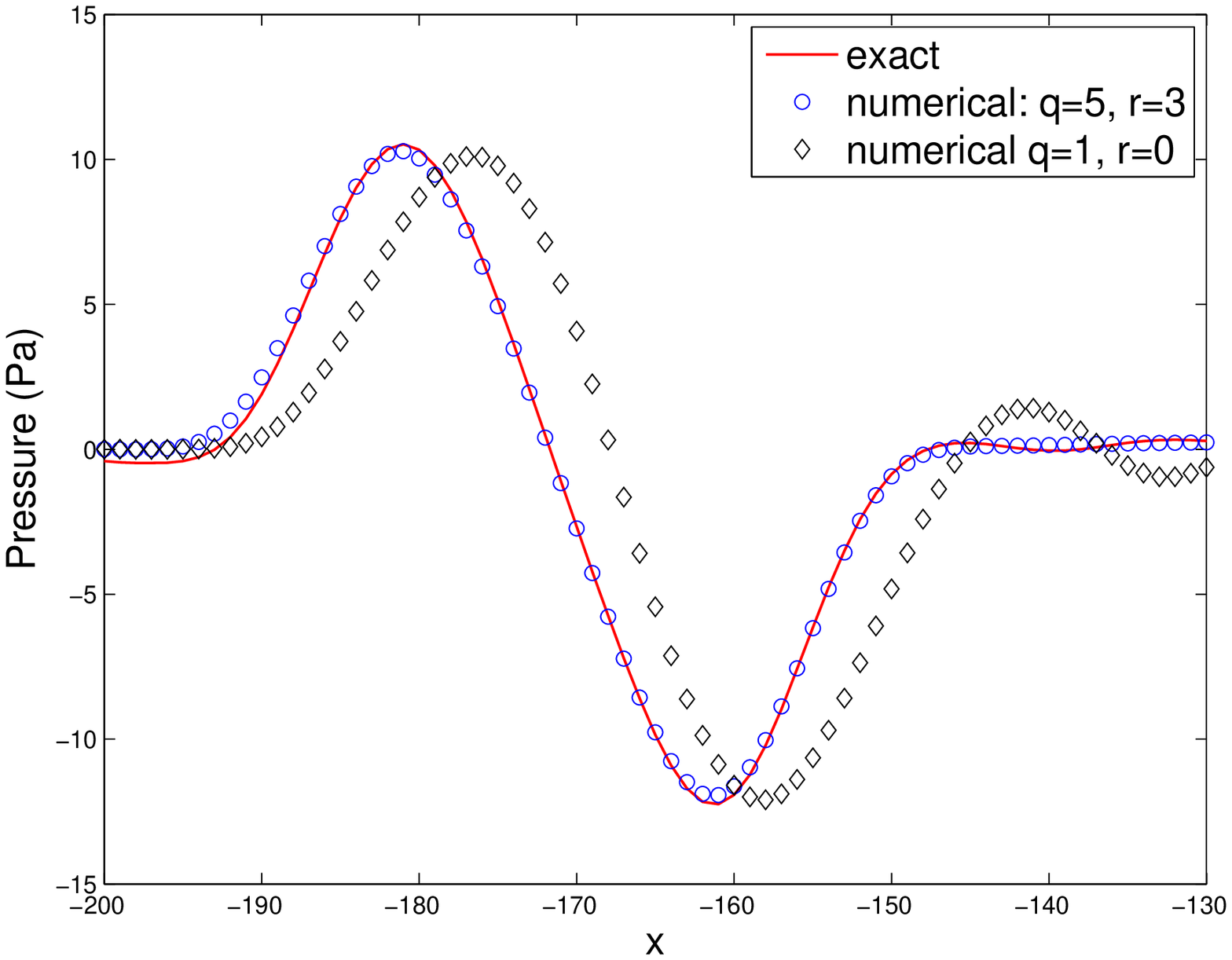} \\
(c) & (d)\\
\includegraphics[scale=0.35]{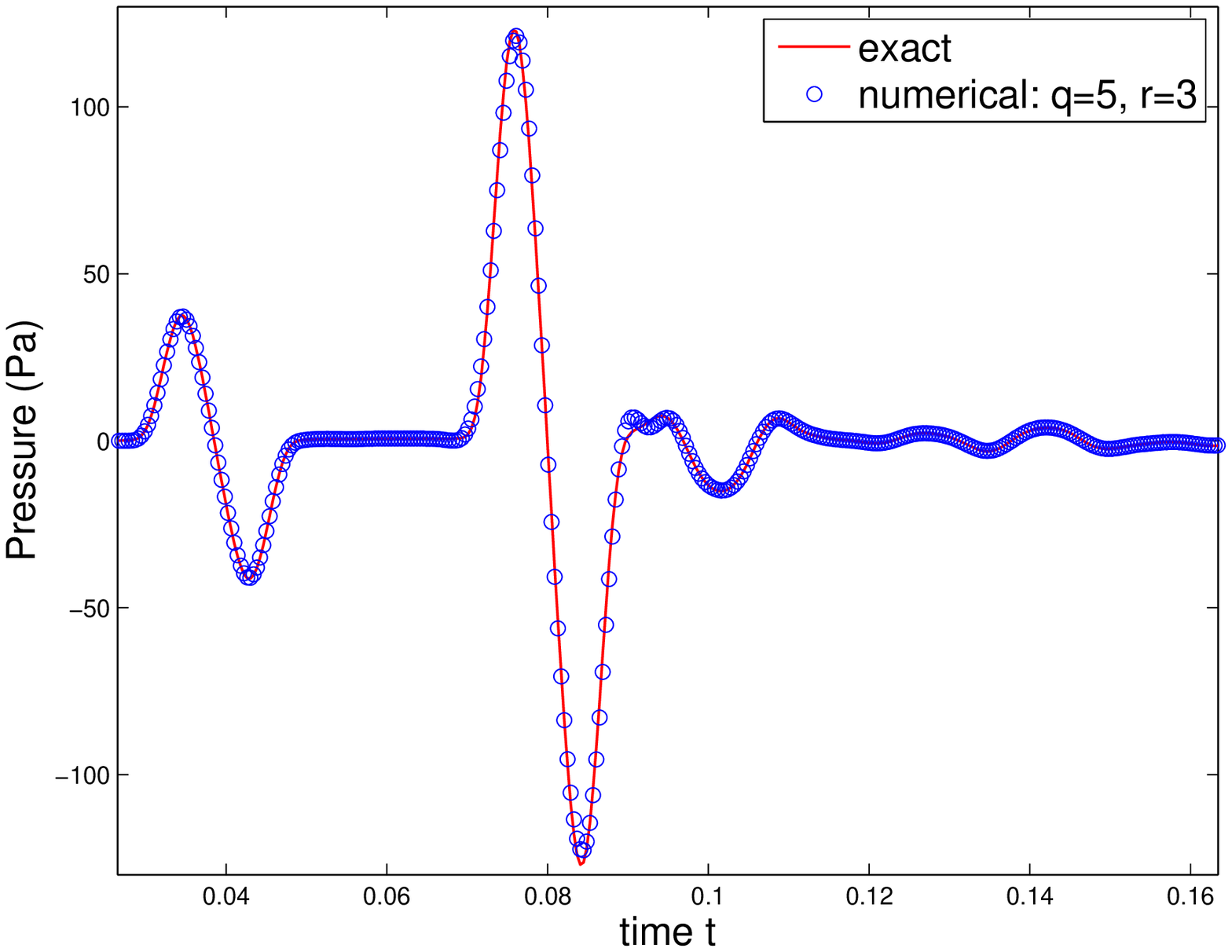}& 
\includegraphics[scale=0.35]{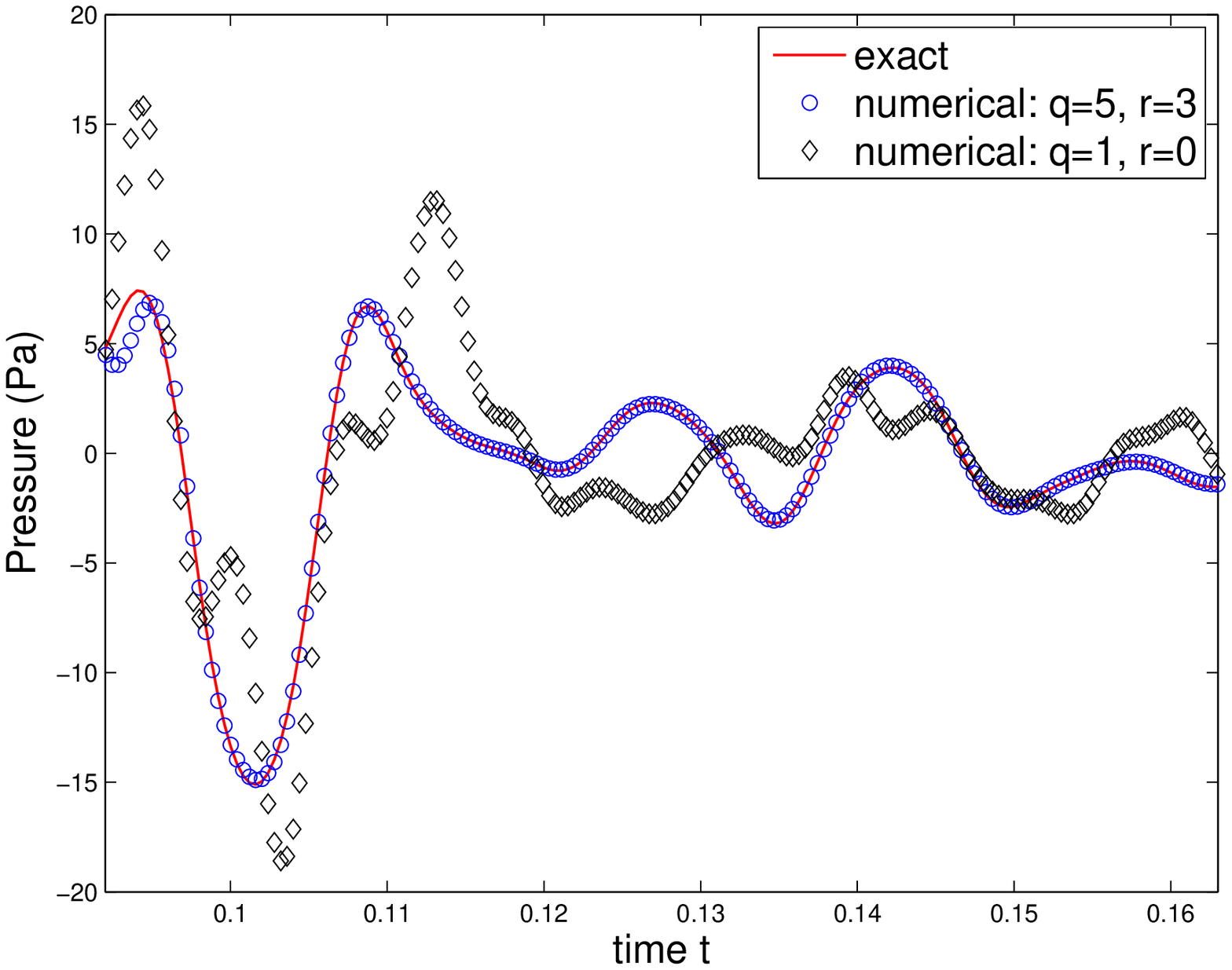}
\end{tabular}
\end{center}
\caption{Test 4, inviscid fluids. (a): $p$ on the line $y=0$ after 180 iterations. (b): zoom around the reflected fast compressional wave. (c): time evolution of $p$ recorded at receiver $(x\,=\,-60,y\,=\,60)$ from $t=0.028$ s to the final instant. (d) zoom on the time evolution of scattered waves.}
\label{FigCoupeNonvisq}
\end{figure}

Without refinement nor interface method ($q=1$, $r=0$, left column), the waves created during the interaction with the scatterer are polluted by spurious numerical artifacts. With $q=5$ and $r=3$ (right column), these non-physical perturbations disappear and the reflected-transmitted waves are correctly computed. Comparisons with the exact solution presented on figure \ref{FigCoupeNonvisq} confirm the accuracy of the simulation. Without mesh refinement nor immersed interface method, inaccurate results are obtained, especially for the reflected fast compressional wave where a shift of about 5 m is observed (figure \ref{FigCoupeNonvisq}-(b)).

In figure \ref{FigCoupeNonvisq} -(c) and (d), the time evolution of the pressure registered at the point $(-60,\,60)$ is presented from $t=0.028$ s, in order to avoid the incident initial wave. We observe the fast reflected wave, followed by the slow wave and a combination of diffracted fast and slow waves. Once again, the full strategy captures accurately all the temporal variations of the solution, while the basic algorithm gives very poor results, especially concerning the waves diffracted by the scatterer (see figure \ref{FigCoupeNonvisq}-(d)).\\

\begin{figure}[htbp]
\begin{center}
\begin{tabular}{cc}
(a) & (b)\\
\includegraphics[scale=0.37]{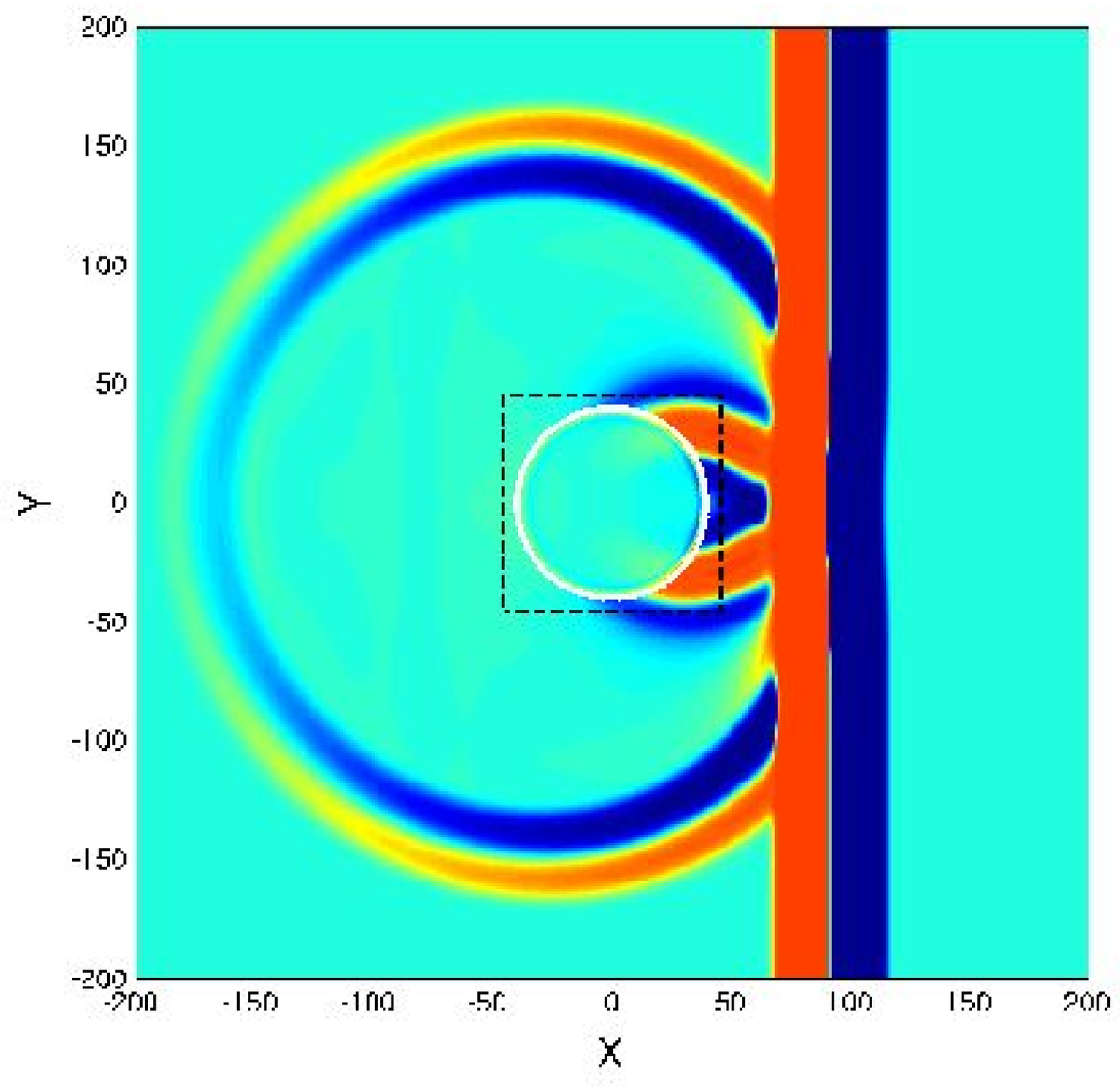}&
\includegraphics[scale=0.37]{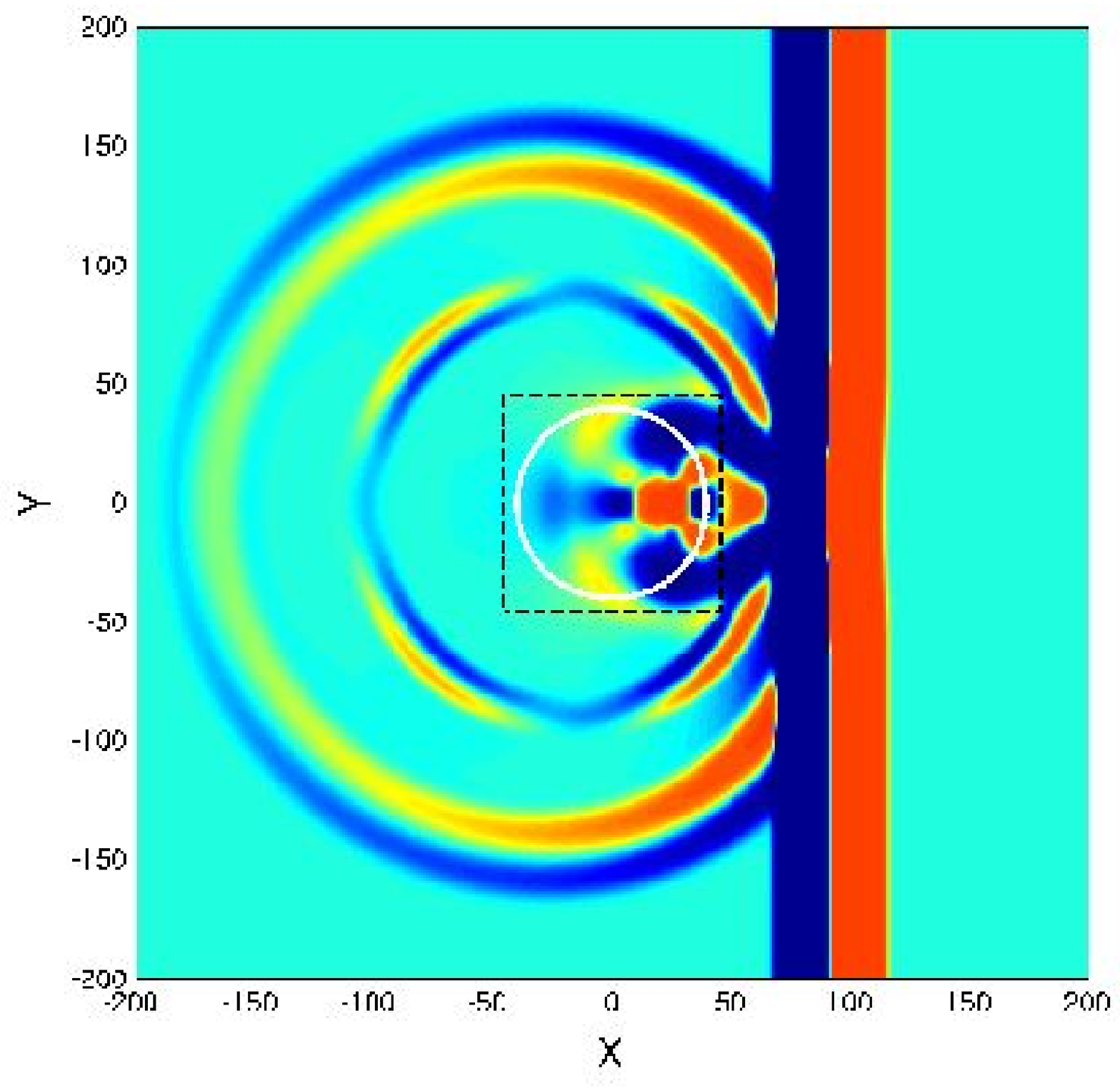}\\
(c) & (d)\\
\includegraphics[scale=0.37]{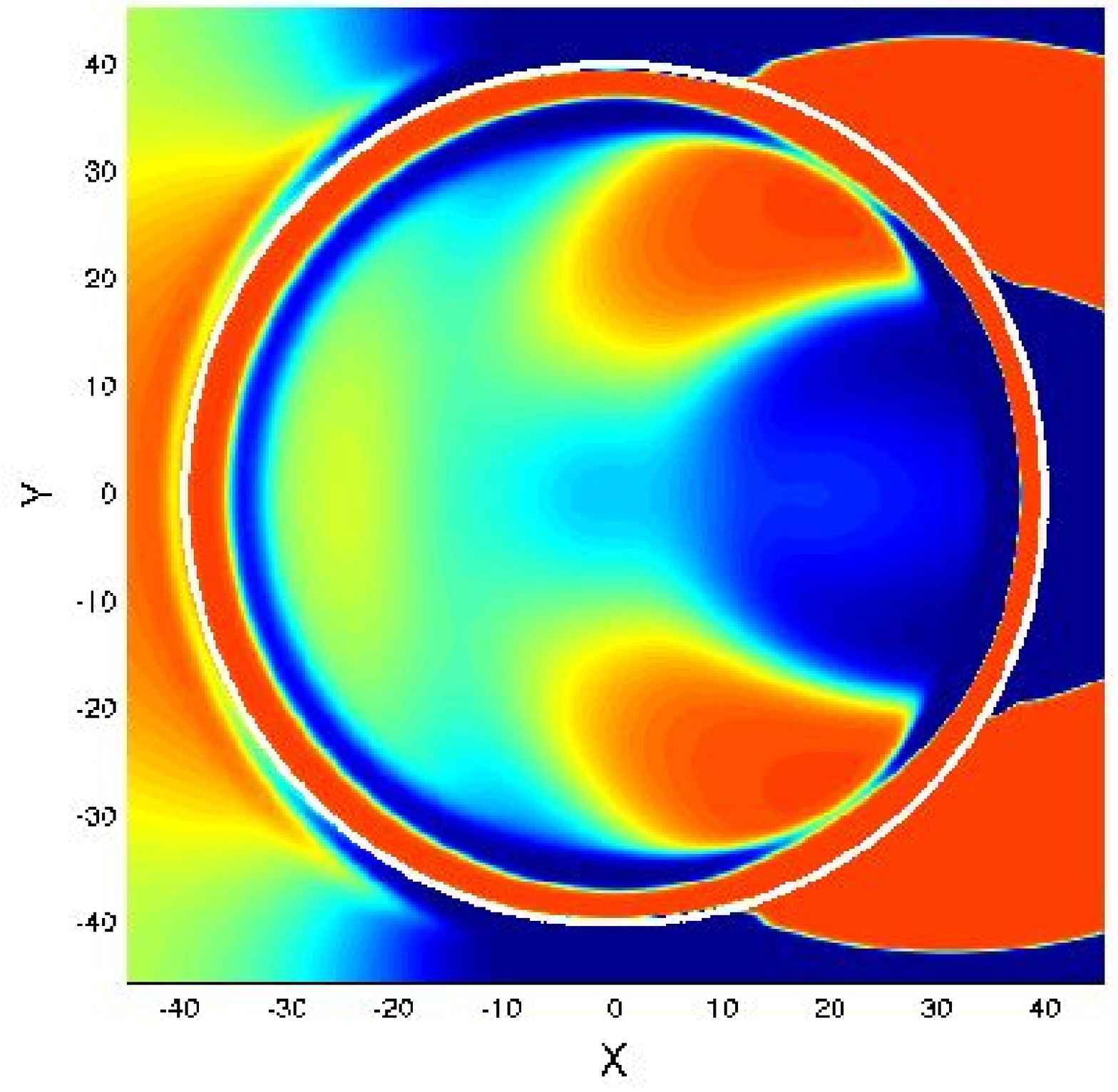}&
\includegraphics[scale=0.37]{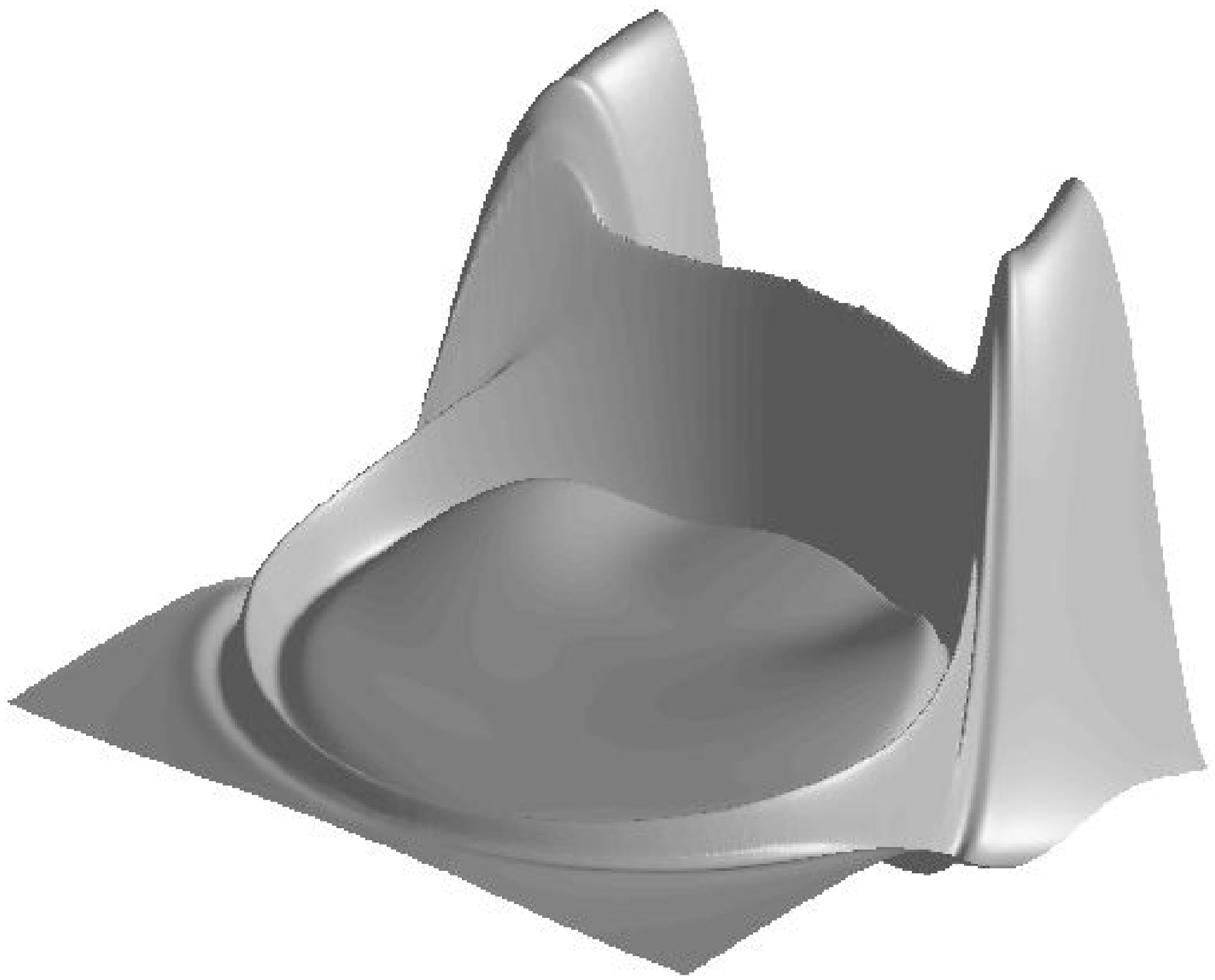}\\
(e)&(f)\\
\includegraphics[scale=0.31]{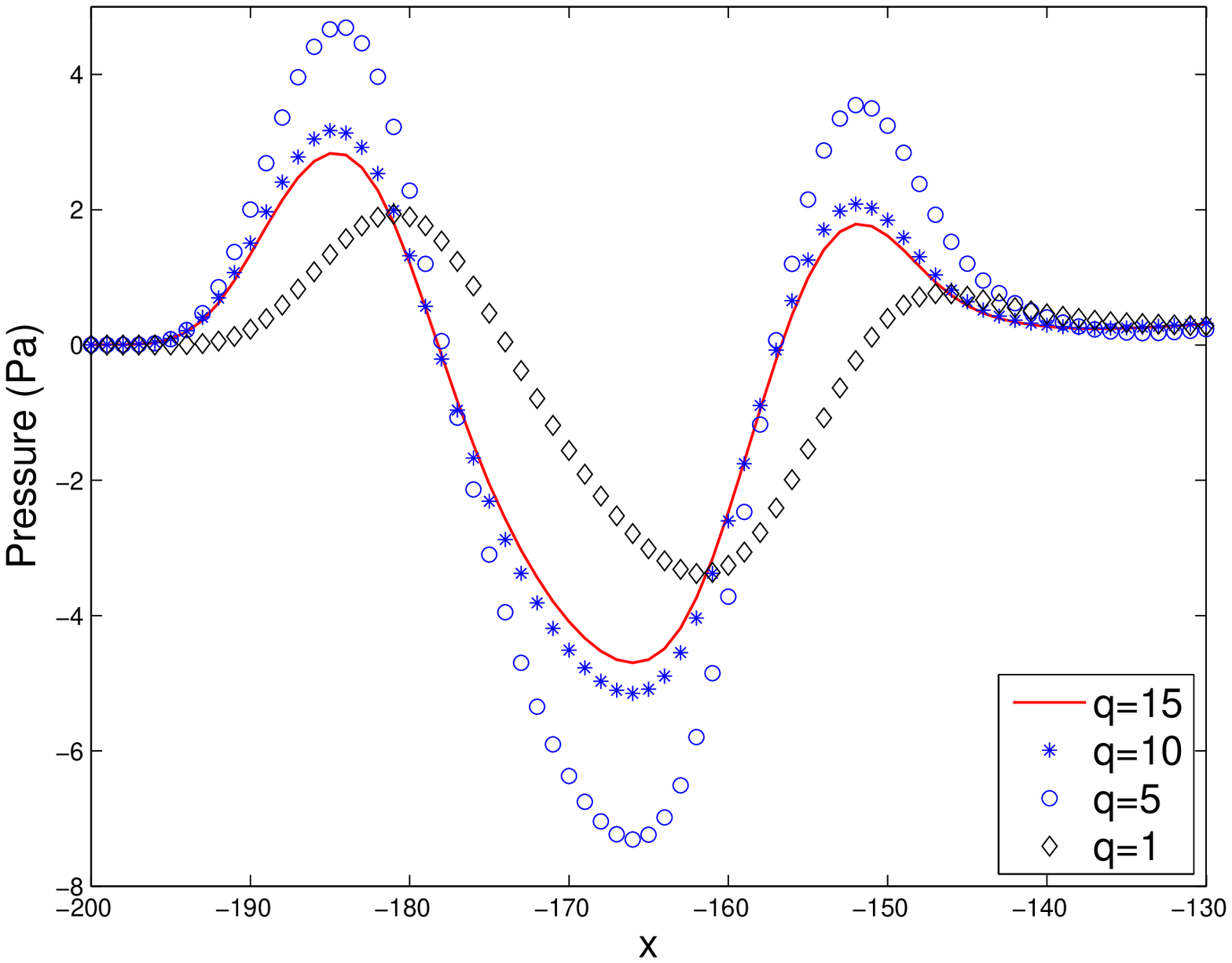}& 
\includegraphics[scale=0.31]{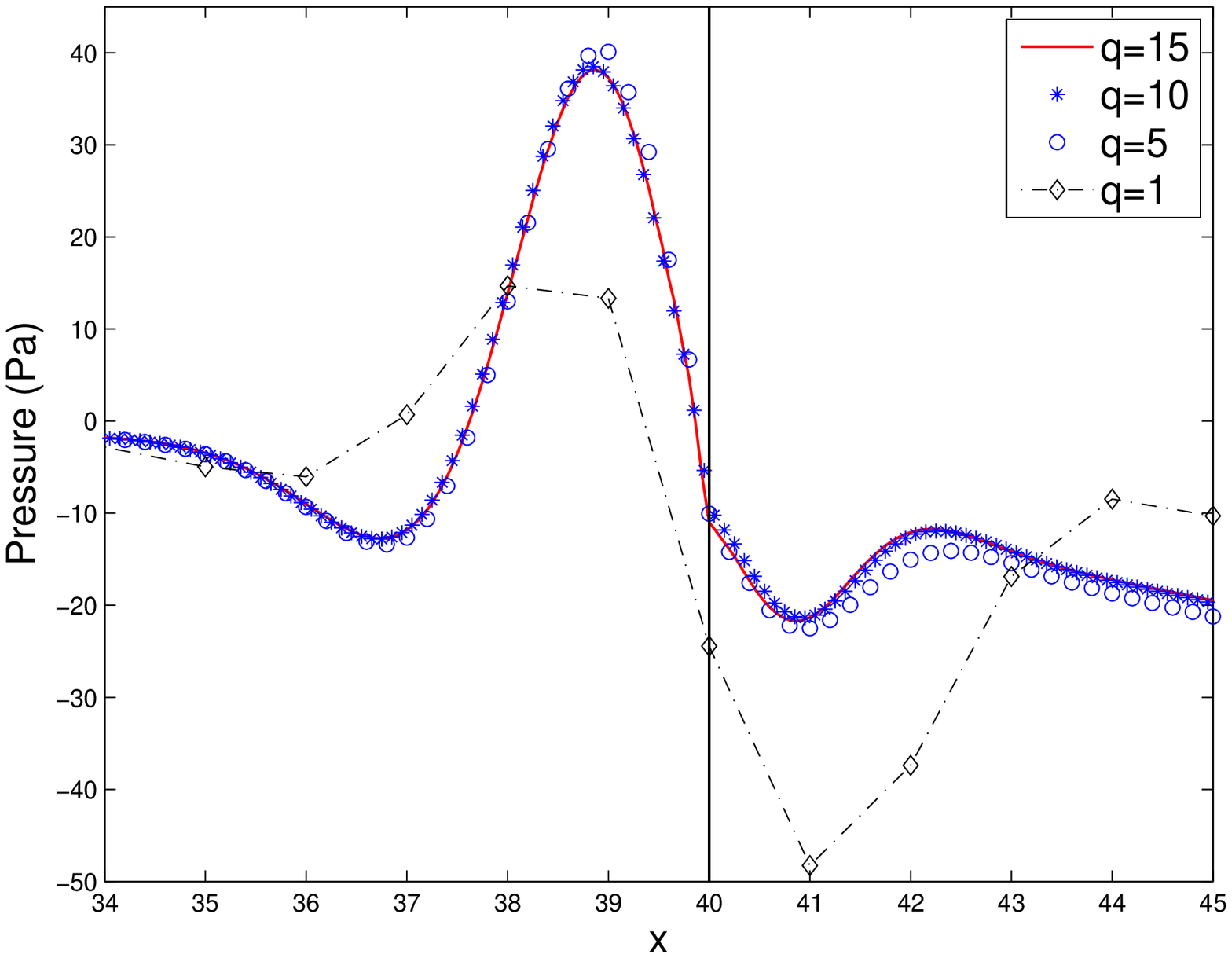}
\end{tabular}
\end{center}
\caption{Test 4, viscous fluids. Snapshot of $p$ (a) and $\sigma_{11}$ (b) after 180 iterations, with a refinement factor $q=15$ and a third order interface method ($r=3$). Snapshot of $p$ around the scatterer (c), (d). Zoom of $p$ on the line $y=0$ around the reflected fast compressional wave (e) and around the diffusive slow wave around the right interface (f).}
\label{FigCercleVisq}
\end{figure}

The same configuration is now considered by taking into account the viscosity of the saturating fluids (see table \ref{TabParametres}). Based on the dispersion analysis performed in section \ref{SecPhysDispersion}, the values of the phase velocity at $f_0=40$ Hz give a refinement factor 
$q \approx 22$ in the grid ${\cal G}_1$ to satisfy our refinement criterion  (\ref{factq}). Snapshots of $p$ and $\sigma_{11}$ after 180 iterations are given on figure \ref{FigCercleVisq}, showing different structures than in the inviscid case presented in figure \ref{FigCercleNonvisq}. The diffracted fast compressional and shear waves propagate, while the slow compressional wave remains localized around the scatterer. 
 
The spatial structure of this diffusive wave is clearly visible in figure \ref{FigCercleVisq}-(c),(d). Figure \ref{FigCercleVisq}-(e) focuses on the pressure component of the slow wave around the interface. It is observed that the numerical method applied without mesh refinement nor interface treatment gives inaccurate results about this static wave. This is also the case of the propagative reflected compressional wave shown in figure \ref{FigCercleVisq}-(f). The computed fields converge when the refinement factor in ${\cal G}_1$ increases. With refinement factors $q$ larger than 15, the results are mostly indistinguishable, and hence they are not represented here. With $q=15$, the refined grid ${\cal G}_1$ contains $1365^2$ discretization points and involves 19184 irregular points in the immersed interface algorithm.

\begin{figure}[htbp]
\begin{center}
\includegraphics[scale=0.42]{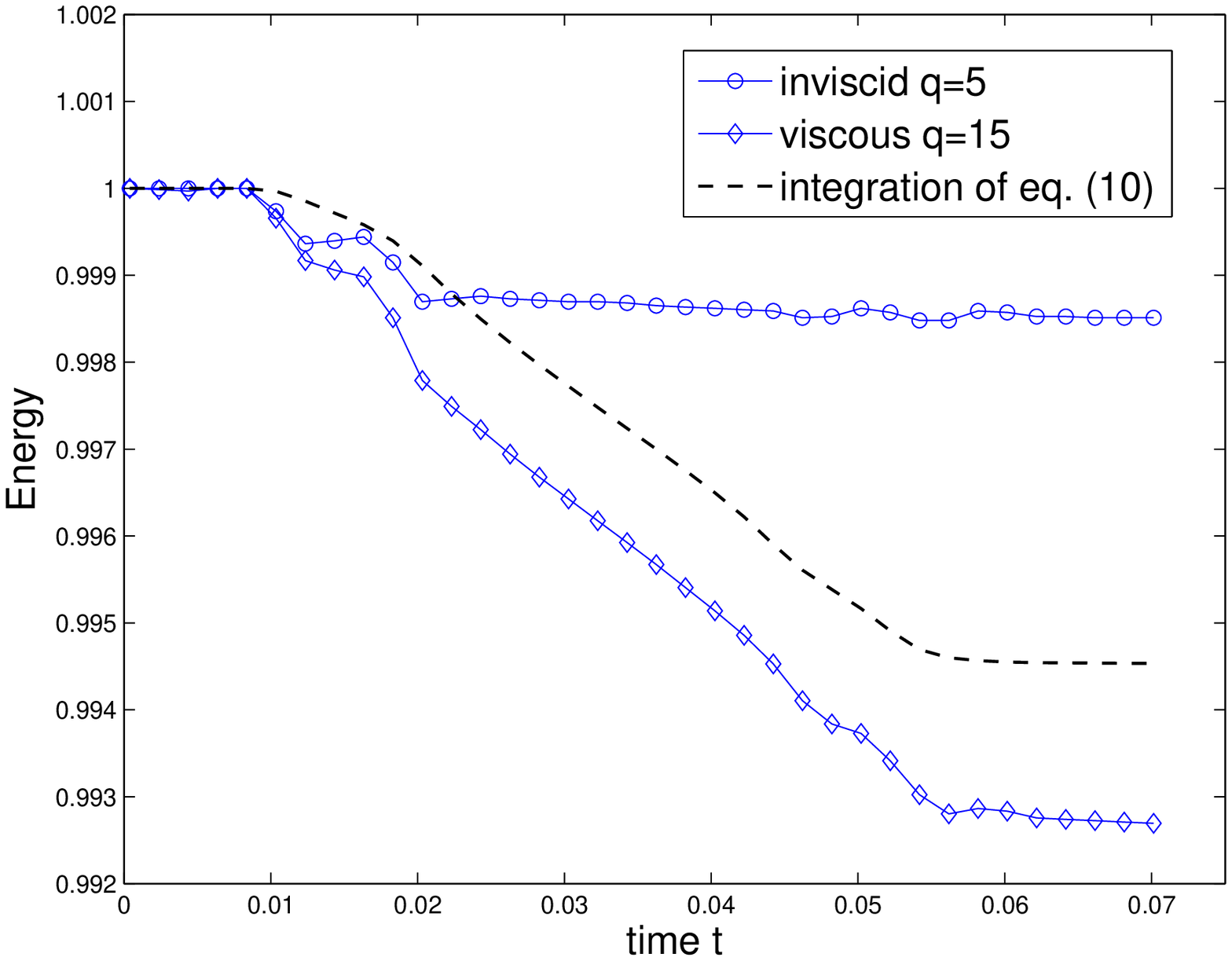}
\end{center}
\caption{Test 4. Time variation of the computed mechanical energy $E(t)/E(0)$ defined by relation (\ref{DefEnergie})  for inviscid and viscous case. Numerical integration of relation
(\ref{VariationNRJ}) is also represented by the dashed curve.}
\label{figvarenergy}
\end{figure}

The time evolution of the mechanical energy (\ref{DefEnergie}) is illustrated in figure \ref{figvarenergy}. The times $t_i=0.008$ s and $t_f \simeq 0.066$ s correspond respectively to the instants when the plane wave begins to interact with the scatterer and when it has crossed the scatterer entirely. In the case of inviscid saturating fluids (blue circles on figure \ref{figvarenergy}), the energy is almost conserved, as expected, and only a decrease of $0.15 \,\%$ is observed. In the viscous case (blue diamond on figure \ref{figvarenergy}), the energy decreases when the plane wave interacts with the scatterer, and remains constant otherwise. This behavior is logical, since the rate of decrease of energy in (\ref{VariationNRJ}) is governed by ${\bf w}$. In homogeneous medium, ${\bf w}$ is extremely small compared with other fields, and hence $\frac{d\,E}{d\,t}\approx 0$. On the other hand, during the interaction between the wave and the scatterer, the filtration velocity of the slow wave generated at the interface has an important amplitude. The theoretical decrease of energy, obtained by a numerical integration of equation (\ref{VariationNRJ}), is shown using a dotted line in figure  \ref{figvarenergy} and is very close to the observed decrease of $E$. This confirms that our numerical strategy accurately models the dissipation of mechanical energy.

%------------------------------------------------------------------------------------------

\subsection{Test 5: multiple scatterers}\label{SecNumTest5}
\begin{figure}[htbp]
\begin{center}
\begin{tabular}{cc}
$t_1$ & $t_1$\\
\includegraphics[scale=0.42]{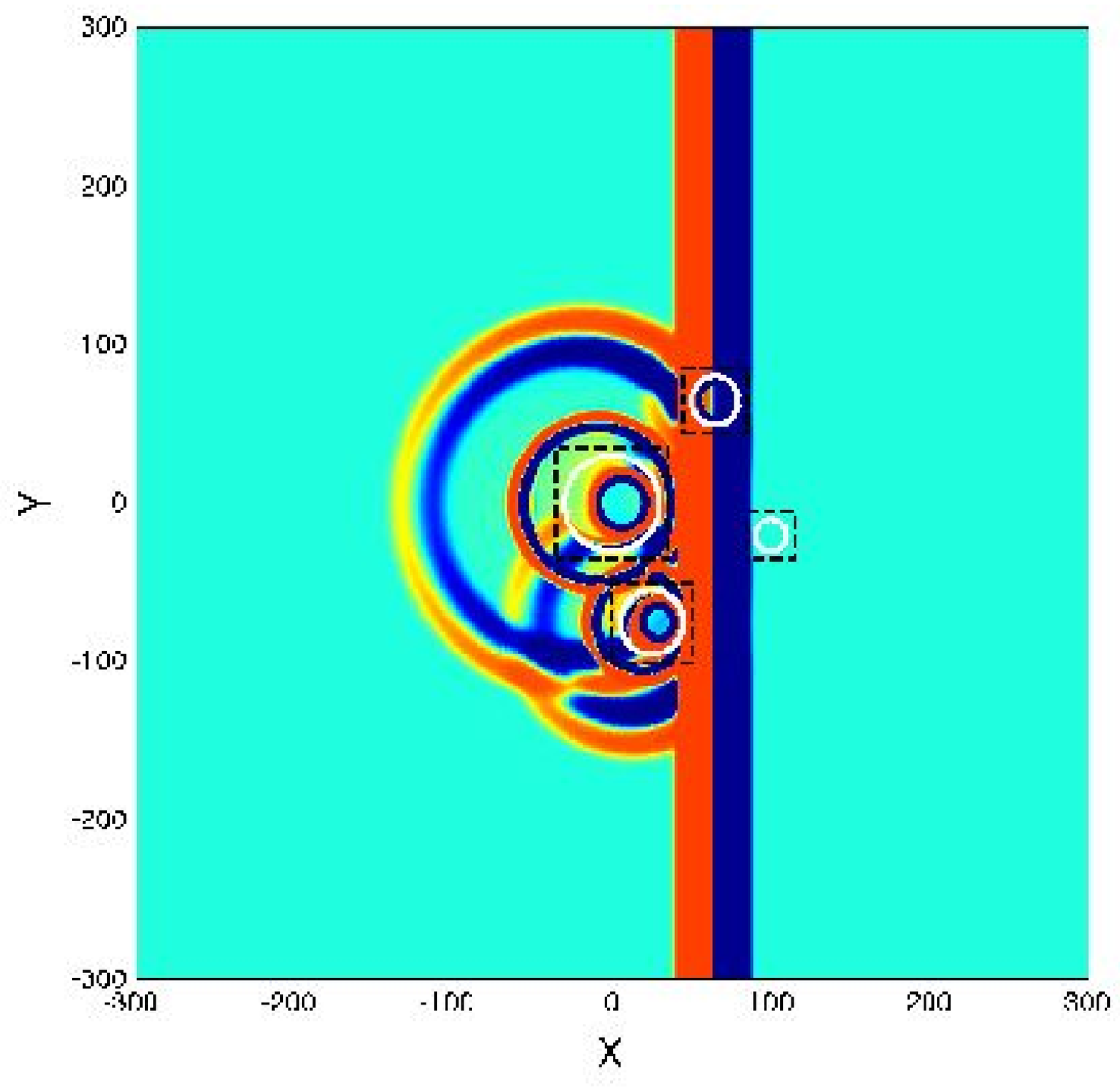}&
\includegraphics[scale=0.42]{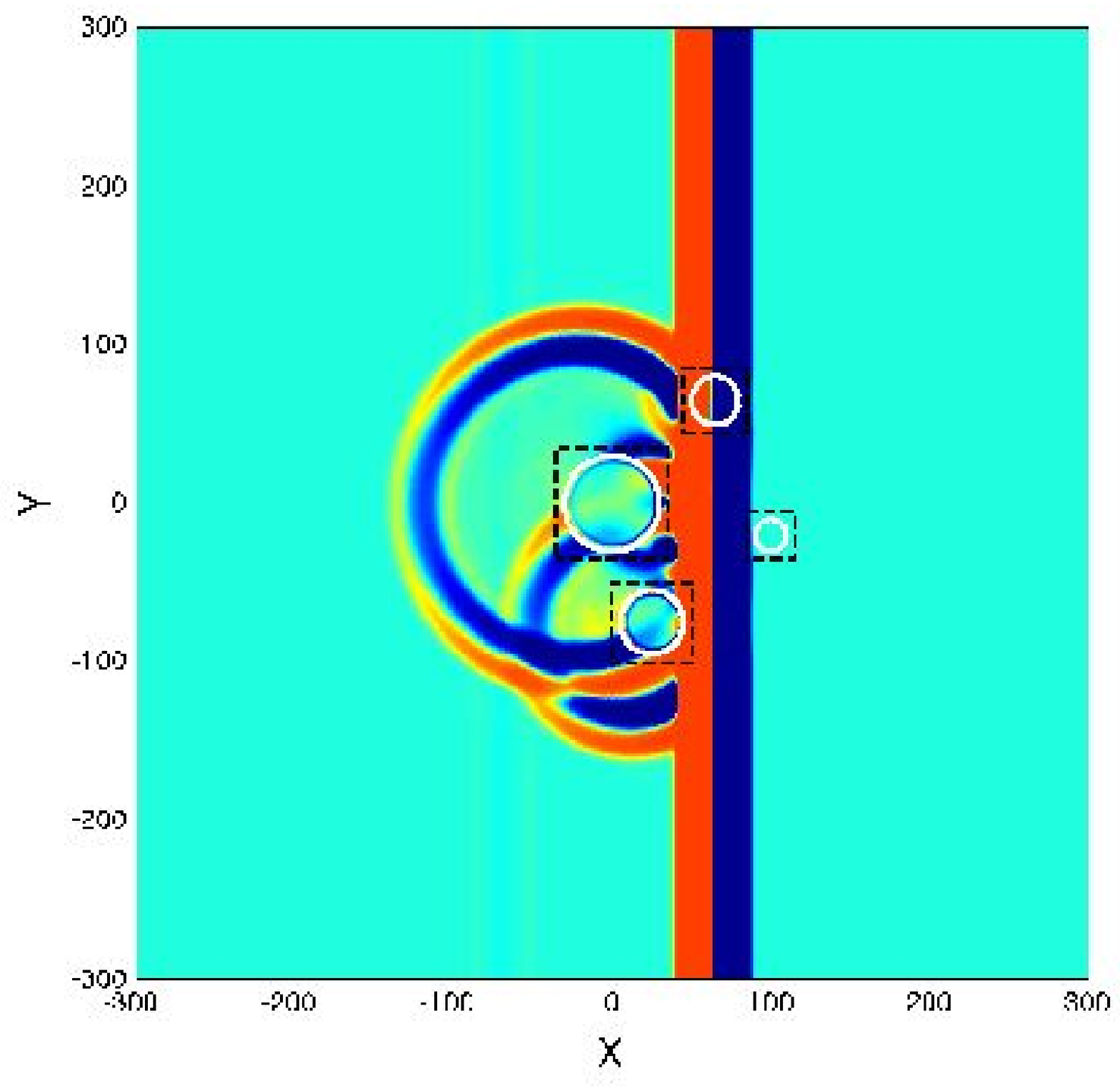}\\
$t_2$ & $t_2$\\
\includegraphics[scale=0.42]{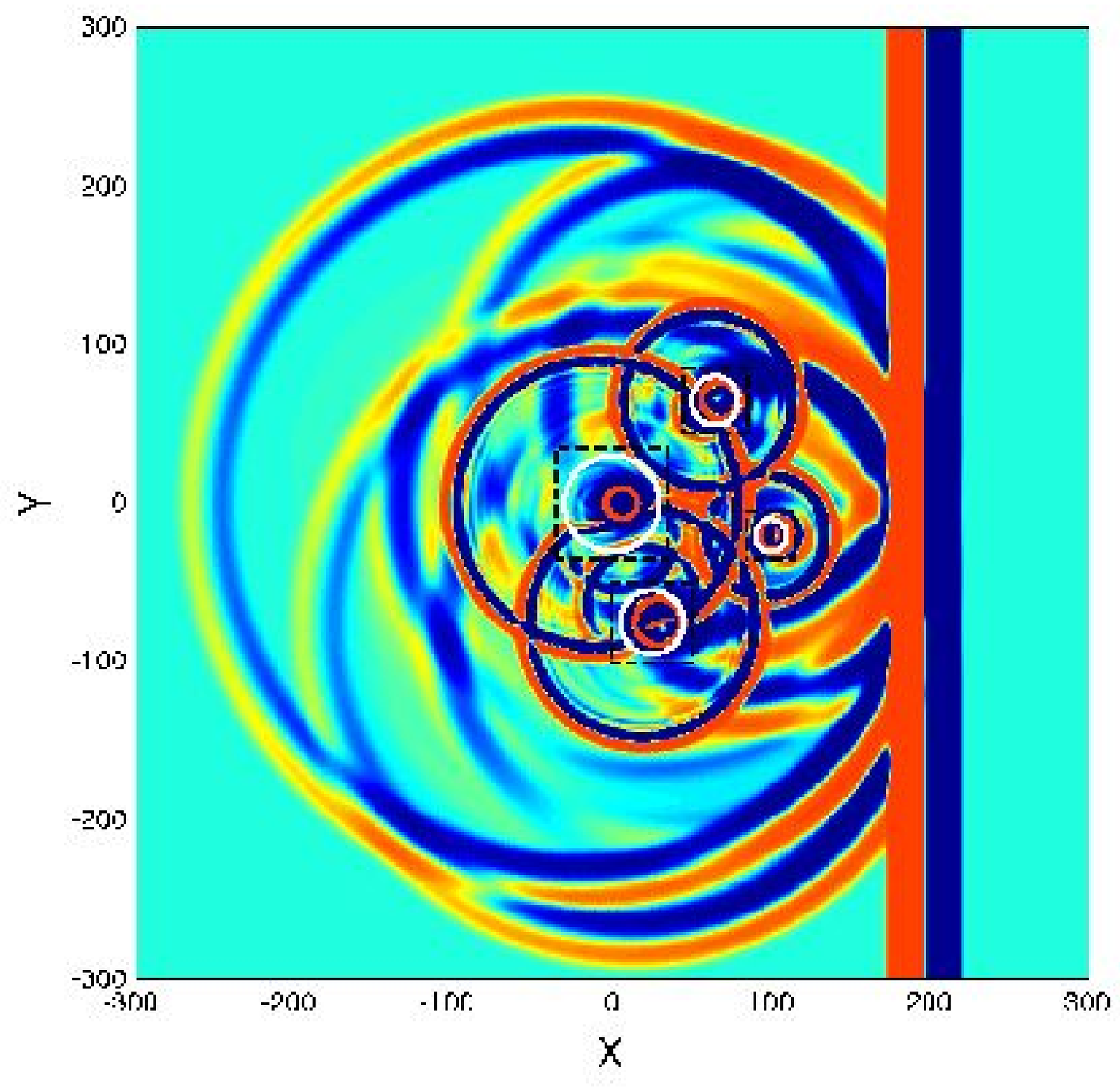}&
\includegraphics[scale=0.42]{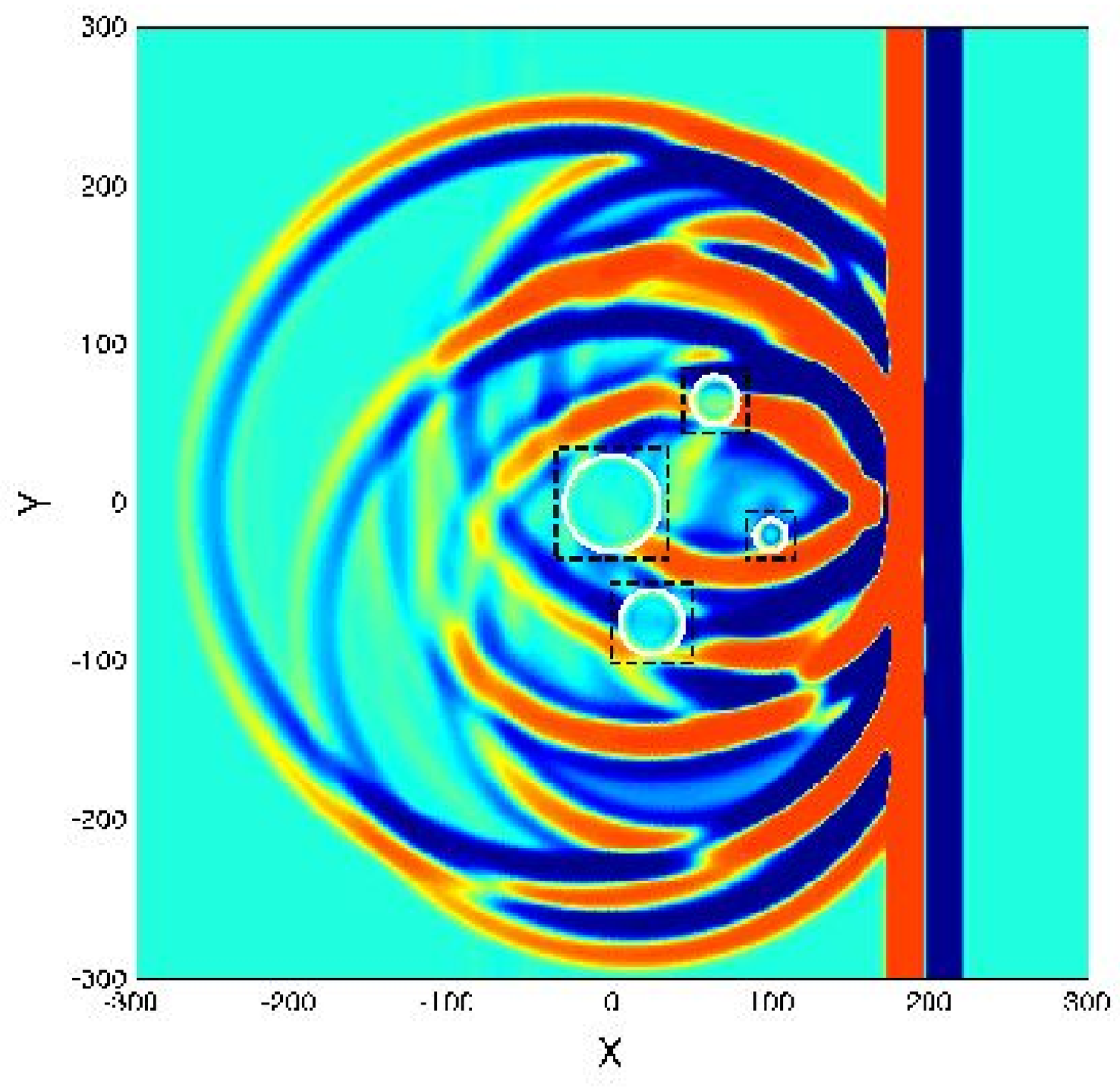}
\end{tabular}
\end{center}
\caption{Test 5. Snapshots of $p$ at two different times $t_1$ ant $t_2$, corresponding to 140 and 280 iterations. Left column: inviscid fluids; right column: viscous saturating fluids.}
\label{MultiCercles}
\end{figure}

In the last test, the ability of the proposed numerical strategy to handle complex geometries is illustrated. Four cylindrical scatterers of medium $\Omega_1$ are inserted in the matrix $\Omega_0$.  Each cylinder is surrounded by a refined grid. The refinement factor is $q=5$ in the inviscid case, and $q=10$ in the case of viscous saturating fluids. The pressure field is represented in figure \ref{MultiCercles}. The behavior of the fast compressional waves is qualitatively the same in both cases, unlike the slow waves:
\begin{itemize}
\item in the inviscid case (figure \ref{MultiCercles}, left column), the slow waves propagate and interact with the other scatterers, which generates new sets of reflected-transmitted waves. Accurate computation of these successive interactions is obtained thanks to the combination of the local refinement procedure and the immersed interface method;
\item in the viscous case (figure \ref{MultiCercles}, right column), the slow waves remain localized around the interfaces, and hence they do not participate directly to the scattering process. At time $t_2$, they have mostly disappeared due to their physically diffusive behavior. As indicated by test 4, a fine modeling of these waves is nevertheless necessary in order to compute accurately the propagative diffracted waves. 
\end{itemize}

%------------------------------------------------------------------------------------------
%------------------------------------------------------------------------------------------

\section{Conclusion}\label{SecConclusion}

Numerical modeling of 2-D poroelastic waves was addressed numerically in the time-domain. The evolution equations were issued from Biot's theory, assuming viscous efforts of Poiseuille type, which is valid essentially in the low-frequency range. Three numerical tools were combined: a fourth-order scheme with time-splitting, a space-time mesh refinement, and an immersed interface method. The resulting method provides highly accurate simulations of wave propagation in realistic configurations. Numerical experiments have indicated that even if the slow waves remain localized around sources or interfaces, the manner in which they are computed influences the propagative waves, that are measured in practical applications. Academic cylindrical geometries have been considered here; however, more complex smooth geometries can be handled without any restrictions, for example cubic splines. Various extensions of the present work are suggested:
\begin{itemize}
\item The numerical methods presented here make it possible to simulate relevant  physical experiments. We have especially in mind the modeling of multiple scattering in random media. Based on simulated data, the properties of the {\it effective medium} equivalent to the disordered medium under study can be deduced \cite{CHEKROUN09}. This numerical approach compares advantageously with the methods usually followed by physicists: real experiments are expensive, and analytical methods are restricted to very small concentrations of scatterers \cite{LUPPE08}. The example given in section \ref{SecNumTest5} is obviously preliminary: interaction of a plane wave with hundreds of scatterers needs to be addressed, which requires the parallelization of the algorithms. 
\item Poroelastic media in perfectly bonded contact are considered here. More realistic conditions can be studied and properly enforced by the immersed interface method, for instance sliding and imperfect bonding \cite{LOMBARD06}, or imperfect hydraulic contact \cite{BOURBIE87}. Comparison of numerical simulation with experimental results could help to validate or improve the models of contact, which constitutes a current issue in poroelasticity. Generalizing our approach to the interface between a poroelastic medium and a fluid is another direction of work with large applications, for instance in biomechanics. 
\item Incorporation of attenuation in the elastic skeleton is required to model the real processes of dissipation \cite{CARCIONE07}. For this purpose, memory variables need to be introduced in the evolution equations (\ref{LC}).  
\item Lastly, the numerical modeling of the transient Biot equations in the full range of validity of poroelasticity constitutes a natural extension of the present work. At frequencies greater than $f_c$ in (\ref{Fc}), a correction of the viscosity proportional with the square root of frequency needs to be introduced, as described e.g. by the JKD model \cite{JKD87,DUPUY11}. In the time domain, the new evolution equations  involve fractional derivatives of order $1/2$, whose efficient numerical evaluation is a major challenge \cite{LU05,MATIGNON07,MASSON10}. 
\end{itemize} 

%------------------------------------------------------------------------------------------
%------------------------------------------------------------------------------------------

\appendix

\section{Matrices involved in system (\ref{SystHyp})} \label{AnnexeSystem}

Based on (\ref{LC}) and (\ref{DefU}), the matrices in (\ref{SystHyp}) write
\begin{equation}
\begin{array}{l}
%\label{matA}
{\bf A}=\left(\begin{array}{c|c}  
  0 & \begin{array}{cccc}  -\rho_w/\chi & 0 & 0& -\rho_f/\chi \\
                                             0 &  -\rho_w/\chi &0 &0 \\
                                              \rho_f/\chi & 0 &0 & \rho/\chi \\
                                              0 & \rho_f/\chi &0 &0 \\
                                              \end{array}  \\
\hline                                            
  \begin{array}{cccc}  -(\lambda_f+2\,\mu)& 0 & -\beta\, m& 0 \\
                                             0 &-\mu &0 &0 \\
                                              -\lambda_f & 0 &-\beta\,m &0 \\
                                              \beta\, m & 0 &m &0 \\
                                              \end{array}             & 0\\                       
\end{array} \right),\\
\\
%\end{equation}
%
%\begin{equation}
%\label{matB}
{\bf B}=\left(\begin{array}{c|c}  
  0 & \begin{array}{cccc}  0 &-\rho_w/\chi & 0 & 0\\
                                             0 &  0 & -\rho_w/\chi &-\rho_f/\chi \\
                                             0 & \rho_f/\chi &0 &0 \\
                                              0& 0 & \rho_f/\chi& \rho/\chi \\
                                              
                                              \end{array}  \\
\hline                                            
  \begin{array}{cccc}         0& -\lambda_f& 0& -\beta\, m\\
                                             -\mu&0 &0 &0 \\
                                              0& -(\lambda_f+2\,\mu)& 0 & -\beta \,m \\
                                               0&  \beta \,m & 0 &m  \\
                                              \end{array}             & 0\\                       
\end{array} \right),\\
\\
%\end{equation}
%
%\begin{equation}
%\label{matS}
\displaystyle
{\bf S}=\frac{\eta}{\kappa} \left(\begin{array}{c|c}  
 \begin{array}{cccc}  0 &0 & -\rho_f/\chi & 0\\
                                             0 &  0 &0 &-\rho_f/\chi \\
                                             0 &0&\rho/\chi &0 \\
                                              0& 0 & 0& \rho/\chi \\                              
                                              \end{array}  & 0 \\
\hline                                  
0&0 \\          
\end{array} \right).
\end{array}
\end{equation}
 
%------------------------------------------------------------------------------------------

\section{Algorithm for the Beltrami-Michell conditions}\label{AnnexeBeltrami}

The following algorithm is proposed to compute the non-zero components of matrices ${\bf G}_k^r$ ($k=0,1$) involved in the Step 2 of the immersed interface method (see section \ref{SecEsim}):
\begin{equation}
\left|
\begin{array}{l}
\alpha=-1,\quad \beta=-1,\\[5pt]
%\\
\mbox{for } \gamma=0,...,r, \mbox{ for } \delta=0,...,\gamma\\[5pt]
%\\
\hspace{1cm} \mbox{if } \delta=0 \mbox{ then for }\varepsilon=1,...,8\\[5pt]
%\\
\hspace{2cm} \alpha=\alpha+1,\quad  \beta=\beta+1,\,\hspace{1cm}{\bf G}_k^r[\alpha,\beta]=1\\[5pt]
%\\
\hspace{1cm} \mbox{if } \gamma \neq 0 \mbox{ and } \delta \neq 0 \mbox{ and } \gamma \neq \delta \mbox{ then}\\[5pt]
%\\
\hspace{2cm} \mbox{if } \gamma = 2 \mbox{ then }\nu=0,\,\eta=0,\\[5pt]
%\\
\hspace{2cm} \mbox{else if } \delta = 1 \mbox{ then }\nu=0,\,\eta=1,\\[5pt]
%\\
\hspace{2cm} \mbox{else if } \delta = \gamma-1 \mbox{ then }\nu=1,\,\eta=0,\\[5pt]
%\\
\hspace{2cm} \mbox{else } \nu=1,\,\eta=1,\\[5pt]
%\\
\hspace{2cm} \mbox{for }\varepsilon=1,...,5\\[5pt]
%\\
\hspace{3cm} \alpha=\alpha+1,\quad  \beta=\beta+1,\,\hspace{0cm}{\bf G}_k^r[\alpha,\beta]=1\\[5pt]
%\\
\hspace{2cm} \alpha=\alpha+1,\quad \beta=\beta-8+\nu,\hspace{0.4cm} {\bf G}_k^r[\alpha,\beta]=\theta_0\\[5pt]
%\\
\hspace{4.4cm} \beta=\beta+2-\nu,  \hspace{0.4cm} {\bf G}_k^r[\alpha,\beta]=\theta_1\\[5pt]
%\\
\hspace{4.4cm} \beta=\beta+1,      \hspace{1.15cm} {\bf G}_k^r[\alpha,\beta]=\theta_2\\[5pt]
\hspace{4.4cm} \beta=\beta+12,     \hspace{0.95cm} {\bf G}_k^r[\alpha,\beta]=\theta_1\\[5pt]
%\\
\hspace{4.4cm} \beta=\beta+2-\eta, \hspace{0.42cm} {\bf G}_k^r[\alpha,\beta]=\theta_0\\[5pt]
%\\
\hspace{4.4cm} \beta=\beta+1,      \hspace{1.15cm} {\bf G}_k^r[\alpha,\beta]=\theta_2\\[5pt]
\hspace{2cm} \alpha=\alpha+1,\quad \beta=\beta-9+\eta, \hspace{0.4cm} {\bf G}_k^r[\alpha,\beta]=1\\[5pt]
%\\
\hspace{2cm} \alpha=\alpha+1,\quad \beta=\beta+1,\hspace{1.15cm} {\bf G}_k^r[\alpha,\beta]=1\\[5pt]
%\\
\hspace{1cm} \mbox{if } \gamma \neq 0 \mbox{ and } \gamma = \delta \mbox{ then for }\varepsilon=1,...,8\\[5pt]
%\\
\hspace{2cm} \alpha=\alpha+1,\quad  \beta=\beta+1,\hspace{1.15cm} {\bf G}_k^r[\alpha,\beta]=1.
\end{array}
\right.
\label{AlgoG}
\end{equation}

%------------------------------------------------------------------------------------------

\section{Implementation of sources}\label{SecSources}

Two sources are considered. The first one involves a plane right-going fast compressional wave, whose wavevector ${\bf k}$ makes an angle $\theta$ with the horizontal $x$-axis. Its time evolution is 
\begin{equation}
h(t)=
\left\{
\begin{array}{l}
\displaystyle
\displaystyle \sum_{m=1}^4 a_m\,\sin(\beta_m\,\omega_0\,t)\quad \mbox{ if  }\, 0<t<\frac{\textstyle 1}{\textstyle f_0},\\
[12pt]
0 \,\mbox{ otherwise}, 
\end{array}
\right.
\label{JKPS}
\end{equation}
where $\beta_m=2^{m-1}$, $\omega_0=2\pi\,f_0$; the coefficients $a_m$ are: $a_1=1$, $a_2=-21/32$, $a_3=63/768$, $a_4=-1/512$, ensuring $C^6$ smoothness. The support of the incident plane wave lies initially in $\Omega_0$. If $\eta\neq 0$, this wave is slightly dispersive and its time-domain expression follows from a Fourier synthesis:
\begin{equation}
{\hat {\bf U}}({\bf r},\,\omega)=\gamma
\left(
\begin{array}{c}
-\cos \theta\\
[6pt]
-\sin \theta\\
[6pt]
\phi\,(1-Y_{pf})\,\cos \theta\\
[6pt]
\phi\,(1-Y_{pf})\,\sin \theta\\
[6pt]
\displaystyle
\frac{\textstyle k_{pf}}{\textstyle \omega}\,\left(\lambda_f+2\,\mu\,\cos^2 \theta+\beta\,m\,\phi\,(Y_{pf}-1)\right)\\
[6pt]
\displaystyle
\frac{\textstyle k_{pf}}{\textstyle \omega}\,2\,\mu\,\sin \theta\,\cos\,\theta\\
[6pt]
\displaystyle
\frac{\textstyle k_{pf}}{\textstyle \omega}\,\left(\lambda_f+2\,\mu\,\sin^2 \theta+\beta\,m\,\phi\,(Y_{pf}-1)\right)\\
[6pt]
\displaystyle
-\frac{\textstyle k_{pf}}{\textstyle \omega}\,m\,\left(\beta+\phi\,(Y_{pf}-1)\right)
\end{array}
\right)\,e^{i\,\left(\omega\,t-{\bf k}_{pf}.{\bf r}\right)}\,{\hat h}(\omega),
\label{OndePlaneFourier}
\end{equation}
where $\gamma$ is an amplitude factor, $k_{pf}$ is the wavenumber (\ref{DispersionP}) and
\begin{equation}
Y_{pf}(\omega)=\frac{\textstyle \left((1-\phi)\,\rho_s+\rho_f\,\beta\,(a-1)\right)\,\omega^2-\left(\lambda_f+2\,\mu-m\,\beta^2\right)\,k_{pf}^2+i\,\omega\,\phi\,\beta\,\frac{\textstyle \eta}{\textstyle \kappa}}{\textstyle \rho_f\,(a\,\beta-\phi)\,\omega^2-i\,\omega\,\phi\,\beta\,\frac{\textstyle \eta}{\textstyle \kappa}}.
\label{Yp1}
\end{equation}
When $\eta=0$, $k_{pf}$ depends linearly on $\omega$, and consequently $Y_{pf}$ and the vector column in (\ref{OndePlaneFourier}) no more involve $\omega$: a straightforward time-domain expression of the incident plane wave can be obtained.

As a second source, we also implement force densities acting on $\sigma_{12}$ in (\ref{LC}). The only non-null component in (\ref{DefU}) is
\begin{equation}
f_{\sigma_{12}}=g(x,y)\,h(t)
\label{SourcePonctuel}
\end{equation}
where $g$ is a truncated gaussian centered at point $(x_s,\,y_s)$: 
\begin{equation}
g(x,y)=
\left\{
\begin{array}{l}
\displaystyle
\displaystyle \gamma\,e^{-\left(\frac{r}{\zeta}\right)^2}   \quad \mbox{ if  }\, r=\sqrt{(x-x_s)^2+(y-y_s)^2} \leq r_0,\\
[12pt]
0 \,\mbox{ otherwise}, 
\end{array}
\right.
\label{gaussian}
\end{equation}
and $h$ is a Ricker signal:
\begin{equation}
h(t)=
\left\{
\begin{array}{l}
\displaystyle
\displaystyle  \left(2\,\pi^2f_0^2\left(t-\frac{\textstyle 1}{\textstyle f_0}\right)^2-1\right)\  \exp{\left(-2\,\pi^2\,f_0^2\left(t-\frac{\textstyle 1}{\textstyle f_0}\right)^2\right)} \, \mbox{ if  }\, 0<t<\frac{\textstyle 2}{\textstyle f_0},\\
[12pt]
0 \,\mbox{ otherwise}. 
\end{array}
\right.
\label{RICKER}
\end{equation}
This source generates cylindrical waves of all types: fast and slow compressional waves, shear waves. 

%------------------------------------------------------------------------------------------
%------------------------------------------------------------------------------------------

\end{document}